\documentclass[preprint,amsmath,amssymb,prd,superscriptaddress]{revtex4}
\pdfoutput=1 
\usepackage[toc,page]{appendix}
\usepackage{xcolor}
\usepackage[utf8]{inputenc}
\definecolor{darkred}{rgb}{0.4,0.0,0.0}
\definecolor{darkgreen}{rgb}{0.0,0.4,0.0}
\definecolor{darkblue}{rgb}{0.0,0.0,0.4}
\usepackage{graphicx}
\usepackage{pdfpages}
\usepackage{float}
\usepackage{multirow}
\usepackage{array}
\usepackage{amsmath,bm}
\usepackage{booktabs,siunitx}
\usepackage{dcolumn}     % Align table columns on decimal point
\usepackage{multirow}    % Required for multirows
\newcommand{\p}{\partial}
\newcommand{\pslash}{p\kern-1ex /}
\newcommand{\lslash}{l\kern-1ex /}
\newcommand{\kslash}{k\kern-1ex /}
\newcommand{\dslash}{\p\kern-1.2ex /}
\newcommand{\Dslash}{{\cal D}\kern-1.5ex /}
\newcommand{\Tr}{{\rm Tr}}
\newcommand{\tr}{{\rm tr}}
\newcommand{\re}{{\rm Re}}

\def\Id{\mbox{1\hspace{-1.2mm}I} }

\newcommand{\Dodwf}{\mathcal{D}}
\newcommand{\diag}{{\rm diag}}

\newcommand{\bea}{\begin{eqnarray}}
\newcommand{\eea}{\end{eqnarray}}

\newcommand{\nn}{\nonumber\\}

\newcommand{\BAN}{\begin{eqnarray*}}
\newcommand{\EAN}{\end{eqnarray*}}
\def\g5{\gamma_5}
\def\g4{\gamma_4}
\def\g3{\gamma_3}
\def\g2{\gamma_2}
\def\g1{\gamma_1}

\def\u{{\bf u}}
\def\d{{\bf d}}
\def\s{{\bf s}}
\def\c{{\bf c}}
\def\b{{\bf b}}
\def\t{{\bf t}}
\def\q{{\bf q}}

\def\cbar{{\bf \bar c}}
\def\bbar{{\bf \bar b}}

\def\qbar{{\bf \bar q}}

\begin{document}

\newcommand{\NTNU}{Physics Department, National Taiwan Normal University, Taipei, Taiwan~11677, Republic of China}

\newcommand{\ASIOP}{Institute of Physics, Academia Sinica, Taipei, Taiwan~11529, Republic of China}

\newcommand{\NTU}{Physics Department, National Taiwan University, Taipei, Taiwan~10617, Republic of China}

\preprint{NTUTH-20-505B}

\title{Beauty mesons in $N_f=2+1+1+1 $ lattice QCD \\ with exact chiral symmetry}

\author{Ting-Wai~Chiu}
\affiliation{\NTNU}
\affiliation{\ASIOP}
\affiliation{\NTU}

%\collaboration{TWQCD Collaboration}
%\noaffiliation

%\noindent Keywords: Lattice QCD, Domain-Wall/Overlap Fermion, Heavy Quarks, Charmed Mesons, Beauty Mesons 

\begin{abstract}

We present the first study of $N_f=2+1+1+1$ lattice QCD with domain-wall quarks. 
%on the $ 40^3 \times 64 $ lattice. 
The $(\b, \c, \s)$ quarks are physical, 
%which are fixed by the  
%masses of the vector mesons $\Upsilon(9460)$, $ J/\psi(3097), $ and $ \phi(1020) $ respectively, 
while the $(\u, \d) $ quarks are heavier than their physical masses, 
with the pion mass $ \sim 700 $ MeV. 
The gauge ensemble is generated by hybrid Monte Carlo simulation with  
the Wilson gauge action for the gluons, 
%at $ \beta = 6/g_0^2 = 6.70 $,  
and the optimal domain-wall fermion action for the quarks.  
%(with $N_s = 16$, $m_0 = 1.3 $, $\lambda_{max}/\lambda_{min}=6.20/0.05 $). 
Using point-to-point quark propagators, we measure the time-correlation functions 
of quark-antiquark meson interpolators with quark contents 
$ \bbar\b$, $\bbar\c$, $\bbar\s$, and $\cbar\c$, and obtain the masses of the low-lying mesons. 
They are in good agreement with the experimental values, plus some predictions which 
have not been observed in experiments. Moreover, we also determine the masses of $(\b, \c, \s) $ quarks.   

\end{abstract}

\maketitle

\section{Introduction}

In 2007, we performed the first study of treating valence $(\u, \d, \s, \c, \b) $ quarks
as Dirac fermions in quenched lattice QCD with exact chiral symmetry \cite{Chiu:2007km,Chiu:2007bc}.
The low-lying mass spectra of mesons
with quark contents $ \bbar\b $, $ \bbar\c $, $\bbar\s $, and $ \cbar\c $
were determined, together with the pseudoscalar decay constants. 
Some of our results (e.g., the masses of $ \eta_b $ and $h_b$) were theoretical predictions 
at the time of publication, which turn out to be in good agreement with later experimental 
results. This asserts that it is feasible to treat $(\u,\d,\s,\c,\b)$ valence quarks 
as Dirac fermions, in lattice QCD with exact chiral symmetry. 

\begin{figure}[!h]
\begin{center}
\includegraphics*[width=9cm,clip=true]{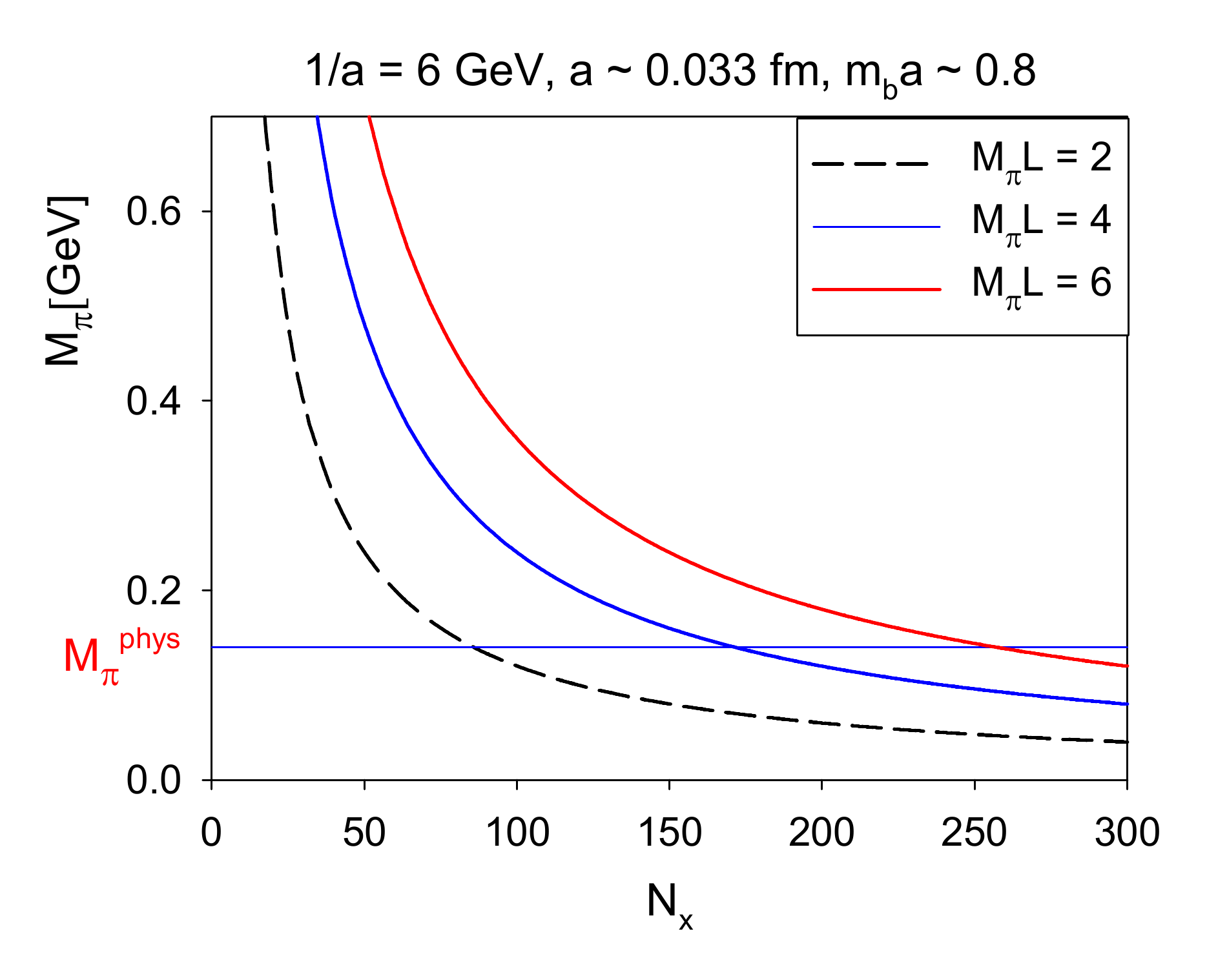}
\caption{The design of lattice QCD with physical $(\u,\d,\s,\c,\b)$ quarks.}
\label{fig:design_LQCD_udscb}
\end{center}
\end{figure}

Now the question is whether one can simulate dynamical $(\u,\d,\s,\c,\b)$ quarks 
in lattice QCD with exact chiral symmetry. This motivates the present study.  
Since the $ \b $ quark is heavy, with mass $ m_b \sim 4500 $ MeV/$c^2$, 
it requires a fine lattice spacing such that the condition $ m_b a < 1 $ is well satisfied 
in order to keep the discretization error under control. 
On the other hand, to keep the finite-volume error of the light hadrons under control, 
the lattice size $ L $ has to be sufficiently large such that $ M_\pi L \gg 1 $. 
These two constraints ($ a \sim 0.033 $~fm and $ M_\pi L \sim 4-6 $)
together give the lattice size $ \sim 170^4 - 260^4 $ (see Fig. \ref{fig:design_LQCD_udscb}), 
which is beyond the capability of the present generation of supercomputers.

Nevertheless, even before the next generation of Exaflop supercomputers 
will be available $\sim 2022$, one may use a smaller lattice to investigate whether the
$(\b,\c,\s)$ quarks with physical masses can be dynamically simulated on the lattice, 
while keeping $\u$ and $\d$ quarks heavier than their physical masses. 
If the pion mass is kept at $ \sim 700 $ MeV/$c^2$, then both constraints 
$ M_\pi L > 4 $ and $ m_b a < 1 $ can be satisfied by the $ 40^3 \times 64 $ lattice. 
For domain-wall fermion with the extent $ N_s = 16 $ in the fifth dimension,
the entire hybrid Monte Carlo (HMC) simulation \cite{Duane:1987de} 
on the $40^3 \times 64 \times 16 $ lattice can be performed  
by one GPU with at least 19 GB device memory, provided that 
the exact one-flavor pseudofermion action (EOFA) for domain-wall fermion \cite{Chen:2014hyy} is used. 
In this study, we use two Nvidia GTX-TITAN-X GPU cards (each of 12 GB device memory) 
for each stream of HMC simulation, with the peer-to-peer communication between 2 GPUs 
through the PCIe bus on the motherboard.   

The outline of this paper is as follows. 
In Sec. II, we recall the basics of lattice QCD with exact chiral symmetry, 
and discuss what is a viable framework to perform HMC simulation of 
lattice QCD with both heavy and light domain-wall quarks such that all topological sectors 
are sampled ergodically and also the chiral symmetry can be peserved to a high precision, i.e., 
the residual mass of any heavy/light quark flavor is negligible in comparison with its bare mass.
In Sec. III, we describe our lattice setup. 
In Sec. IV, we determine the low-lying mass spectra of mesons with valence quark contents 
$\bbar\b$, $\bbar\c$, $\bbar\s$ and $\cbar\c$. In Sec. V, we determine the masses of 
$ (\b, \c, \s) $ quarks. In Sec. VI, we conclude with some remarks.

\section{Simulation of lattice QCD with exact chiral symmetry}    

\subsection{Preliminaries}

Since all quarks in QCD are excitations of Dirac fermion fields,  
it is vital to preserve this essential feature in lattice QCD. 
The most theoretically appealing lattice fermion scheme 
is the domain-wall/overlap fermion \cite{Kaplan:1992bt,Neuberger:1997fp,Narayanan:1994gw}, 
which preserves the exact chiral symmetry at finite lattice spacing, 
thus provides a proper formulation of QCD on the lattice. 

To implement the exact chiral symmetry on the lattice, 
we use the optimal domain-wall fermion \cite{Chiu:2002ir}, 
of which the lattice fermion operator can be written as   
\BAN
\label{eq:D_odwf}
[\Dodwf(m_q)]_{xx';ss'}(m_q) &=&
  (\omega_s D_w + 1)_{xx'} \delta_{ss'}
 +(\omega_s D_w - 1)_{xx'} L_{ss'},
\EAN
where $ \{ \omega_s, s = 1, \cdots, N_s \} $ are the exact solutions
such that the effective 4-dimensional lattice Dirac operator possesses 
the optimal chiral symmetry for any finite $ N_s $.    
The indices $ x $ and $ x' $ denote the lattice sites on the 4-dimensional lattice,
and $ s $ and $ s' $ the indices in the fifth dimension,
while the Dirac and color indices have been suppressed.
Here $D_w$ is the standard Wilson Dirac operator plus a negative parameter $-m_0 \; (0 < m_0 < 2)$
($m_0$ is usually called the domain-wall height), 
\BAN
(D_w)_{xx'} = (4-m_0) -\frac{1}{2} \sum_{\hat\mu=1}^4 \left[
  (1-\gamma_\mu)U_\mu(x)\delta_{x+\hat{\mu},x'}
 +(1+\gamma_\mu)U^\dagger_\mu(x')\delta_{x-\hat{\mu},x'} \right],
\EAN
where $U_\mu(x)$ denotes the link variable pointing from $ x $ to $ x + \hat\mu $.
The operator $ L $ is independent of the gauge field, and it can be written as
\bea
\label{eq:L}
L = P_+ L_+ + P_- L_-, \quad P_\pm = (1\pm \gamma_5)/2,
\eea
and
\bea
\label{eq:Lpm}
(L_+)_{ss'} = (L_-)_{s's} = \left\{
    \begin{array}{ll}
      - (m_q/m_{PV}) \delta_{N_s,s'}, & s = 1,  \\
      \delta_{s-1,s'}, & 1 < s \leq N_s
    \end{array}\right.,
\eea
where $ m_q $ is the bare quark mass, and $ m_{PV} = 2 m_0 $ is the Pauli-Villars mass for the optimal DWF.
Note that the matrices $ L_{\pm} $ satisfy $ L_\pm^T = L_\mp $, and $ R_5 L_\pm R_5 = L_\mp $,
where $ R_5 $ is the reflection operator in the fifth dimension,
with elements $ (R_5)_{ss'} = \delta_{s',N_s+1-s} $. Thus $ R_5 L_\pm $ is real and symmetric. 

Then the pseudofermion action for the optimal DWF can be written as
\BAN
\label{eq:S_nf1}
S = \phi^\dagger \frac{\Dodwf(m_{PV})}{\Dodwf(m_q)} \phi, \hspace{4mm} m_{PV}=2 m_0,   
\EAN
where $ \phi $ and $ \phi^\dagger $ are complex scalar fields carrying the same quantum numbers
(color, spin) of the fermion fields.
Integrating the pseudofermion fields in the fermionic partition function 
gives the fermion determinant of the effective 4-dimensional lattice Dirac operator $D_{N_s}(m_q)$, 
i.e., 
\BAN
\label{eq:Z_nf1}
\int [d\phi^{\dagger}][d\phi] \exp\left\{ -\phi^\dagger \frac{\Dodwf(m_{PV})}{\Dodwf(m_q)} \phi \right\}
= \det \frac{\Dodwf(m_q)}{\Dodwf(m_{PV})}
= \det  D_{N_s}(m_q), 
\EAN
where  
\BAN
\label{eq:odwf_4d}
\begin{aligned}
D_{N_s}(m_q) &= m_q + \frac{1}{2}(m_{PV}-m_q) [1+ \gamma_5 S_{N_s}(H_w) ], \hspace{2mm} H_w=\gamma_5 D_w  \\
S_{N_s}(H_w) &= \frac{1-\prod_{s=1}^{N_s} T_s}{1 + \prod_{s=1}^{N_s} T_s}, \hspace{4mm}
T_s = \frac{1-\omega_s H_w}{1+ \omega_s H_w}.
\end{aligned}
\EAN
In the limit $ N_s \to \infty $, $ S_{N_s}(H_w) \to H_w/\sqrt{H_w^2} $, and $ D_{N_s}(m_q) $ goes to 
\BAN
\label{eq:Dmq_Hw}
D(m_q) = m_q + \frac{1}{2} \left( m_{PV} - m_q \right) \left[1+ \gamma_5 \frac{H_w}{\sqrt{H_w^2}} \right]. 
\EAN
In the massless limit $ m_q = 0 $, $D(0)$ is equal to the overlap-Dirac operator \cite{Neuberger:1997fp},  
and it satisfies the Ginsparg-Wilson relation \cite{Ginsparg:1981bj}
\bea
\label{eq:GW}
D(0) \gamma_5 + \gamma_5 D(0) = \frac{2}{m_{PV}} D(0) \gamma_5 D(0) \Longleftrightarrow
D^{-1} \gamma_5  +  \gamma_5 D^{-1} = \frac{2}{m_{PV}} \gamma_5 \Id,
\eea
where the chiral symmetry is broken by a contact term, 
i.e., the exact chiral symmetry at finite lattice spacing.  
Note that (\ref{eq:GW}) does not guarantee that any Ginsparg-Wilson Dirac operator $D$
must possess exact zero modes in topologically nontrivial gauge background, 
not to mention to satisfy the Atiyah-Singer index theorem, $ Q_{t} = n_+ - n_- $, 
where $ Q_{t} $ is the topological charge of the gauge background, 
and $ n_\pm $ is the number of exact zero modes of $D$ with $\pm $ chirality.  
For example, the lattice Dirac operator constructed in Ref. \cite{Chiu:2001bg} 
satisfies the Ginsparg-Wilson relation and possesses the correct axial anomaly 
in the continuum limit \cite{Chiu:2001ja}, but its index is always zero in any gauge background. 
So far, the overlap Dirac operator is the only lattice Dirac operator  
to possess topologically exact zero modes satisfying the Atiyah-Singer index theorem 
on a finite lattice. 

However, to perform HMC simulation of lattice QCD 
with the overlap Dirac operator is prohibitively expensive even for a small lattice 
(e.g., $ 16^3 \times 32 $), since it requires to compute the change of the number of 
exact zero modes $ n_\pm $ at each step of the molecular dynamics \cite{Fodor:2003bh}.
Moreover, the discontinuity of the fermion determinant at the topological boundary highly suppresses
the crossing rate between different topological sectors, thus renders HMC failing to sample 
all topological sectors ergodically.
These difficulties can be circumvented by using DWF with finite $ N_s $. 
First, any positive lattice Dirac operator satisfying 
$ \gamma_5 $-Hermiticity ($ \gamma_5 D \gamma_5 = D^\dagger $) 
possesses a positive-definite pseudofermion action, without explicit dependence on $ n_\pm $.
Second, the step function of the fermion determinant at the topological boundary 
can be smoothed out by using DWF with finite $ N_s $ (e.g., $ N_s = 16 $),  
then the HMC on the 5-dimensional lattice can sample all topological sectors ergodically
and also keep the chiral symmetry to a high precision with the optimal DWF \cite{Chiu:2002ir,Chiu:2015sea}. 
This has been demonstrated for $N_f=2$ \cite{Chiu:2011bm}, 
$N_f=1+1$ \cite{Chen:2014hyy}, 
$N_f=2+1+1$ \cite{Chen:2017kxr}, 
and also $N_f=2+1+1$ lattice QCD at the physical point \cite{Chiu:2020ppa}.

\subsection{Domain-wall fermion for heavy and light quarks}

In this subsection, we discuss which variant of DWF is more capable in capturing 
the quantum fluctuations of both heavy and light quarks in lattice QCD. 

Unlike other lattice fermions, DWF has the mass cutoff, i.e., the Pauli-Villars mass $ m_{PV}$, 
and any quark mass has to satisfy the constraint $ m_q \ll m_{PV} $.  
Otherwise, if $ m_q \sim m_{PV} $, then $ \det(m_q)/\det(m_{PV}) \sim 1 $,  
the internal quark loops are highly suppressed, and the quantum fluctuations of the quark field 
become mostly quenched.     
In general, the Pauli-Villars mass is equal to $ m_{PV} a = 2 m_0 (1 - d m_0) $,  
where $ d $ is a parameter depending on the variant of DWF. 
For the Shamir\cite{Shamir:1993zy}/M\"obius\cite{Brower:2004xi} DWF, 
$ d = 1/2 $ and $ m_{PV} a = m_0 (2-m_0) < 1 $, 
since $ m_0 $ has to be greater than 1 ($\sim 1.3-1.8$) 
in order for its effective 4-dimensional Dirac operator to be able 
to detect the topology of a gauge configuration with nonzero topological charge.  
This imposes an upper-bound on the mass of Shamir/M\"obius heavy quark on the lattice,  
which is more severe than the common constraint $ m_q a < 1 $ for all lattice fermions. 
In other words, the Shamir/M\"obius DWF is not well-suited for studying lattice QCD with heavy quarks. 
On the other hand, for the Borici\cite{Borici:1999zw}/Optimal\cite{Chiu:2002ir} DWF, 
$ d=0 $ and $ m_{PV} a = 2 m_0 \gg 1 $, 
thus provides the highest ceiling for accommodating the heavy quarks on the lattice, as well as    
the minimal lattice artifacts due to the mass cutoff. 
This can be seen by comparing the eigenvalues 
of their effective 4D Dirac operators in the limit $N_s \to \infty$, 
which is exactly equal to the overlap Dirac operator with the kernel 
$ H = c H_w (1+ d \gamma_5 H_w)^{-1} $ in the sign function, 
\bea
\label{eq:Dmq_H}
D(m_q) = m_q + \frac{1}{2} (m_{PV} - m_q) \left(1+ \gamma_5  \frac{H}{\sqrt{H^2}} \right), 
\hspace{4mm} m_{PV}= 2 m_0 (1 - d m_0), 
\eea
where $ c=d=1/2 $ for the Shamir/M\"obius DWF, while $ c=1 $ and $ d = 0 $ for the Borici/Optimal DWF. 
The eigenvalues of (\ref{eq:Dmq_H}) are lying on a circle in the complex plane 
with radius $ (m_{PV} - m_q)/2 $, and center at $ m_q + (m_{PV} - m_q)/2 $ on the real axis. 
%\bea
%\label{eq:lambda}
%\lambda(m_q) = m_q + \frac{1}{2} (m_{PV} - m_q) \left(1+ e^{i \theta} \right), 
%\hspace{4mm} m_{PV}= 2 m_0 (1 - d m_0), 
%\eea

\begin{figure*}[!ht]
\begin{center}
\begin{tabular}{@{}c@{}c@{}}
\includegraphics*[width=8.6cm,clip=true]{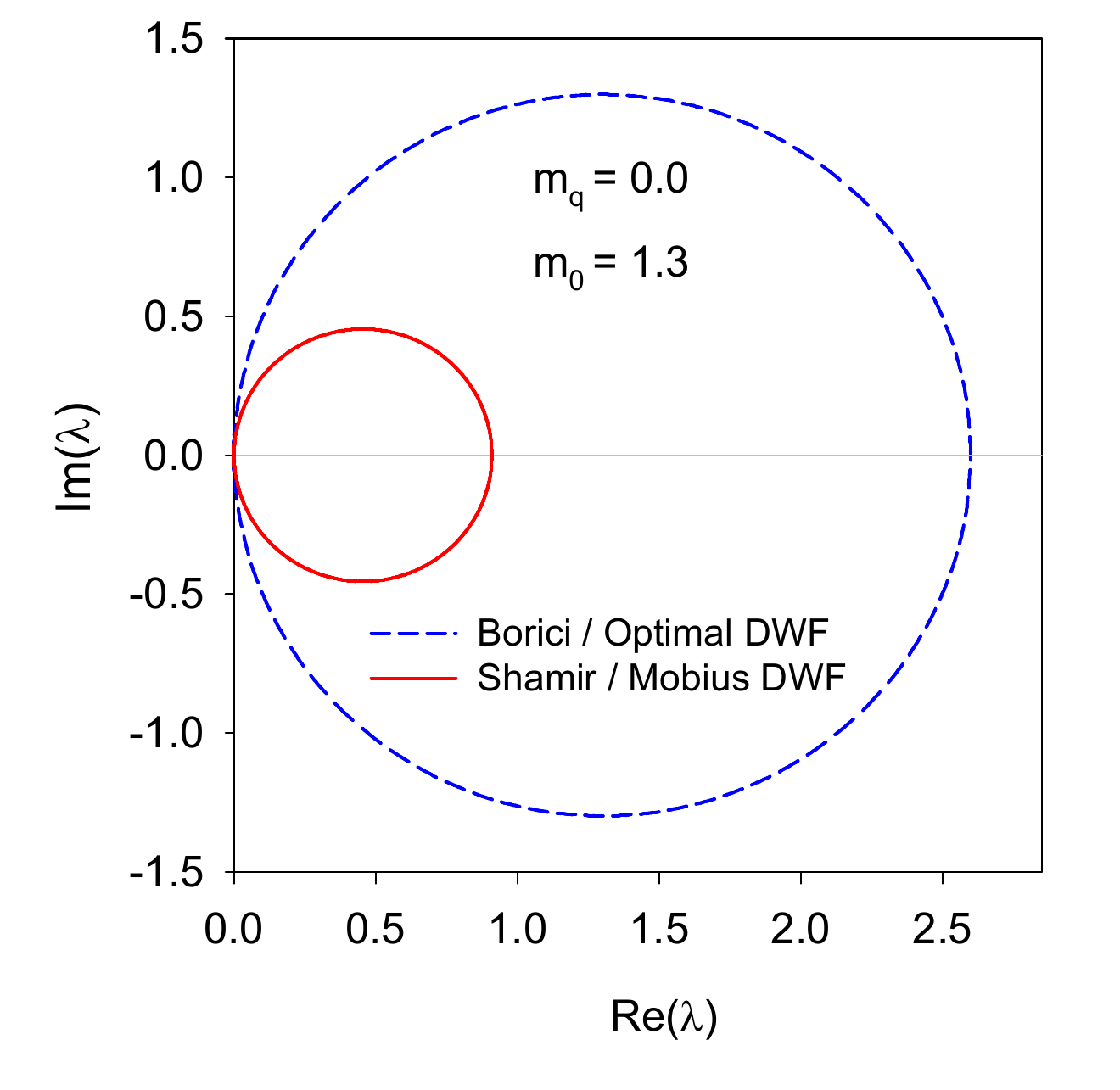}
&
\includegraphics*[width=8.0cm,clip=true]{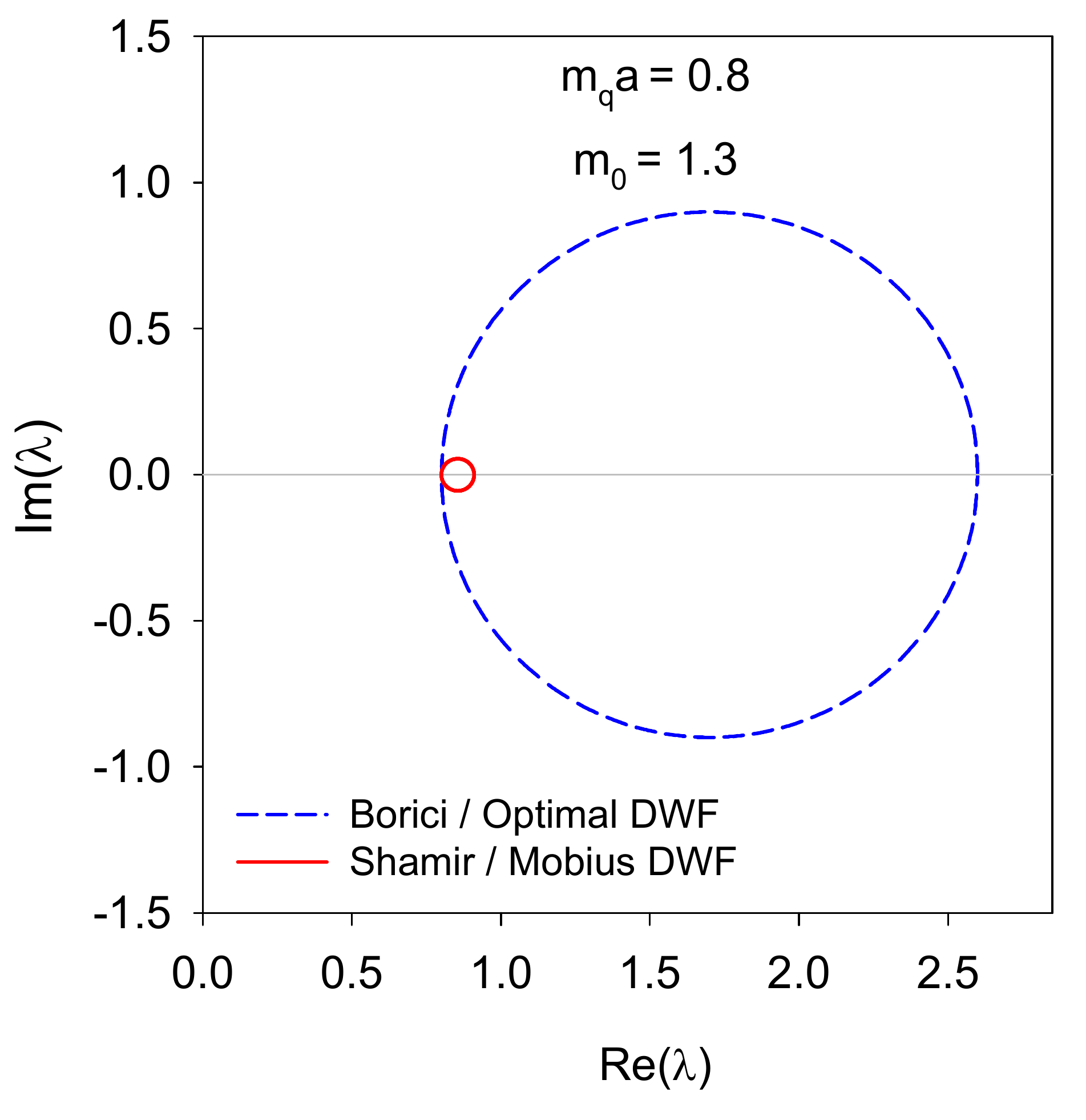}
%\\ (a) & (b)
\end{tabular}
\caption{Comparing the eigenvalue spectra of the effective 4D Dirac operator 
         of the Borici/Optimal DWF and the Shamir/M\"obius DWF for $ m_q = 0 $ (left panel)
         and $ m_q a = 0.8 $ (right panel).  
}
\label{fig:lambda}
\end{center}
\end{figure*}

For example, fixing $ m_0 = 1.3 $, then $ m_{PV} a = 2 m_0 = 2.6 $
for the Borici/Optimal DWF, while $ m_{PV} a = m_0 (2-m_0) = 0.91 $ for the Shamir/M\"obius DWF. 
In Fig. \ref{fig:lambda}, the eigenvalues of (\ref{eq:Dmq_H}) are plotted 
for $ m_q = 0 $ (left panel) and $ m_q a = 0.8 $ (right panel).  
Evidently, for the Shamir/M\"obius DWF, the radius $ (m_{PV}-m_q)/2 $ 
of the eigenvalue circle for a heavy quark with 
$ m_q a = 0.8 $ (right panel) is rather small due to $ (m_{PV} - m_q) a = 0.11 $, 
and it shrinks to zero in the limit $ m_q a \to m_{PV} a = 0.91 $.  
On the other hand, the Borici/Optimal DWF has $ m_{PV} a = 2 m_0 = 2.6 $, 
and $ (m_{PV}-m_q) a > 1 $ for any $ m_q a < 1 $, thus the eigenvalues of $ D(m_q) $ 
are not restricted to a very small circle even for the heavy quark. 
Moreover, in the chiral limit (left panel),  
the radius of the eigenvalue circle for the Borici/Optimal DWF is 
more than 2 times of that of the Shamir/M\"obius DWF. 
This implies that the Borici/Optimal DWF is more capable than the Shamir/M\"obius DWF  
in capturing the short-distance quantum fluctuations of the QCD vacuum, 
for both light and heavy quarks.

\subsection{Zolotarev optimal rational approximation and optimal domain-wall fermion}

For any numerical simulation of lattice QCD with DWF, an important question is 
what is the optimal chiral symmetry for any finite $N_s$ in the fifth dimension, 
in the sense how its effective 4D lattice Dirac operator can be exactly equal to the 
Zolotarev optimal rational approximation of the overlap Dirac operator. 
The exact solution to this problem is given in Ref. \cite{Chiu:2002ir}, 
with the optimal $\{ \omega_s \} $  
\bea
\label{eq:omega}
\omega_s = \frac{1}{\lambda_{min}} \sqrt{ 1 - \kappa'^2 \mbox{sn}^2
                  \left( v_s ; \kappa' \right) }, \hspace{4mm} s = 1, \cdots, N_s,
\eea
where $ \mbox{sn}( v_s; \kappa' ) $ is the Jacobian elliptic function
with argument $ v_s $ (see Eq. (13) in Ref. \cite{Chiu:2002ir}) and
modulus $ \kappa' =\sqrt{ 1 - \lambda_{min}^2 / \lambda_{max}^2 } $.  
%
%Here $ \lambda_{max}^2 $ and $ \lambda_{min}^2 $ are the upper-bound and lower-bound 
%for the eigenvalues of $ H_w^2 $.
%It should be emphasized that $ \lambda_{min} $ and $ \lambda_{max} $ have to be fixed properly for
%each set of simulations, depending on the parameters $ \beta = 6/g^2 $, quark masses, and lattice size,
%such that the desired precision of chiral symmetry can be attained with the minimal cost of the simulation.
%
Then $ S_{N_s}(H_w) $ is exactly equal to the 
Zolotarev optimal rational approximation of $ H_w/\sqrt{H_w^2} $, 
i.e., the approximate sign function $ S_{N_s}(H_w) $  
satisfying the bound $ | 1-S_{N_s}(\lambda) | \le d_Z $ 
for $ \lambda^2 \in [\lambda_{min}^2, \lambda_{max}^2] $,  
where $ d_Z $ is the maximum deviation $ | 1- \sqrt{x} R_Z(x) |_{\rm max} $ of the 
Zolotarev optimal rational polynomial $ R_Z(x) $ of $ 1/\sqrt{x} $ for 
$ x \in [1, \lambda_{max}^2/\lambda_{min}^2] $, with degree $(n-1,n)$ for $ N_s = 2n $.

Nevertheless, the optimal weights $ \{ \omega_s \} $ in (\ref{eq:omega}) do not satisfy 
the $ R_5 $ symmetry ($ \omega_s = \omega_{N_s -s + 1 } $) 
which is required for the exact one-flavor pseudofermion action for DWF \cite{Chen:2014hyy}.
The optimal $ \{ \omega_s \} $ satisfying $ R_5 $ symmetry is obtained in Ref. \cite{Chiu:2015sea}. 
For $N_s = 2n $, the optimal $ \{ \omega_s \} $ satisfying $ R_5 $ symmetry are written as   
\bea
\label{eq:omega_sym_even}
\omega_s=\omega_{N_s+1-s}=\frac{1}{\lambda_{min}} 
                          \sqrt{1-{\kappa'}^2 \mbox{sn}^2 \left( \frac{(2s-1) K'}{N_s}; \kappa' \right)},
\hspace{4mm} s = 1, \cdots, N_s/2,
\eea
where $ \mbox{sn}(u ; \kappa') $ is the Jacobian elliptic function 
with modulus $ \kappa' =\sqrt{ 1 - \lambda_{min}^2 / \lambda_{max}^2 } $,   
and $ K' $ is the complete elliptic function of the first kind with modulus $ \kappa' $.
Then the approximate sign function $ S_{N_s}(H_w) $  
satisfies the bound $ 0 \le 1-S_{N_s}(\lambda) \le 2 d_Z $ 
for $ \lambda^2 \in [\lambda_{min}^2, \lambda_{max}^2] $,  
where $ d_Z $ is defined above. 
Note that $\delta(\lambda)=1-S(\lambda) $ does not satisfy the criterion that the maxima and minima 
of $\delta(\lambda)$ all have the same magnitude but with the opposite sign 
($ \delta_{min} = -\delta_{max} $). However,    
the most salient features of the optimal rational approximation of degree $(m,n) $ are preserved, namely,
the number of alternate maxima and minima is $(m+n+2)$, with $ (n+1) $ maxima and $ (m+1) $ minima,
and all maxima (minima) are equal to $ 2 d_Z $ ($0$).
This can be regarded as the generalized optimal rational approximation (with a constant shift).

\begin{figure*}[!ht]
\begin{center}
\begin{tabular}{@{}c@{}c@{}}
\includegraphics*[width=8.0cm,clip=true]{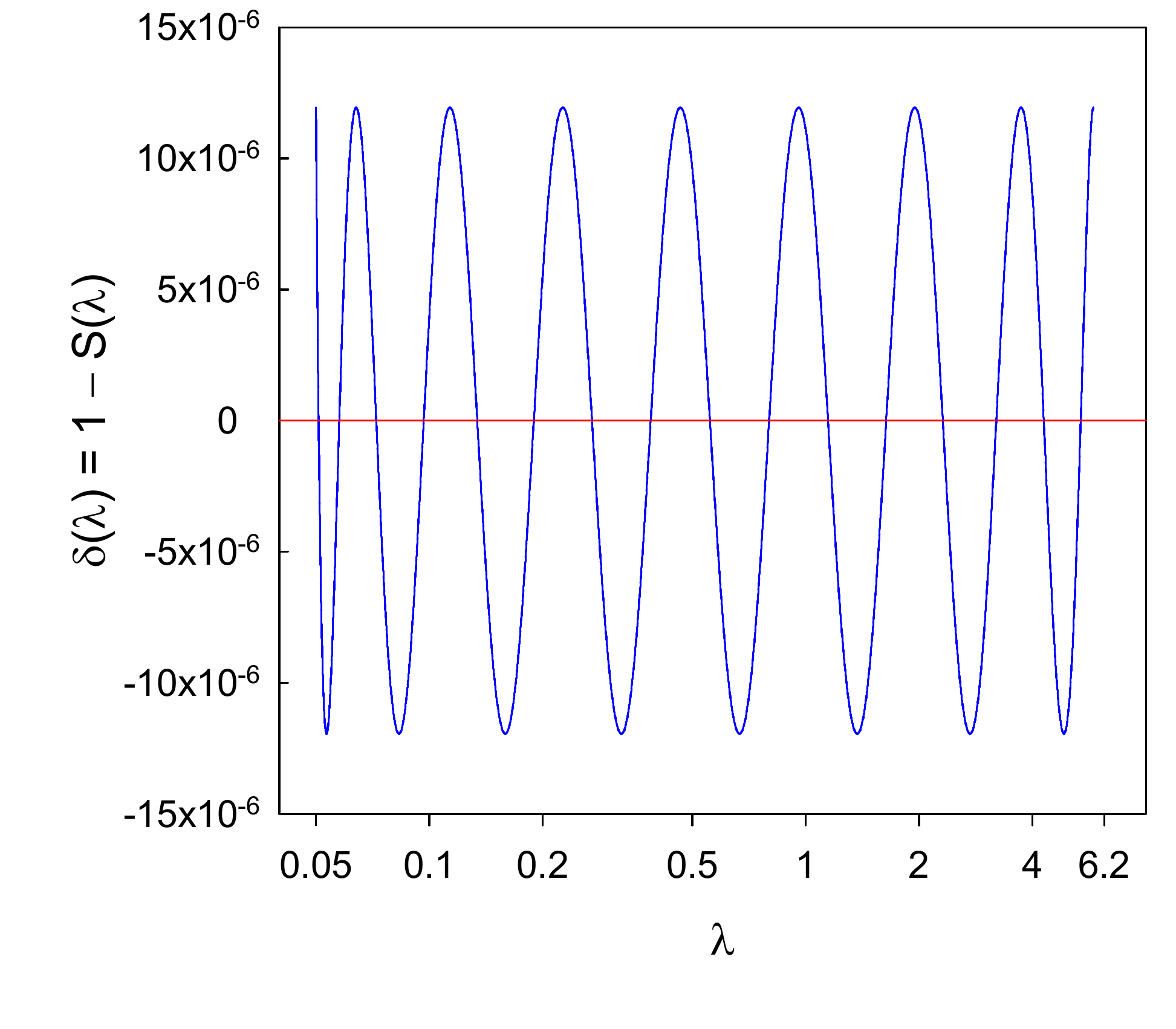}
&
\includegraphics*[width=8.0cm,clip=true]{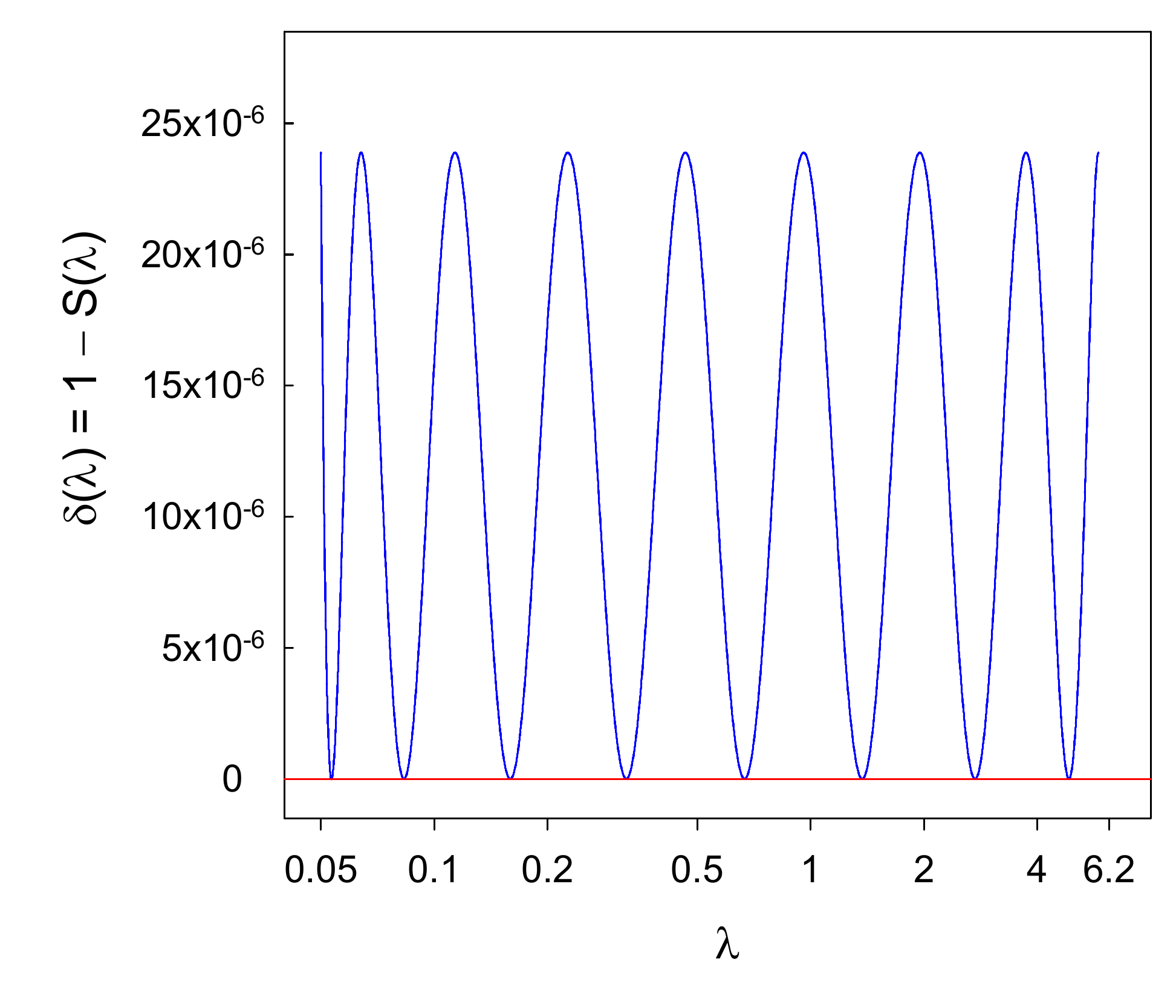}
\\ (a) & (b)
\end{tabular}
\caption{The deviation $ \delta(\lambda) = 1 - S(\lambda) $ of the optimal DWF 
         with $ N_s=2n=16 $ and $ \lambda_{max}/\lambda_{min} = 6.20/0.05 $, 
         for (a) without $ R_5 $ symmetry, and (b) with $R_5$ symmetry.
%         where the weights are computed according to (\ref{eq:omega_sym_even}).
%         Here $ 1-S(x) $ has $ 2n+1 = 17 $ alternate maxima and minima in the interval 
%         $ [\lambda_{min}, \lambda_{max}] $,
%         with $ 8 $ minima and $ 9 $ maxima, and $ |1-S(\lambda)|_{max} = 2 d_Z \simeq $, where $ d_Z $
%         is the maximum deviation $ | 1-\sqrt{x} R_Z^{(7,8)}|_{max} $ of the 
%         Zolotarev optimal rational polynomial.
}
\label{fig:odwf_sym_Ns16}
\end{center}
\end{figure*}

In this study, the parameters for the pseudofermion action are: $ m_0 = 1.3 $, $ N_s = 2n = 16 $, 
$ \lambda_{max}/\lambda_{min} = 6.20/0.05 $, and the optimal weights $ \{ \omega_s, s=1,\cdots,N_s \} $ 
for the 2-flavor parts are obtained with (\ref{eq:omega}),  
while for the one-flavor parts with (\ref{eq:omega_sym_even}). 
In Fig. \ref{fig:odwf_sym_Ns16}, the deviation of the sign function, $\delta(\lambda) = 1-S(\lambda)$,  
is plotted versus $\lambda$, for (a) without the $ R_5 $ symmetry, and (b) with the $ R_5 $ symmetry.
Here $ \delta(\lambda) $ has $ 2n+1=17 $ alternate maxima and minima in the interval 
$ [\lambda_{min}, \lambda_{max}] = [0.05, 6.2] $,
with $ 9 $ maxima and $ 8 $ minima, for (a), satisfying $ -d_Z \le 1-S(\lambda)\le  d_Z $,  
while for (b), $ 0 \le 1-S(\lambda)\le 2 d_Z $,  
where $ d_Z $ is the maximum deviation $ | 1-\sqrt{x} R_Z^{(7,8)}|_{\rm max} $ of the 
Zolotarev optimal rational polynomial.

\section{Generation of the Gauge Ensemble}    

In this section, we give the details of the actions, the algorithms,  
and the parameters to perform the HMC simulations in this study. 
Moreover, for the initial 257 trajectories generated by a single node (with 2 Nvidia GTX-TITAN-X GPU cards), 
the topological charge fluctuation is measured, and the HMC characteristics are presented. 
Details of the lattice setup are given as follows.

\subsection{The actions}
\label{subsection:actions}

In the following, we present the details of the fermion actions and the gauge action 
in our HMC simulations.  

As noted in Ref. \cite{Chen:2017kxr}, for domain-wall fermions (DWF),   
to simulate $ N_f = 2+1+1 $ amounts to simulate $ N_f = 2+2+1 $. 
Similarly, to simulate $ N_f = 2+1+1+1 $ amounts to simulate $ N_f = 2+2+1+1 $, i.e.,  
\bea
&& \left(\frac{\det \Dodwf(m_{u/d})}{\det \Dodwf(m_{PV})}\right)^2   
   \frac{\det \Dodwf(m_s)}{\det \Dodwf(m_{PV})}   
   \frac{\det \Dodwf(m_c)}{\det \Dodwf(m_{PV})}    
   \frac{\det \Dodwf(m_b)}{\det \Dodwf(m_{PV})}   \nn 
&=& \left(\frac{\det \Dodwf(m_{u/d})}{\det \Dodwf(m_{PV})}\right)^2   
    \left( \frac{\det \Dodwf(m_c)}{\det \Dodwf(m_{PV})} \right)^2  
    \frac{\det \Dodwf(m_s)}{\det \Dodwf(m_{c})}       
    \frac{\det \Dodwf(m_b)}{\det \Dodwf(m_{PV})},    
\label{eq:Nf2p1p1p1}       
\eea
where only one of the 6 possible possibilities for $ N_f = 2+2+1+1 $ is written.    
Note that on the rhs of Eq. (\ref{eq:Nf2p1p1p1}),  
the 2-flavor simulation with $ (\det \Dodwf(m_c)/\det \Dodwf(m_{PV}))^2 $ is more efficient 
than its counterpart of one-flavor with $ (\det \Dodwf(m_c)/\det \Dodwf(m_{PV})) $ on the lhs.
Moreover, the one-flavor simulation with $ \det \Dodwf(m_s)/\det \Dodwf(m_{c}) $ on the rhs
is more efficient than the original one with $ \det \Dodwf(m_s)/\det \Dodwf(m_{PV}) $ on the lhs. 
Thus, we perform the HMC simulation with the expression on the rhs of Eq. (\ref{eq:Nf2p1p1p1}). 

For the two-flavor parts, $ \left(\det \Dodwf(m_{u/d})/\det \Dodwf(m_{PV})\right)^2 $  
and $ \left( \det \Dodwf(m_c)/\det \Dodwf(m_{PV}) \right)^2 $,  
we use the $N_f=2$ pseudofermion action which has been using since 2011 \cite{Chiu:2011bm}, 
and it can be written as 
\bea
\label{eq:S_old2F}
S(m_q, m_{PV}) = \phi^\dagger C^\dagger(m_{PV}) \{ C(m_q) C^\dagger(m_q) \}^{-1} C(m_{PV}) \phi, 
\hspace{4mm} m_{PV} = 2 m_0, 
\eea
where
\BAN
\label{eq:C_def}
C(m_q) &=& 1 - M_5(m_q) D_w^{\text{OE}} M_5(m_q) D_w^{\text{EO}}, \nn
\label{eq:M_5}
M_5(m_q) &=& \{ 4 - m_0 + \omega^{-1/2} [1-L(m_q)][(1+L(m_q)]^{-1} \omega^{-1/2} \}^{-1},  
\EAN
and $ L(m_q) $ is defined in (\ref{eq:L}) and (\ref{eq:Lpm}).
Here $\omega \equiv \diag \{\omega_1, \omega_2, \cdots, \omega_{N_s} \}$ 
is a diagonal matrix in the fifth dimension, and 
$ D_w^{\text{EO}/\text{OE}} $ denotes the part of $ D_w $ with gauge links pointing from 
even/odd sites to odd/even sites after even-odd preconditioning on the 4-dimensional lattice.  

For the two-flavor part of $\u$ and $\d$ quarks, we turn on the mass-preconditioning 
\cite{Hasenbusch:2001ne} by introducing an auxiliary heavy fermion field with mass 
$ m_H a = 0.1$. Then the $N_f=2$ pseudofermion action (\ref{eq:S_old2F}) is replaced with 
\BAN
\label{eq:S_ud}
&& S(m_q, m_H) + S(m_H, m_{PV}) \nn
&=& \phi^{\dagger} C(m_H)^{\dagger} \{ C(m_q) C(m_q)^{\dagger} \}^{-1} C(m_H) \phi +
    \phi_H^{\dagger} C^{\dagger}(m_{PV}) \{ C(m_H) C(m_H)^{\dagger} \}^{-1} C(m_{PV}) \phi_H, \hspace{6mm}
\EAN
which gives the partition function (fermion determinant) exactly the same as that of (\ref{eq:S_old2F}).

For the one-flavor parts, $ \det\Dodwf(m_s)/\det\Dodwf(m_{c}) $ and 
$ \det\Dodwf(m_b)/\det\Dodwf(m_{PV}) $, we use the exact one-flavor pseudofermion action (EOFA)
for DWF \cite{Chen:2014hyy}. For the optimal DWF, it can be written as ($m_1 < m_2 $) 
\bea
\label{eq:detDodwf}
  \frac{\det\Dodwf(m_1)}{\det\Dodwf(m_2)} 
= \frac{\det D_T(m_1)}{\det D_T(m_2)}    
= \int d \phi_\pm^\dagger d \phi_\pm \exp\left( - \phi_+^\dagger G_+(m_1,m_2) \phi_+ 
                                                - \phi_-^\dagger G_-(m_1,m_2) \phi_- \right), \hspace{4mm}
\eea
where $ \phi_\pm $ and $ \phi_\pm^\dagger $ are pseudofermion fields (each of two spinor components) 
on the 4-dimensional lattice, and 
\bea
\label{eq:Gp}
G_-(m_1,m_2) &=& P_- \left[I- k(m_1,m_2)\omega^{-1/2} v_-^T \frac{1}{H_T(m_1)} v_- \omega^{-1/2}\right] P_-, \\ 
G_+(m_1,m_2) &=& P_+ \left[I+ k(m_1,m_2) \omega^{-1/2} v_+^T 
                             \frac{1}{H_T(m_2)-\Delta_+(m_1,m_2) P_+} v_+ \omega^{-1/2}\right] P_+.
\label{eq:Gm}
\eea
Here
\BAN
D_{T}(m_i) &=& D_w + M(m_i), \hspace{4mm} i=1,2\\  
M(m_i) &=& \omega^{-1/2}[1-L(m_i)][1+L(m_i)]^{-1}\omega^{-1/2}
        = P_+ M_+(m_i) + P_- M_-(m_i),  \\ 
H_T(m_i) &=& R_5 \gamma_5 D_T(m_i),  \\
\Delta(m_1,m_2) &=& R_5 \left[ M(m_2)-M(m_1) \right] = P_+ \Delta_+(m_1,m_2) + P_- \Delta_-(m_1,m_1), \\
\Delta_\pm(m_1,m_2) &=& k(m_1,m_2) \omega^{-1/2} v_\pm v_\pm^T \omega^{-1/2}, \\ 
\label{eq:k}
              k(m_1,m_2) &=& \frac{m_2 - m_1}{m_2 + m_1}, \\ 
\label{eq:vpm}
v_+^{T} &=& (-1, 1, \cdots, (-1)^{N_s}), \hspace{2mm} v_{-} = -v_+.
\EAN

For the gluon fields, we use the Wilson plaquette gauge action \cite{Wilson:1974sk}
at $ \beta = 6/g_0^2 = 6.70 $.  
\BAN
S_g(U) = \frac{6}{g_0^2} \sum_{plaq.}\left\{1-\frac{1}{3} \re \Tr (U_p) \right\}, 
\EAN
where $ g_0 $ is the bare coupling. 

The bare mass of $\u/\d$ quarks is set to $ m_{u/d} = 0.01$ such that $ M_\pi L > 4 $, 
while the bare masses of $ (\b, \c, \s) $ are tuned to 
$ \{m_b, m_c, m_s\} = \{ 0.850(5), 0.200(5), 0.150(2) \} $ such that they give
the masses of the vector mesons $\Upsilon(9460)$, $ J/\psi(3097), $ and $ \phi(1020) $ 
respectively. The tuning process is outlined as follows.

With $ \beta = 6.70 $ and $ m_{u/d} = 0.01 $, the tuning amounts to search for the physical point 
in the 3-dimensional space of $(m_b, m_c, m_s) $.
Basically it is a trial-and-error method, with every trial
in the 3-dimensional space involving a HMC simulation,
plus the computation of quark propagators and the determination of meson masses.
This could be a very slow process if one performs the search iteratively
starting from one point in this 3-dimensional space. 
Our strategy to speed up the search process is to use many GPUs 
to perform the search simultaneously, each with a different set of parameters.
Thus all searches together cover a domain in this 3-dimensional space, 
with a resolution up to the total number of GPUs and the total number of batches.
Moreover, the search is first performed on a small lattice $10^3 \times 32$, 
then move on to a larger lattice $20^3 \times 32$, 
and finally to the $ 40^3 \times 64$ lattice.
At the completion of the search for each lattice size, the optimal physical parameters 
for this lattice size are obtained, which are then used as the input to the next search 
on a larger lattice, and also to reduce the domain of search by eliminating 
the most unphysical parameters. The entire search process took about one year, using 200 GPUs
of various specifications, i.e., each of them can perform the HMC on the $ 10^3 \times 32$ 
and the $ 20^3 \times 32 $ lattices, but only 32 of them (each with 12 GB device memory)
can be grouped into 16 pairs to run 16 independent streams of HMC 
on the $ 40^3 \times 64 $ lattice.
 
The algorithm for simulating 2-flavor action for optimal domain-wall quarks has been 
outlined in Ref. \cite{Chiu:2011bm}, while that for simulating 
the exact one-flavor pseudofermion action (EOFA) 
of domain-wall fermion has been presented in Refs. \cite{Chen:2014hyy,Chen:2014bbc}.
In the molecular dynamics, we use the Omelyan integrator \cite{Omelyan:2001abc}, 
the multiple-time scale method \cite{Sexton:1992nu}, 
and the mass-preconditioning \cite{Hasenbusch:2001ne}.

\begin{figure}[!ht]
\begin{center}
\includegraphics[width=12cm,clip=true]{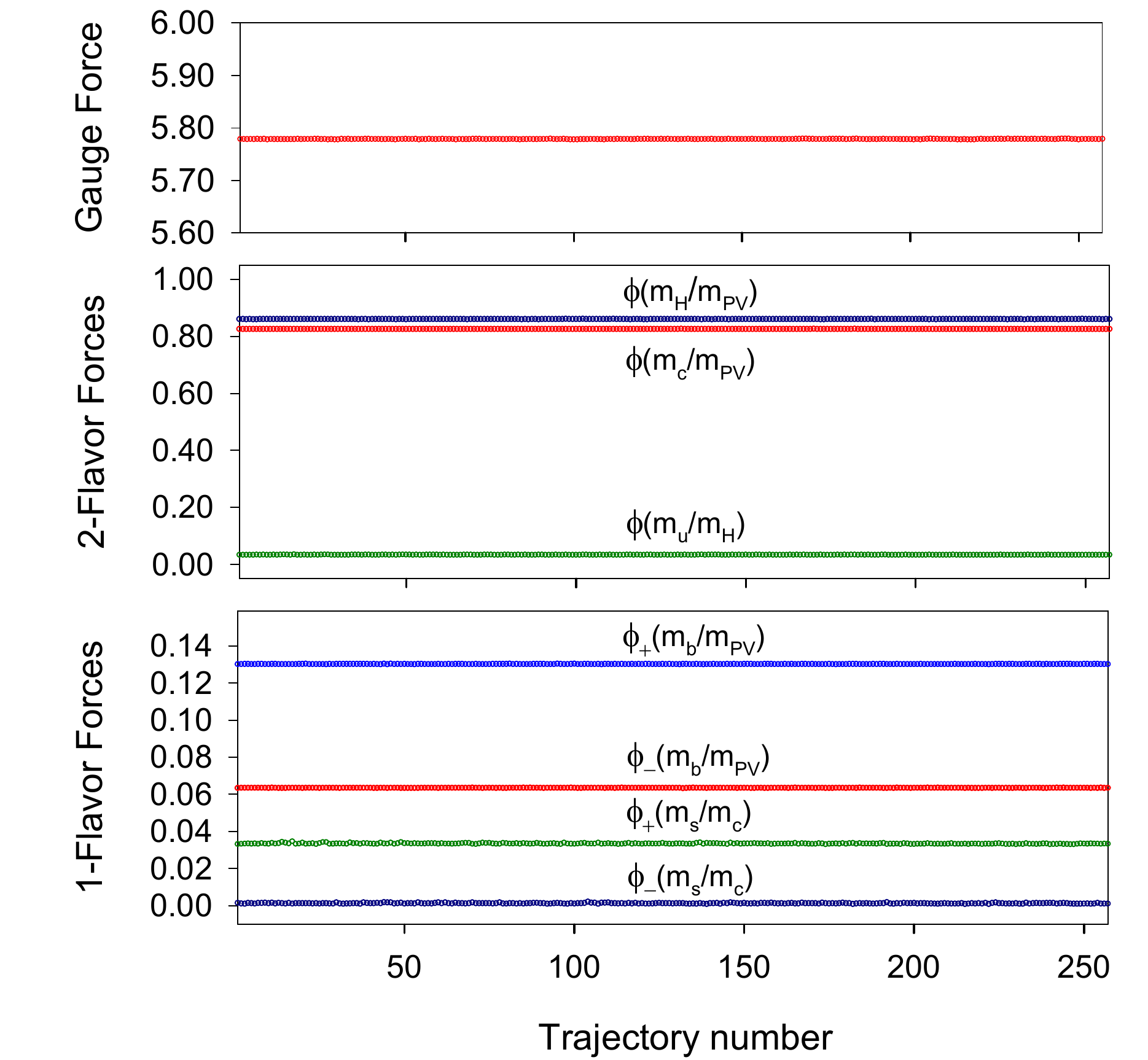}
\caption{The maximum forces of the gauge field, the 2-flavor pseudofermion fields, 
and the one-flavor pseudofermion fields versus the HMC trajectory in the HMC simulations of
the lattice QCD with $ N_f=2+1+1+1 $ optimal DWF.
}
\label{fig:HMC_forces}
\end{center}
\end{figure}

\subsection{HMC simulations}

Following the common strategy to reduce the thermalization time 
for a large lattice such as $ 40^3 \times 64 $,
we first perform the thermalization on a smaller lattice $ 20^3 \times 32 $ 
with the same set of parameters $(\beta, m_{u/d}, m_s, m_c, m_b)$. 
Then the thermalized gauge configuration on the $ 20^3 \times 32 $ lattice 
is used to construct the initial gauge configuration on the $ 40^3 \times 64 $ lattice by 
doubling the size of the lattice in each direction with the periodic extension.  
With this initial gauge configuration, we generate the first 257 trajectories 
on the $40^3 \times 64 $ lattice with two Nvidia GTX-TITAN-X GPU cards, each with device memory $ 12 $~GB. 
After discarding the initial 187 trajectories for thermalization, we sample one configuration
every 5 trajectories, resulting 14 ``seed" configurations. 
Then we use these seed configurations as the initial configurations for 14 independent simulations 
on 14 nodes, each of two Nvidia GTX-TITAN-X GPU cards.  
Each node generates $\sim 40$ trajectories independently, and   
all 14 nodes accumulate a total of 535 trajectories. 
We sample one configuration every 5 trajectories in each stream, 
and obtain a total of $103$ configurations for physical measurements.

\begin{figure}[!ht]
\begin{center}
\includegraphics[width=10cm,clip=true]{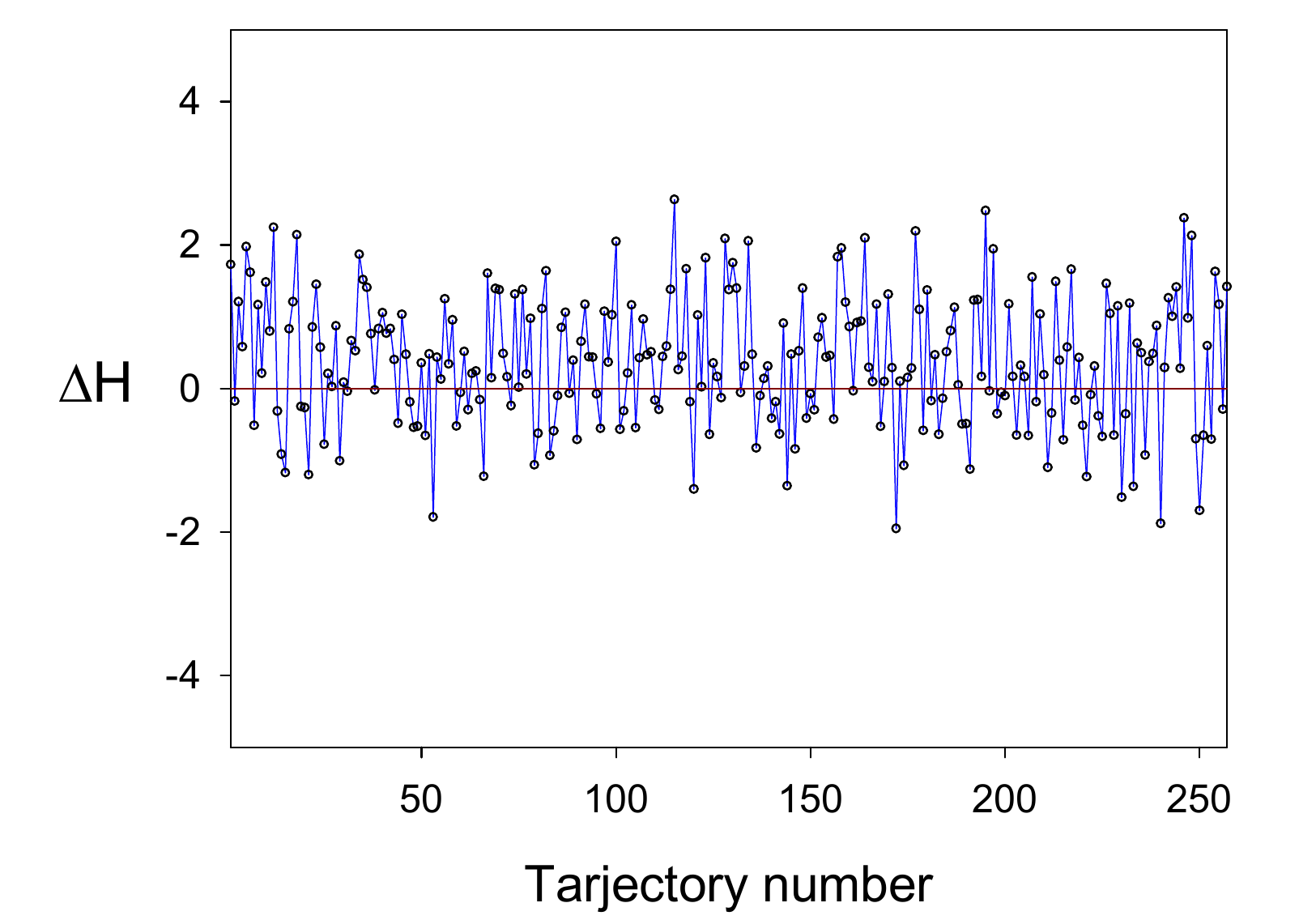}
\caption{The change of the Hamiltonian $ \Delta H $ versus the trajectory in the HMC simulations
of lattice QCD with $ N_f=2+1+1+1 $ optimal DWF.
The line connecting the data points is only for guiding the eyes.}
\label{fig:HMC_dH}
\end{center}
\end{figure}

In the following, we summarize the HMC characteristics of the first 257 trajectories. 
In Fig. \ref{fig:HMC_forces}, we plot the maximum force (averaged over all links)
among all momentum updates in each trajectory, for the gauge force, the 2-flavor pseudofermion forces,
and the one-flavor pseudofermion forces respectively, where $ \phi(m_1/m_2) $ denotes 
the two-flavor fermion force due to the pseudofermion action $ S(m_1, m_2) $, 
and $\phi_\pm(m_1/m_2) $ denotes the one-flavor pseudofermion force due to the exact one-flavor action  
with $\pm$ chirality, $S_{\pm}(m_1, m_2) = \phi_\pm^\dagger G_\pm(m_1, m_2) \phi_\pm $.
From the sizes of various forces in Fig. \ref{fig:HMC_forces}, 
the multiple timescales can be designed in the momentum update with the gauge force 
and the pseudofermion forces. 
With the length of the HMC trajectory equal to one, we use 4 different time scales
for the momentum updates with 
(1) the gauge force; (2) the two-flavor fermion forces 
associated with $ \phi(m_c/m_{PV}) $ and $ \phi(m_H/m_{PV})$; 
(3) the two-flavor force associated with $ \phi(m_u/m_H) $ and 
the one-flavor fermion force associated with $ \phi_+(m_b/m_{PV}) $;  
(4) the one-flavor fermion forces associated with $ \phi_-(m_b/m_{PV}) $, 
$\phi_+(m_s/m_c)$, and $\phi_-(m_s/m_c)$,    
which correspond to the step sizes $1/(k_1 k_2 k_3 k_4)$, $1/(k_2 k_3 k_4)$, $1/(k_3 k_4)$,  
and $1/k_4$ respectively. In our simulation, we set $(k_1, k_2, k_3, k_4)=(10, 2, 2, 5)$.

In Fig. \ref{fig:HMC_dH}, the change of Hamiltonian $ \Delta H $ versus the
HMC trajectory is plotted for the first 257 trajectories, with $ \left< \Delta H \right> = 0.376(57)$.
The number of accepted trajectories is 173, giving the acceptance rate $0.673(29)$. 
Using the measured value of $ \left< \Delta H \right> = 0.376(57) $, we can obtain
the theoretical estimate of the acceptance rate with the formula
$ P_{\rm acc} = {\rm erfc} \left( \sqrt{ \left< \Delta {\cal H} \right>}/2 \right)$ \cite{Gupta:1990ka}, 
which gives 0.664(24), in good agreement with the measured acceptance rate 0.673(29).
Moreover, we measure the expectation value of $ \exp(-\Delta H) $, to check whether
it is consistent with the theoretical formula $ \left< \exp(-\Delta H) \right> = 1 $
which follows from the area-preserving property of the HMC simulation \cite{Creutz:1988wv}.
The measured value of $ \left< \exp(-\Delta H) \right> $ is 1.026(66), 
in good agreement with the theoretical expectation value. The summary of the HMC characteristics
for the initial 257 trajectories is given in Table \ref{tab:HMC_summary_257}.

\begin{table}[h!]
\begin{center}
\caption{Summary of the HMC characteristics for the first 257 trajectories in the 
simulation of $N_f = 2+1+1+1 $ lattice QCD with the optimal DWF. }
\setlength{\tabcolsep}{4pt}
\vspace{2mm}

\begin{tabular}{|ccccccc|}
\hline
  $N_{\rm traj}$ 
  & Time(s)/traj
  & Acceptance
  & $ \left< \Delta H \right> $
  & $P_{\rm acc} = {\rm erfc}(\sqrt{\left< \Delta H \right>}/2)$
  & $ \left< \exp(-\Delta H) \right> $
  & $ \left<{\rm plaquette} \right> $ \\
\hline
\hline
257 & 76349(146) & 0.673(29) & 0.376(57) & 0.664(24) & 1.026(66) & 0.63185(1) \\
\hline
\end{tabular}
\label{tab:HMC_summary_257}
\end{center}
\end{table}

\subsection{Topological charge fluctuations}

In this subsection, we examine the evolution of the topological charge $ Q_t $ 
in the first 257 trajectories, and obtain the histogram of its distribution.

\begin{figure}[!ht]
\begin{center}
\begin{tabular}{@{}cccc@{}}
\includegraphics*[width=9cm,height=7.1cm,clip=true]{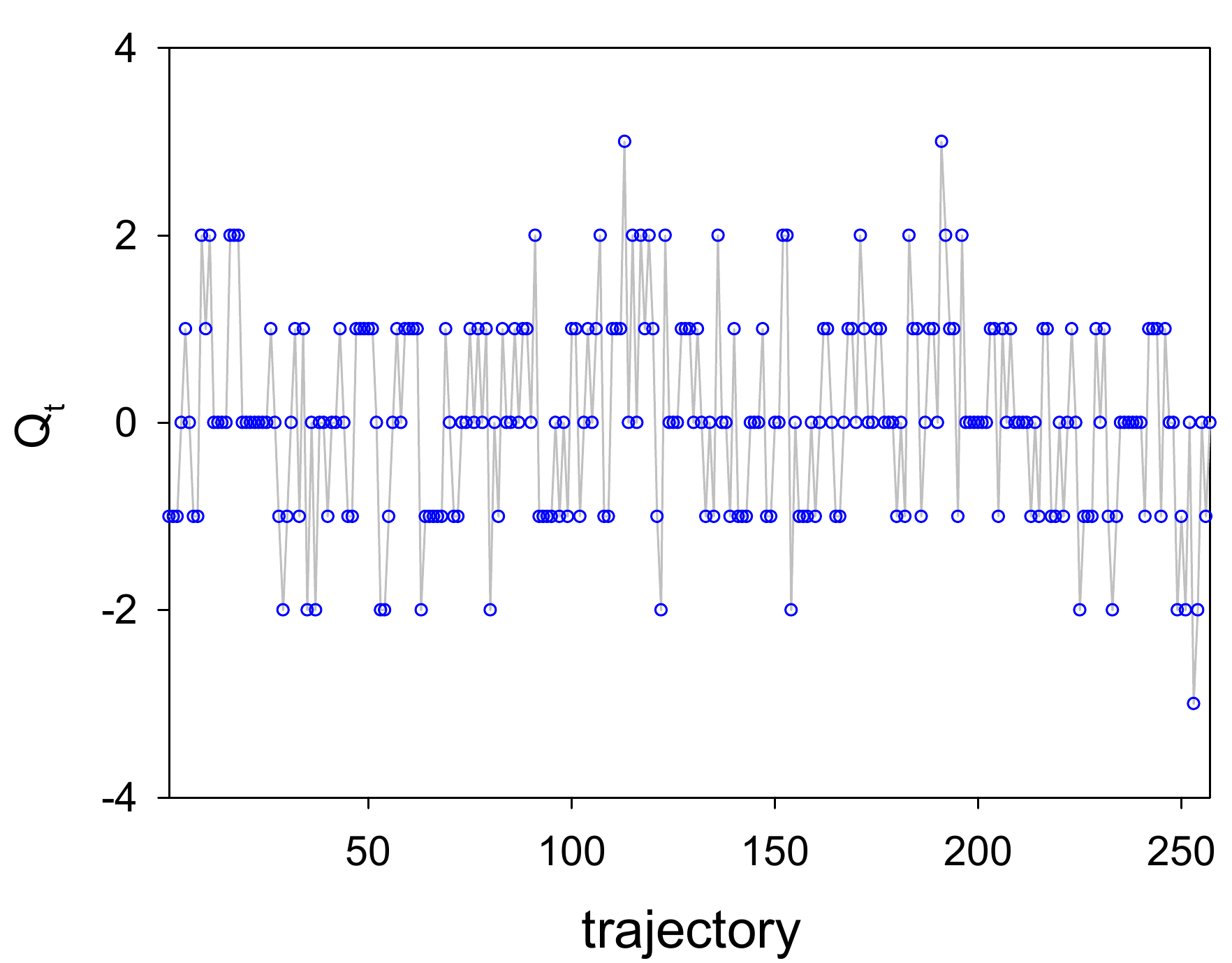}
&
\includegraphics*[width=7.5cm,height=7cm,clip=true]{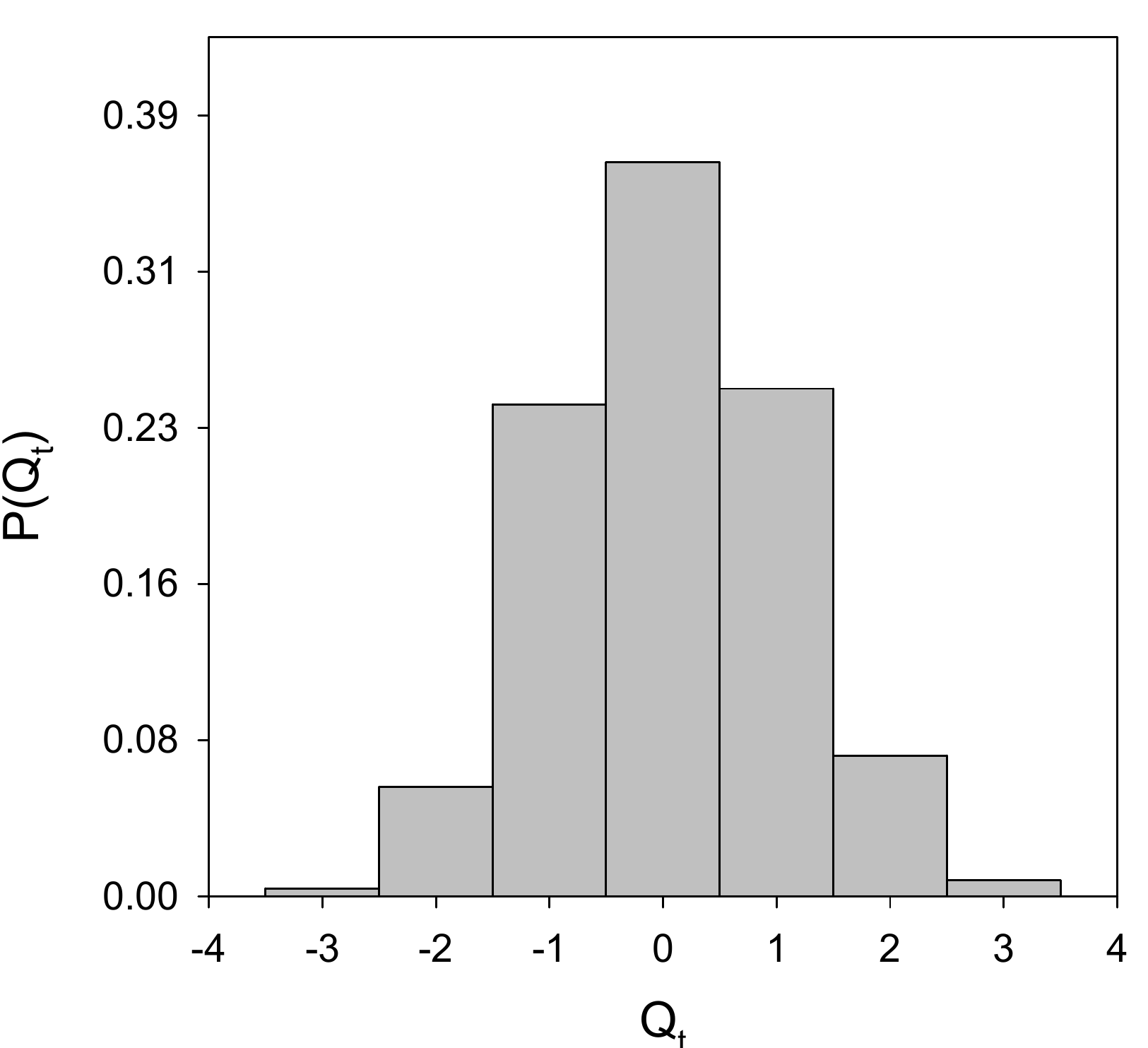}
\\
%\\ (a) & (b)
\end{tabular}
\caption{
(left panel) The evolution of $Q_t$ versus the HMC trajectory. 
             The line connecting the data points is only for guiding the eyes.
(right panel) The histogram of the probability distribution of $Q_t $ for the first 257 HMC trajectories.
Here the topological charge $Q_t$ is sampled at the Wilson flow time $ t/a^2 = 0.4 $ 
(see text for discussions). 
}
\label{fig:Qc_g257_hist}
\end{center}
\end{figure}

In lattice QCD with exact chiral symmetry, the topological charge $ Q_t $ 
can be measured by the index of the massless overlap-Dirac operator, since its index satisfies 
the Atiyah-Singer index theorem, $ \text{index}(D_{\text{ov}}) = n_+ - n_- = Q_t $. 
However, to project the zero modes of the massless overlap-Dirac operator for
the $ 40^3 \times 64 $ lattice is prohibitively expensive.
On the other hand, the clover topological charge
$ Q_{\text{clover}} = \sum_x \epsilon_{\mu\nu\lambda\sigma}
\tr[ F_{\mu\nu}(x) F_{\lambda\sigma}(x) ]/(32 \pi^2) $
is not reliable [where the matrix-valued field tensor $ F_{\mu\nu}(x) $ is obtained from
the four plaquettes surrounding $ x $ on the ($\hat\mu,\hat\nu$) plane],
unless the gauge configuration is sufficiently smooth.
Nevertheless, the smoothness of a gauge configuration
can be attained by the Wilson flow \cite{Narayanan:2006rf,Luscher:2010iy},
which is a continuous-smearing process to average gauge field
over a spherical region of root-mean-square radius $ R_{rms} = \sqrt{8 t} $, where $ t $ is
the flow-time. In this study, the flow equation is numerically integrated from $ t = 0 $  
with $ \Delta t/a^2 = 0.01 $, and measure the $ Q_{\text{clover}} $ at $ t/a^2 = 0.4 $ 
which amounts to averaging the gauge field over a spherical region of 
root-mean-square radius $ R_{rms} = \sqrt{8 t} \sim 1.8 a $. 
Then each gauge configuration becomes very smooth, with $ Q_{\text{clover}} $ close to an integer,  
and the average plaquette greater than 0.997. Denoting the nearest integer of $ Q_{\text{clover}} $ 
by $ Q_t \equiv \text{round}(Q_{\text{clover}}) $, $ Q_t $ is plotted versus the trajectory number 
in the left-panel of Fig. \ref{fig:Qc_g257_hist}, while the right-panel displays the  
histogram of the probability distribution of $ Q_t $ of the first 257 HMC trajectories.   
Evidently, the HMC simulation samples all topological sectors ergodically. 
However, there are some subtle issues which we will discuss in the following.   

\begin{figure}[!h]
\begin{center}
\includegraphics[width=10cm,clip=true]{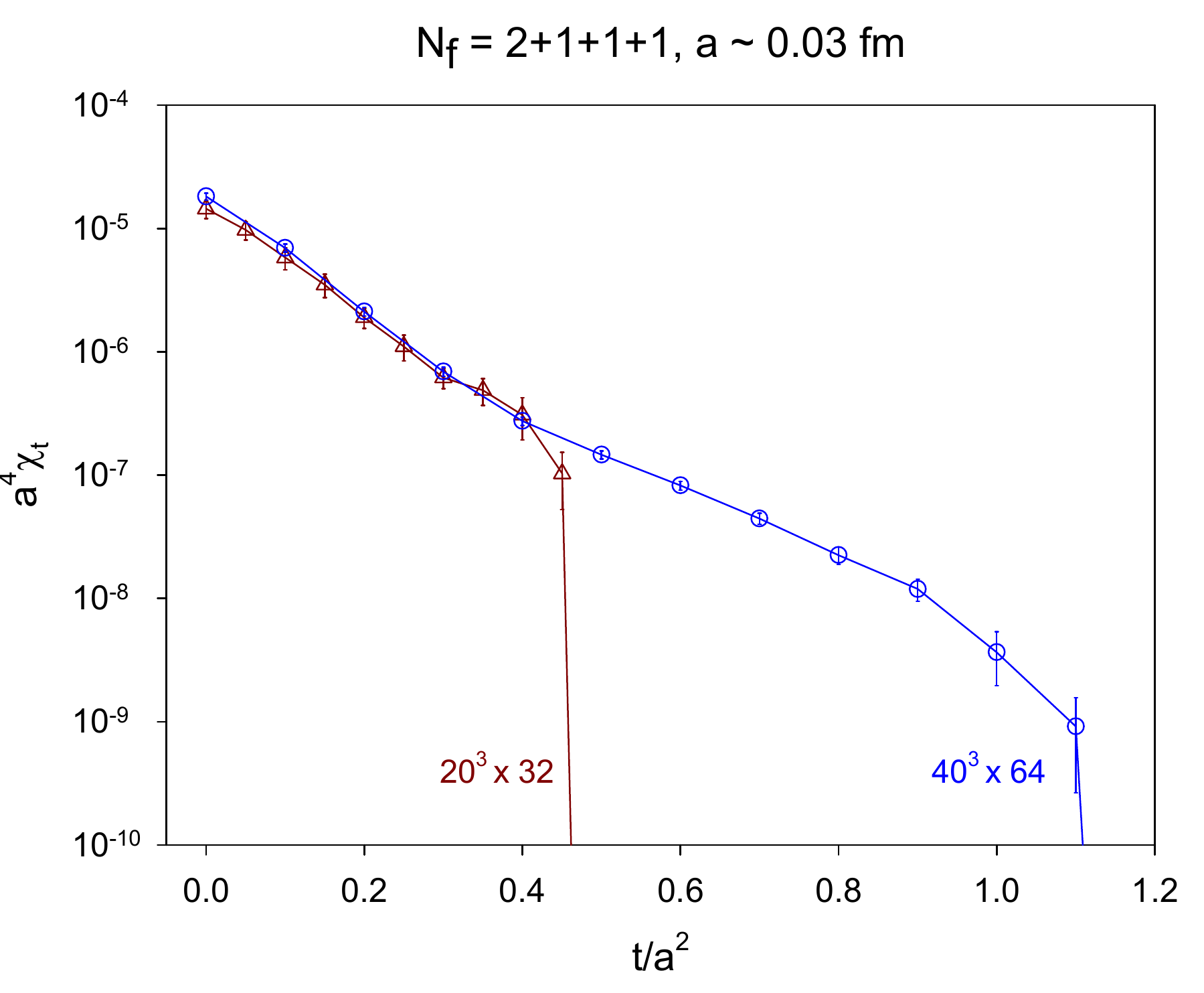}
\caption{The topological susceptibility versus the wilson flow time $t$ for $N_f=2+1+1+1$ lattice QCD 
on the $20^3 \times 32$ and $40^3 \times 64 $ lattices. See text for detailed descriptions.}
\label{fig:chit_Nf2p1p1p1}
\end{center}
\end{figure}

Note that the topological charges are sampled at $ t/a^2 = 0.4 $ which is much smaller than 
the flow time $ t_0/a^2 \sim 22 $ for setting the lattice scale [see Eq. (\ref{eq:t0})]. 
The reason of not using a large $t/a^2 \gg 1$ for measuring $Q_{\text{clover}}$ is because that     
the lattice volume $ V \sim (\text{1.2 fm})^3 \times (\text{1.9 fm}) $ is too small 
to preserve the nonzero topological charge against any scheme for smoothing the gauge configuration.
In other words, for lattice QCD in such a small lattice volume with a fine lattice spacing 
($ a \sim 0.03 $~fm), any gauge configuration must become topologically trivial 
after it has been flowed for a sufficient long time $t/a^2 \gg 1$, thus the  
topologically susceptibility $ \chi_t = \left<Q_t^2\right>/V $ becomes zero for $ t > t_z $, 
where $ t_z $ depends on the relevant parameters (e.g., lattice volume, lattice spacing, 
$N_f$, and the quark masses) in generating the gauge configurations. 
On the other hand, for a sufficiently large lattice volume,  
the topologically susceptibility $\chi_t$ would attain a plateau for the large flow time $ t/a^2 \gg 1 $,  
as shown in the right panel of Fig. 1 in Ref. \cite{Chiu:2020ppa}, 
where the lattice volume is $ V \sim (\text{4 fm})^4 $ for $N_f=2+1+1$ lattice QCD with 
domain-wall quarks at the physical point. 
In the latter case, the topological charge fluctuations can be sampled at any large flow time 
$ t/a^2 \gg 1 $. However, for $ N_f = 2+1+1+1$ lattice QCD  
with the lattice spacing $ a \sim 0.03 $~fm (see the next subsection), 
a sufficiently large lattice volume would exceed the lattice size $ \gtrsim 120^4 $, 
which is beyond our current computational capability. 
Thus, for the small lattice volume $ V \sim (\text{1.2 fm})^3 \times (\text{1.9 fm}) $ in this study,   
$ \chi_t $ cannot attain a plateau at the large flow time $ t/a^2 \gg 1 $, but 
goes to zero at $ t/a^2 \sim 1.1 $, as shown in Fig. \ref{fig:chit_Nf2p1p1p1}.
Assuming that the $ \chi_t $ of the same $N_f=2+1+1+1$ lattice QCD on a large lattice volume
attains a plateau at the large flow time $ t/a^2 \gg 1 $,  
we still do not know whether the topological charge fluctuations 
(as shown in Fig. \ref{fig:Qc_g257_hist}) sampled at the flow time $ t/a^2 = 0.4 \sim t_z/(2a^2) $ 
on this small lattice volume is consistent with the plateau of the $ \chi_t $ on the large lattice volume. 
To answer this question requires to perform the HMC simulation (with the same actions and parameters) 
on a large lattice with size $ \gtrsim 120^4 $, which is beyond the scope of this paper.

\begin{figure}[!h]
\begin{center}
\includegraphics[width=11cm,clip=true]{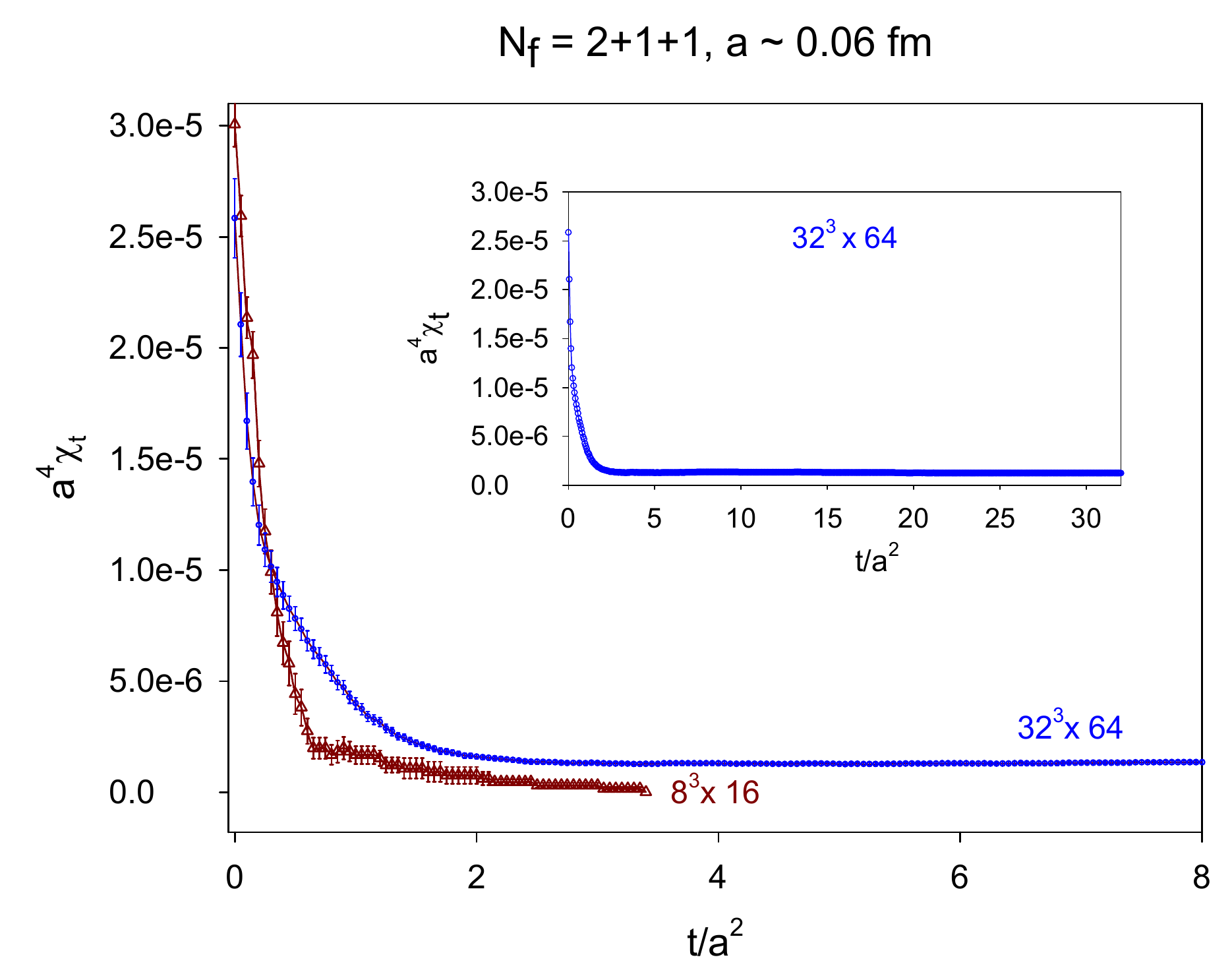}
\caption{The topological susceptibility versus the wilson flow time $t$ for $N_f=2+1+1$ lattice QCD 
on the $8^3 \times 16$ and $32^3 \times 64 $ lattices. See text for detailed descriptions.}
\label{fig:chit_Nf2p1p1}
\end{center}
\end{figure}

At this point, it is interesting to point out that for the small lattices, 
the $ t_z $ (where $\chi_t=0$ for $ t \ge t_z $) increases as
the lattice volume gets larger, as shown in Fig. \ref{fig:chit_Nf2p1p1p1}. 
Here both ensembles ($ 20^3 \times 32 $ and $ 40^3 \times 64 $) are generated with the 
same (gauge and fermion) actions (see the Sec. \ref{subsection:actions}) 
and the same parameters, namely, 
$\beta=6/g_0^2 = 6.70$, $N_s = 16 $, $m_0 = 1.3 $, $ \lambda_{max}/\lambda_{min} = 6.20/0.05 $, 
$ m_{u/d} a = 0.01$, $m_s a = 0.015$, $m_c a = 0.20$, and $m_b a = 0.85 $. 
The number of configurations is $ 252 $ for the $20^3 \times 32$ ensemble, while $ 257 $ for the 
$ 40^3 \times 64 $ ensemble.   
For the $20^3 \times 32 $ lattice with volume $ \sim (\text{0.6 fm})^3 \times (\text{0.96 fm}) $, 
all configurations become trivial and $ \chi_t = 0 $ for $ t/a^2 \ge t_z/a^2 \sim 0.5 $, while for 
the $40^3 \times 64 $ lattice with volume $ \sim (\text{1.2 fm})^3 \times (\text{1.9 fm}) $, 
all configurations become trivial and $ \chi_t = 0 $ for $ t/a^2 \ge t_z/a^2 \sim 1.1 $.
Thus the $ t_z $ of the $40^3 \times 64 $ lattice is more than twice of that of 
the $ 20^3 \times 32 $ lattice. 
This seems to imply that for a sufficiently large lattice volume, say, 
$V \gtrsim (\text{4 fm})^4 $, $ \chi_t $ would attain a plateau for the large flow time $ t/a^2 \gg 1 $, 
similar to the case of $N_f=2+1+1$ lattice QCD, as shown in the right panel 
of Fig. 1 in Ref. \cite{Chiu:2020ppa}. 

If the above scenario is true in general, then there must be at least one example in lattice QCD 
to show that its $\chi_t$ on a small lattice volume goes to zero at the large flow-time, 
but its counterpart on a large lattice volume attains a plateau at the large flow-time. 
To this end, we consider the $N_f=2+1+1$ lattice QCD with the (gauge and fermion) actions 
and the same parameters as given in Ref. \cite{Chen:2014hyy}, 
for the $ 8^3 \times 16 $ and $ 32 \times 64 $ lattices. 
The results of $ \chi_t $ versus the flow time $ t/a^2 $ are plotted in Fig. \ref{fig:chit_Nf2p1p1}.  
Here both $ 8^3 \times 16 $ and $ 32^3 \times 64 $ ensembles are generated with the 
same (gauge and fermion) actions and the same parameters, namely, 
$\beta=6/g_0^2 = 6.20$, $N_s = 16 $, $m_0 = 1.3 $, $ \lambda_{max}/\lambda_{min} = 6.20/0.05 $, 
$ m_{u/d} a = 0.005$, $m_s a = 0.04$, and $m_c a = 0.55$.  
The $ 32^3 \times 64 $ ensemble is exactly the same as that in Ref. \cite{Chen:2014hyy}, 
while the $ 8^3 \times 16 $ ensemble is generated in the present study.  
The number of configurations is $ 401 $ for the $ 32^3 \times 64 $ ensemble,    
while $ 800 $ for the $8^3 \times 16$ ensemble.  
For the $8^3 \times 16 $ lattice with volume $ \sim (\text{0.5 fm})^3 \times (\text{1.0 fm}) $, 
$ \chi_t $ becomes zero for $ t/a^2 \gtrsim 3.4 $, while for the
$32^3 \times 64 $ lattice with volume $ \sim (\text{2.0 fm})^3 \times (\text{4.0 fm}) $, 
its $ \chi_t a^4 $ attains a plateau ($ \sim 1.0 \times 10^{-6} $) 
for $ t/a^2 \gtrsim 3 $ (see also the subpanel in Fig. \ref{fig:chit_Nf2p1p1}),  
where $ t/a^2 = 32 $ is the maximum flow-time in this study. 
This example shows how the topological charge fluctuations depend on the lattice volume, 
and it also implies that only in the large lattice volume limit, 
the topological charge fluctuations of the QCD vacuum can be captured properly.

Now we conjecture that even for a lattice with very fine lattice spacing,   
the gauge configurations generated in the HMC simulation might not suffer from the topology freezing,    
provided that the lattice volume is kept sufficiently large, e.g., $V \gtrsim (\text{4 fm})^4 $.

Before closing this section, we discuss the role of heavy quarks in enhancing the topological fluctuations
of the QCD vacuum. First, we recall the relationship between the topological susceptibility 
$ \chi_t = \langle Q_t^2 \rangle / V $ (where $V$ is the 4-dimensional volume) and the quark condensates, 
which holds for lattice QCD with exact chiral symmetry, and for any number of heavy/light quark flavors. 
%and for the zero/finite temperature QCD. 
For lattice QCD with $(\u,\d,\s,\c,\b)$ quarks, 
in the chiral limit of $\u$ and $\d$ quarks ($m_{u/d} \to 0 $), 
it can be shown that (see the Appendix of Ref. \cite{Mao:2009sy})
\bea
\label{eq:chit_sigma}
\chi_t = \left(   \frac{\Sigma_u}{m_u} + \frac{\Sigma_d}{m_d} + \frac{\Sigma_{s}}{m_s} 
                + \frac{\Sigma_{c}}{m_c}  + \frac{\Sigma_{b}}{m_b} \right)
         \left( \frac{1}{m_u} +\frac{1}{m_d} + \frac{1}{m_s} + \frac{1}{m_c} + \frac{1}{m_b} \right)^{-2}, 
\eea    
where the quark condensate is defined as  
\BAN
\Sigma_q = \lim_{V \to \infty} \frac{1}{V} \left< \Tr (D_c + m_q)^{-1} \right>.
\EAN
It should be emphasized that the derivation of (\ref{eq:chit_sigma}) only relies on the 
exact chiral symmetry on the lattice, without using the chiral perturbation theory (ChPT) at all, 
thus it holds for any number of heavy/light quark flavors. 
In the limit of $N_f = 2$ QCD, it reproduces the Leutwyler-Smilga relation 
at the leading order of the ChPT. 
In general, for a sufficiently large lattice, empirically, we have 
$ \Sigma_q /m_q \sim \text{constant} $. This implies that on the rhs of (\ref{eq:chit_sigma}),  
each quark flavor contributes almost equally to the numerator (the first factor), 
while the heavy flavors ($\c$ and $\b$) are highly suppressed in the denominator (the second factor). 
Thus the topological susceptibility is enhanced by including more heavy quark flavors 
in the sea. This asserts that the heavy quark flavors indeed play an important role 
in enhancing the topological charge fluctuations of the QCD vacuum.

\subsection{Lattice scale}

First, we recap the generation of the gauge ensemble.
From the initial 257 trajectories generated by a single node, 
we discard the first 187 trajectories for thermalization, 
and sample one configuration every 5 trajectories, resulting 14 ``seed" configurations. 
Then we use these seed configurations as the initial configurations for 14 independent simulations 
on 14 nodes, each of two Nvidia GTX-TITAN-X GPU cards.  
Each node generates $\sim 40$ trajectories independently, and   
all 14 nodes accumulate a total of 535 trajectories. 
We sample one configuration every 5 trajectories in each stream, 
and obtain a total of $103$ configurations for physical measurements. 

To determine the lattice scale, we use the Wilson flow \cite{Narayanan:2006rf,Luscher:2010iy} 
with the condition
\BAN
\label{eq:t0}
\left. \{ t^2 \langle E(t) \rangle \} \right|_{t=t_0} = 0.3,
\EAN
and obtain $ \sqrt{t_0}/a = 4.6884(36) $ for the 103 configurations for physical measurements.
Using $ \sqrt{t_0} =  0.1416(8) $~fm obtained by 
the MILC Collaboration for the $(2+1+1)$-flavors QCD \cite{Bazavov:2015yea}, 
we have $ a^{-1} = 6.503 \pm 0.037 $~GeV. The lattice spacing is $ a = \text{0.0303(2) fm} $, 
giving the spatial volume  $\sim (\text{1.213 fm})^3 $, which is too small 
for studying physical observables involving the light quarks.

\subsection{Quark propagator} 

We compute the valence quark propagator of the effective 4D Dirac operator
with the point source at the origin, and with the mass and other parameters 
exactly the same as those of the sea quarks.  
The boundary conditions are periodic in space and antiperiodic in time.
First, we solve the following linear system with mixed-precision conjugate gradient algorithm,
for the even-odd preconditioned ${\cal D} $ \cite{Chiu:2011rc}
\bea
\label{eq:DY}
{\cal D}(m_q) |Y \rangle = {\cal D}(m_{PV}) B^{-1} |\mbox{source vector} \rangle,
\eea
where $ B^{-1}_{x,s;x',s'} = \delta_{x,x'}(P_{-}\delta_{s,s'}+P_{+}\delta_{s+1,s'}) $
with periodic boundary conditions in the fifth dimension.
Then the solution of (\ref{eq:DY}) gives the valence quark propagator
\bea
\label{eq:v_quark}
(D_c + m_q)^{-1}_{x,x'} = \left( m_{PV} - m_q \right)^{-1} \left[ (BY)_{x,1;x',1} - \delta_{x,x'} \right], 
\hspace{4mm} m_{PV} = 2 m_0. 
\eea
Each column of the quark propagator is computed by a single node with 2 Nvidia GTX-TITAN-X GPU cards,  
which attains more than 1000 Gflops/sec (sustained).

\subsection{Residual masses}

To measure the chiral symmetry breaking due to finite $N_s$, we compute the residual mass
according to \cite{Chen:2012jya}, 
\bea
\label{eq:Mres}
m_{res}=\left< \frac{\tr(D_c + m_q)^{-1}_{0,0}}
                    {\tr[\gamma_5 (D_c + m_q) \gamma_5 (D_c+m_q)] ^{-1}_{0,0}} \right> - m_q,
\eea
where
$ (D_c + m_q)^{-1} $ denotes the valence quark propagator with $ m_q $ equal to the sea-quark mass,
tr denotes the trace running over the color and Dirac indices, 
and the brackets $ \left< \cdots \right> $ denote the averaging over the gauge ensemble.
In the limit $ N_s \to \infty $, $ D_c $ is exactly chiral symmetric and the first term on the rhs 
of (\ref{eq:Mres}) is exactly equal to $ m_q $, thus the residual mass $ m_{res} $ is exactly zero, 
and the quark mass $ m_q $ is well-defined for each gauge configuration. 
On the other hand, for any finite $N_s$ with nonzero residual mass, 
the quark mass is not well-defined for each gauge configuration, 
but its impact on any physical observable can be roughly estimated by  
the difference due to changing the valence quark mass from $m_q$ to $ m_q + m_{res} $.

\begin{table}[H]
%  \small
\centering
\caption{The residual masses of $ \u/\d $, $\s $, $\c$, and $ \b $ quarks.}
\label{tab:mres}
\begin{ruledtabular}
\begin{tabular}{cccc}
%\toprule
   quark & $m_q a $ & $ m_{res} a $ & $ m_{res}$~[MeV]   \\
%\midrule
\hline
   $\u/\d$ & 0.010  & $ 7.93(52) \times 10^{-7} $ & 0.0052(3)   \\
   $\s$    & 0.015  & $ 8.21(52) \times 10^{-7} $ & 0.0053(3)   \\
   $\c$    & 0.200  & $ 9.43(54) \times 10^{-7} $ & 0.0061(4)   \\
   $\b$    & 0.850  & $ 1.06(6)  \times 10^{-6} $ & 0.0069(4)   \\
%\bottomrule
\end{tabular}
\end{ruledtabular}
\end{table}

For the 103 gauge configurations generated by HMC simulation of lattice QCD with
$ N_f = 2 + 1 + 1 +1 $ optimal domain-wall quarks, the residual masses
of $ \u/\d$, $\s $, $\c$, and $\b $ quarks are listed in Table \ref{tab:mres}.
We see that the residual mass of any quark flavor is less than $0.007$~MeV, 
which should be negligible in comparison with other systematic uncertainties.

In the following, we discuss the relationship between the residual mass (\ref{eq:Mres}) and  
the effective residual mass (a function of time)   
\bea
\label{eq:mres_t}
m_{\text res}(t) = \frac{\sum_{\vec{x}} \left< J_{5}(\vec{x},t;N_s/2) \qbar(0)\gamma_5 \q(0) \right>} 
       {\sum_{\vec{x}} \left< \qbar(\vec{x},t) \gamma_5 \q(\vec{x},t) \qbar(0) \gamma_5 \q(0) \right>},
\eea
where $ J_5(x;N_s/2) $ is the pseudoscalar density at the center of the fifth dimension, as 
defined in Ref. \cite{Chen:2012jya}. Note that both (\ref{eq:Mres}) and (\ref{eq:mres_t}) 
can be obtained from the axial Ward identity. The only difference between them is 
whether the axial Ward identity is summed over $ x=(\vec{x},t) $ or $\vec{x}$,  
before the residual mass is extracted. That is, in (\ref{eq:mres_t}), if summing over 
all $t$ in both the numerator and the denominator respectively, then it recovers (\ref{eq:Mres}), 
as shown in Ref. \cite{Chen:2012jya}. The denominator of (\ref{eq:mres_t}) is exactly 
the time-correlation function of the pseudoscalar (PS), 
which behaves as $ \sim [ \exp\{-m_{\text PS}t\} + \exp\{-m_{\text PS}(T-t)\} ] $ at large $ t $, 
say for $ 1 \ll t_1 < t < T/2 $.
If the numerator of (\ref{eq:mres_t}) also behaves similar to the denominator at large $t$, 
then their ratio $ m_{\text res}(t) $ would attain a plateau in the range $ t_1 < t < T/2 $. 
The RBC/UKQCD Collaboration has been taking the plateau value of $ m_{\text res}(t) $ 
as the residual mass, which should be compatible with that computed with (\ref{eq:Mres}).  
On the other hand, if the numerator decays much slower than the exponential function at large $ t $, 
then $ m_{\text res}(t) $ would behave like a monotonically-increasing function of $ t $, 
resulting a peak at $ t = T/2 $, as observed by the RBC/UKQCD Collaboration in the case 
of M\"obius DWF with $ m_q a = 0.45 $ \cite{Boyle:2016imm}. Such anomalous behavior of the numerator 
of (\ref{eq:mres_t}) at heavy quark masses implies that the physical modes are not exponentially local 
to the boundaries of the fifth dimension, thus the M\"obius DWF has difficulties 
to treat heavy quarks. In this case, if one uses (\ref{eq:Mres}) to measure the residual mass, 
then one would also observe a dramatic increase of the residual mass for heavy quark masses,  
e.g., the residual mass would increase $\sim 3-4 x$ by changing $m_q a $ from 0.40 to 0.45,   
a rough estimate using the data in the left-panel of Fig. 2 in Ref. \cite{Boyle:2016imm}. 
In other words, the anomalous behavior of $ J_5(x,N_s/2) $ in DWF 
can be observed by both definitions of residual mass, (\ref{eq:Mres}) and (\ref{eq:mres_t}).  
Since the residual masses in Table \ref{tab:mres} are almost the same for $ m_q a = 0.01-0.85 $, 
it rules out the possibility that $ m_{\text res}(t) $ for the optimal DWF could have any anomalous 
behavior with heavy quarks.
%
%\begin{figure}[H]
%\begin{center}
%\includegraphics*[width=10cm,clip=true]{mres_t.pdf}
%\caption{The effective residual mass $m_{\text res}(t)$ 
%of $(\u/\d,\s,\c,\b)$ quarks in $N_f=2+1+1+1$ lattice QCD with optimal domain-wall fermion.}
%\label{fig:mres_t}
%\end{center}
%\end{figure}
%

\section{Mass spectra of beauty mesons}

In the following, we determine the masses of the low-lying mesons
with valence quark contents $\bbar\b$, $\bbar\c$, $\bbar\s$, and $\cbar\c$.  
We construct the quark-antiquark meson interpolators
and measure their time-correlation functions using the
point-to-point quark propagators computed with the same parameters 
($N_s = 16 $, $m_0 = 1.3 $, $ \lambda_{max}/\lambda_{min} = 6.20/0.05 $)  
of the sea quarks, for the quark masses 
($ m_{u/d} a = 0.01$, $m_s a = 0.015$, $m_c a = 0.20$, $m_b a = 0.85 $), 
where $m_b$, $m_c$ and $ m_s $ are fixed by the masses of the vector mesons
$ \Upsilon(9460) $, $ J/\psi(3097) $, and $ \phi(1020) $ respectively.
Then we extract the mass of the lowest-lying meson state from the time-correlation function.

The time-correlation function of the beauty meson interpolator $ \bbar \Gamma \q $ 
(where $ \q = \{\b,\c,\s \} $) is measured according to the formula
\bea
\label{eq:C}
C_{\Gamma} (t) =
\left<
\sum_{\vec{x}}
\tr\{ \Gamma (D_c + m_b)^{-1}_{x,0} \Gamma (D_c + m_q)^{-1}_{0,x} \}
\right>, 
\eea
%for scalar ($S$), pseudoscalar ($P$), vector ($V$), axial-vector ($A$),
%and tensor ($T$), with Dirac matrix
%$\Gamma=\{\Id,\gamma_5,\gamma_i,\gamma_5\gamma_i,\gamma_5\gamma_4\gamma_i \} $
%$\Gamma=\{\Id,\gamma_5,\gamma_i,\gamma_5\gamma_i,
%          \gamma_5\gamma_4\gamma_i = \epsilon_{ijk}\gamma_j\gamma_k/2 \} $
where $\Gamma=\{\Id,\gamma_5,\gamma_i,\gamma_5\gamma_i, \epsilon_{ijk}\gamma_j\gamma_k \} $,
corresponding to scalar ($S$), pseudoscalar ($P$), vector ($V$), axial-vector ($A$),
and pseudovector ($T$) respectively, and 
the valence quark propagator $(D_c + m_q)^{-1} $ is computed according to the formula (\ref{eq:v_quark}).
Note that $ \qbar \gamma_5\gamma_i \q $ transforms like $ J^{PC} = 1^{++} $, while
$ \qbar \epsilon_{ijk}\gamma_j\gamma_k \q $ like $ J^{PC} = 1^{+-} $.

For the vector meson, we average over $i=1,2,3$ components, namely,
\BAN
\label{eq:CV}
C_V (t) = \left<
\frac{1}{3} \sum_{i=1}^3 \sum_{\vec{x}}
\tr\{ \gamma_i (D_c + m_b)^{-1}_{x,0}
      \gamma_i (D_c + m_q)^{-1}_{0,x} \} \right>.
\EAN
Similarly, we perform the same averaging for the axial-vector and pseudovector mesons.
Moreover, to enhance statistics, we average the forward and the backward time-correlation function.
\BAN
\bar C(t) = \frac{1}{2} \left[ C(t) + C(T-t) \right].
\EAN

The time-correlation function (TCF) and the effective mass of the meson interpolators 
$ \bbar \Gamma \b $, $ \cbar \Gamma \c $, $\bbar \Gamma \c $, and $ \bbar \Gamma \s $ 
are plotted in Figs. \ref{fig:Ct_meff_G55_bb}-\ref{fig:Ct_meff_Gt_bs}, 
in the Appendices \ref{bbar_b}-\ref{bbar_s} respectively.

\subsection{Bottomonium and charmonium}

First of all, we check to what extent we can reproduce the 
bottomonium masses which have been measured precisely
in high energy experiments. 

Our results of the mass spectrum of the low-lying states of 
bottomonium are summarized in Table \ref{tab:bbar_b}.  
The time-correlation function and the effective mass of $ \bbar \Gamma b $ are plotted in 
Appendix \ref{bbar_b}.

The first column in Table \ref{tab:bbar_b} is the Dirac matrix used for  
computing the time-correlation function (\ref{eq:C}). 
The second column is $ J^{PC} $ of the state. 
The third column is the $ [t_1, t_2] $ used for fitting  
the data of $ C_\Gamma(t) $ to the usual formula   
\bea
\label{eq:meson_fit}
\frac{z^2}{2 M a} [ e^{-M a t} + e^{- M a(T-t)} ]  
\eea
to extract the ground state meson mass $ M $, where the excited states have been neglected. 
We use the correlated fit throughout this work.
The fifth column is the mass $ M $ of the meson state, where the first 
error is statistical, and the second is systematic. 
Here the statistical error is estimated using the jackknife method with the bin size of which 
the statistical error saturates, while the systematic error is estimated 
based on all fittings satisfying $ \chi^2/\mbox{dof} < 1.2 $ and $ |t_2 - t_1| \ge 6 $ with
$ t_1 \ge 10 $ and $ t_2 \le 32 $. 
The last column is the experimental state we have identified, 
and its PDG mass value \cite{Tanabashi:2018oca}.

The analysis and the descriptions in the above paragraph apply to all results obtained in this work, 
as given in Table \ref{tab:bbar_b}-\ref{tab:bbar_c}.

\begin{table}[!ht]
\begin{center}
%\LARGE
\caption{The masses of low-lying bottomonium states obtained in this work. 
         The fifth column is the mass of the meson state, where the first 
         error is statistical, and the second is systematic. 
         The last column is the experimental state we have identified, 
         and its PDG mass value \cite{Tanabashi:2018oca}.
         For a detailed description of each column, see the paragraph with Eq. (\ref{eq:meson_fit}). 
}
\vspace{0.5cm}
%\begin{tabular}{c|c|c|c|c|c||c}
\begin{ruledtabular}
\begin{tabular}{cccccc}
$ \Gamma $ & $ J^{PC} $ & $ [t_1,t_2] $ & $\chi^2$/dof & Mass(MeV) & PDG \\
\hline
%
% G11    
%
$ \Id $ & $ 0^{++} $   
                     & [19,29] 
                     & 1.10 
                     & 9859(14)(11) 
                     & $ \chi_{b0}(9859) $ \\
% G55 
%    
$\gamma_5$ & $ 0^{-+} $  
                        & [15,31] 
                        & 1.04 
                        & 9403(4)(5) 
                        & $ \eta_b(9399) $  \\
%
% Gv4   & $ \gamma_4 $ & exotic !  &  &    &   &  \\
%
% Ga4   
%
%$\gamma_5\gamma_4$ & $ 0^{-+} $ & $ 1^1 S_0 $ & & & & $ \eta_b(9300) $  \\
%
% Gv   
%
$\gamma_i$ & $ 1^{--} $  
                        & [21,31] 
                        & 0.51 
                        & 9468(7)(6) 
                        & $ \Upsilon(9460) $ \\
%
% G4v  
%
%$\gamma_4\gamma_i $ & $ 1^{--} $ & $ 1^3 S_1 $ & & & & $ \Upsilon(9460) $ \\
%
% Ga  
%
$\gamma_5\gamma_i$ & $ 1^{++}$  
                               & [19,26] 
                               & 1.15 
                               & 9884(27)(35) 
                               & $ \chi_{b1}(9893) $ \\
%
% Gt 
%
%$\gamma_5\gamma_4\gamma_i$ & $ 1^{+-} $ & $ 1^1 P_1 $ 
$ \epsilon_{ijk} \gamma_j \gamma_k$ & $ 1^{+-} $  
                           & [19,25] 
                           & 0.97 
                           & 9910(20)(25) 
                           &  $ h_b(9899) $ \\
\end{tabular}
\end{ruledtabular}
\label{tab:bbar_b}
\end{center}
\end{table}

Evidently, the masses of bottomonium in Table \ref{tab:bbar_b} 
are in good agreement with the PDG mass values, even though the axial-vector ($1^{--}$) and 
pseudovector ($1^{+-}$) mesons have relatively larger errors than other meson states. 
Note that the theoretical result of the hyperfine splitting ($ 1^3 S_1 - 1^1 S_0 $) is   
$ 65(8)(7) $ MeV, in good agreement with the PDG value $ 61 $ MeV.

\begin{table}[!h]
\begin{center}
\label{tab:cbar_c}
%\LARGE
\caption{The masses of low-lying charmonium states obtained in this work. 
         The fifth column is the mass of the meson state, where the first 
         error is statistical, and the second is systematic. 
         The last column is the experimental state we have identified, 
         and its PDG mass value \cite{Tanabashi:2018oca}.
         For a detailed description of each column, see the paragraph with Eq. (\ref{eq:meson_fit}). 
}
\begin{ruledtabular}
\begin{tabular}{cccccc}
%\toprule
$ \Gamma $ & $ J^{PC} $ & $ [t_1,t_2] $ & $\chi^2$/dof & Mass(MeV) & PDG \\
\hline
%G11  & 
$ \Id $ & $ 0^{++} $  
                     & [14,25] &  1.01
                     & 3403(16)(13) 
                     & $ \chi_{c0}(3415) $ \\
%G55  & 
$ \gamma_5 $ & $ 0^{-+} $ & [15,29] & 1.17 & 2989(6)(4) & $ \eta_c(2984) $ \\ 
%Gv4  & $ \gamma_4 $ & $ 0^{+-} $ & exotic ! &    &   &   &     \\
%Gv   &   
$ \gamma_i $ & $ 1^{--} $ & [15,28] & 0.65 & 3112(7)(5) & $ J/\psi(3097) $ \\
%G4v & $ \gamma_4\gamma_i $ & $ 1^{--} $ & $ 1^3 S_1 $ & & & & $ J/\psi(3097) $ \\
%Ga   & 
$ \gamma_5\gamma_i $ & $ 1^{++} $ 
                                  & [14,21] & 1.13 
                                  & 3513(23)(10) 
                                  & $ \chi_{c1}(3510) $ \\
%Gt   & 
%$ \gamma_5\gamma_4\gamma_i $ & $ 1^{+-} $  
$ \epsilon_{ijk} \gamma_j \gamma_k $ & $ 1^{+-} $  
                             & [17,25] & 0.39 
                             & 3527(14)(19) 
                             & $h_c(3524)$ \\
%\bottomrule
\end{tabular}
\end{ruledtabular}
\end{center}
\end{table}

Next, we turn to the charmonium states extracted from the ground states of $\cbar\Gamma\c$. 
Our results of the masses of the low-lying states of 
charmonium are summarized in Table IV.    
The time-correlation function and the effective mass of 
$ \cbar \Gamma c $ are plotted in Appendix \ref{cbar_c}.
Evidently, the theoretical masses of charmonium in Table IV  
are in good agreement with the PDG values. 
Note that the theoretical result of the hyperfine splitting ($ 1^3 S_1 - 1^1 S_0 $) is   
$ 123(9)(6) $ MeV, in good agreement with the PDG value $ 113 $ MeV.

\subsection{$B_s$ and $B_c$ mesons}

Our results of the masses of the 
low-lying states of $B_s$ mesons   
are summarized in Table \ref{tab:bbar_s}. 
The time-correlation function and the effective mass of $ \bbar \Gamma s $ 
are plotted in Appendix \ref{bbar_s}.
Here we have identified the scalar $ \bbar\s $ meson  
with the state $ B^*_{sJ}(5850) $ observed in high energy experiments, 
due to the proximity of their masses.  
This predicts that $ B_{sJ}^*(5850) $ possesses $ J^P = 0^+ $, 
which can be verified by high energy experiments in the future.  
Moreover, the pseudovector meson (the last entry in Table \ref{tab:bbar_s})
has not been observed in high energy experiments, thus it serves as 
a prediction of $N_f=2+1+1+1$ lattice QCD.

\begin{table}[!h]
\begin{center}
%\LARGE
\caption{The masses of low-lying $ B_s $ meson states obtained in this work.    
         The fifth column is the mass of the meson state, where the first 
         error is statistical, and the second is systematic. 
         The last column is the experimental state we have identified, 
         and its PDG mass value \cite{Tanabashi:2018oca}.
         For a detailed description of each column, see the paragraph with Eq. (\ref{eq:meson_fit}). 
}
\vspace{0.5cm}
\begin{ruledtabular}
\begin{tabular}{cccccc}
$ \Gamma $ & $ J^{P} $  
           & $ [t_1,t_2] $ & $\chi^2$/dof & Mass(MeV) & PDG \\
\hline
%
% G11    
%
$ \Id $ & $ 0^{+} $  
                    & [15,24]  & 0.37 
                    & 5839(30)(18) 
                    &  $B^*_{sJ}(5850)$  \\
% G55 
%    
$\gamma_5$ & $0^{-}$   
           & [23,29] & 0.79 
           & 5406(16)(17) 
           & $ B_s (5367) $ \\ 
%
% Gv4   & $ \gamma_4 $ & $ 0^{+} $ & exotic !  &  &    &   &  \\
%
% Ga4   
%
%$\gamma_5\gamma_4$ & $ 0^{-} $ & $ 1^1 S_0 $ &  &  &  &  \\
%
% Gv   
%
$\gamma_i$ & $ 1^{-} $  
           & [18,29] & 0.66 
           & 5430(17)(18) 
           & $ B_s^*(5415)$ \\ 
%
% G4v  
%
%$\gamma_4\gamma_i $ & $ 1^{-} $ & $ 1^3 S_1 $ &  & & &  \\
%
% Ga  
%
$\gamma_5\gamma_i$ & $ 1^{+} $  
                               & [16,22] & 0.58 
                               & 5839(23)(14) 
                               & $ B_{s1}$(5830) \\
%
% Gt 
%
%$\gamma_5\gamma_4\gamma_i$ & $ 1^{+} $  
$\epsilon_{ijk} \gamma_j\gamma_k$ & $ 1^{+} $ 
                           & [16,23] & 0.56 
                           & 5909(26)(34) 
                           &  \\
\end{tabular}
\end{ruledtabular}
\label{tab:bbar_s}
\end{center}
\end{table}

Finally, we turn to the heavy mesons with beauty and charm. 
In Table \ref{tab:bbar_c}, we summarize our results of the masses of $ B_c $ mesons
extracted from the ground states of $\bbar \Gamma c $. 
The time-correlation function and the effective mass of $ \bbar \Gamma c $ 
are plotted in the Appendix \ref{bbar_c}.
Except for the pseudoscalar meson $ B_c(6275) $, other four meson states have not been
observed in high energy experiments. It is interesting to see to what extent 
the experimental results will agree with our theoretical predictions.

\begin{table}[!h]
\begin{center}
%\LARGE
\caption{The masses of low-lying $ B_c $ meson states obtained in this work.    
         The fifth column is the mass of the meson state, where the first 
         error is statistical, and the second is systematic. 
         The last column is the experimental state we have identified, 
         and its PDG mass value \cite{Tanabashi:2018oca}.
         For a detailed description of each column, see the paragraph with Eq. (\ref{eq:meson_fit}). 
}
\vspace{0.5cm}
\begin{ruledtabular}
\begin{tabular}{cccccc}
$ \Gamma $ & $ J^{P} $  
           & $ [t_1,t_2] $ & $\chi^2$/dof & Mass(MeV) & PDG \\
\hline
%
% G11    
%
$ \Id $ & $ 0^{+} $  
                    & [20,28] & 1.17 
                    & 6766(38)(16) 
                    &  \\
% G55 
%    
$\gamma_5$ & $ 0^{-} $  
                       & [15,31] & 1.02 
                       & 6285(6)(5) 
                       & $ B_c(6275) $ \\
%
% Gv4   & $ \gamma_4 $ & $ 0^{+} $ & exotic ! &  &  &  &  \\
%
% Ga4   
%
%$\gamma_5\gamma_4$ & $ 0^{-} $ &  &  &  &  \\
%
% Gv   
%
$\gamma_i$ & $ 1^{-} $  
                       & [16,31] & 0.68 
                       & 6375(6)(7) 
                       & \\ 
                       %$ B_c^*(?) $  \\
%
% G4v  
%
%$\gamma_4\gamma_i $ & $ 1^{-} $ & $ 1^3 S_1 $ &  & & &  \\
%
% Ga  
%
$\gamma_5\gamma_i$ &$ 1^{+}$  
                             & [21,32] & 0.62  
                             & 6787(34)(28) 
                             &   \\
%
% Gt 
%
$\epsilon_{ijk}\gamma_j\gamma_k$ & $ 1^{+} $  
                           & [19,26] & 0.97 
                           & 6798(33)(17) 
                           &    \\
\end{tabular}
\end{ruledtabular}
\label{tab:bbar_c}
\end{center}
\end{table}

Before we close this section, we would like to point out that the theoretical predictions
of the meson masses in Tables \ref{tab:bbar_b}-{\ref{tab:bbar_c} 
are subject to other systematic uncertainties, e.g., due to the finite lattice spacing,  
and the tuning of $(\b, \c, \s)$ quark masses. Since there is only one lattice spacing 
in this study, it is impossible to extrapolate the meson masses to the continuum limit.
Nevertheless, in view of the fine lattice spacing ($ a \sim 0.03$) and the action is $O(a)$ improved, 
we expect that the discretization uncertainty is negligible in comparison with the combined
statistical and systematic uncertainties in Tables \ref{tab:bbar_b}-{\ref{tab:bbar_c}.
Moreover, we also expect that the systematic uncertainty 
due to the tuning of quark masses (with $ \delta m_q/m_q \lesssim 1\% $) 
is negligible in comparison with the combined statistical and systematic uncertainties 
in Tables \ref{tab:bbar_b}-{\ref{tab:bbar_c}. 

Most importantly, all systematic uncertainties in this study 
(i.e., the unphysical $\u/\d$ quark masses, the residual masses, the tuning of quark masses,  
the finite volume, and the finite lattice spacing) can be systematically reduced/eliminated, i.e.,   
by increasing the lattice volume such that $ M_\pi L \gg 1 $ for the physical pion mass,  
by increasing $N_s$ to reduce the residual masses, by tuning the quark masses to a higher precision, 
and by generating several gauge ensembles with different lattice spacings 
such that the extrapolation to the continuum limit can be performed. 
On the other hand, this is not the case for other approaches not treating 
the $ \b $ and $ \c $ quarks (in the sea/valence) as excitations of Dirac quark fields, 
e.g., with the absence of $\b$/$\c$ quarks in the sea, just using the nonrelativistic approximation,  
the heavy quark effective field theory, or some relativistic action to treat the valence $\b$/$\c$ quarks. 
These approaches often introduce a large number of interaction terms with associated parameters, 
thus largely limit the predictive power of the theory, 
and introduce the systematic errors which cannot be reduced/eliminated 
by going to larger volumes and/or smaller lattice spacings. 
Strictly speaking, results coming from these studies are not theoretical predictions 
from the first principles of QCD (or the Standard Model), 
regardless of whether these results are in good agreement with the HEP experimental results or not.

%%%%%%%%%%%%%%%%%%%%%%%%%%%%%%%%%%%%%%%%%%%%%%%%%%%%

\section{Quark masses of ($\b, \c, \s$)}

The quark masses cannot be measured directly in high energy experiments 
since quarks are confined inside hadrons. 
Therefore, the quark masses can only be determined by comparing theoretical calculations of 
physical observables with the experimental values. 
For any field theoretic calculation, the quark masses depend on the regularization,
as well as the renormalization scheme and scale.
For lattice QCD, the hadron masses can be computed
nonperturbatively from the first principles, and from which
the quark masses can be determined.

We have used the mass of the vector meson $ \Upsilon(9460) $
to fix the bare mass of $ \b $ quark equal to $ m_b = 0.850(5) a^{-1} $. 
To transcribe the bare mass to the corresponding value in the usual renormalization scheme
$ \overline{\mbox{MS}} $ in high energy phenomenology, 
one needs to compute the lattice renormalization
constant $ Z_m = Z_s^{-1} $, where $ Z_s $ is the renormalization
constant for $ \bar\psi \psi $. In general, $ Z_m $ should be 
determined nonperturbatively. However, in this study, 
the lattice spacing is rather small ($ a \simeq 0.03 $ fm), 
thus it is justified to use the one-loop perturbation formula \cite{Alexandrou:2000kj}
\bea
\label{eq:Zs}
Z_s(\mu) = 1 + \frac{ g^2 }{ 4 \pi^2 } \left[ \mbox{ln} ( a^2 \mu^2 ) + 0.17154 \right] 
          \hspace{8mm} (m_0=1.30).
\eea
At $ \beta = 6.70 $, $ a^{-1} = 6.503(37) $ GeV, and $ \mu = 2 $~GeV, (\ref{eq:Zs})
gives $ Z_s = 1.1001(2) $, which transcribes the bare mass $ m_b $ 
to the $\overline{\mbox{MS}} $ mass at $ \mu = 2 $~GeV
\BAN
\label{eq:mb_MS_2}
\overline{m}_{b}(2 \mbox{ GeV}) = m_b Z_m(2 \mbox{ GeV}) = 5.024 \pm 0.025 \mbox{ GeV},   
\EAN
where the error bar combines (in quadrature) the statistical error and the systematic errors 
of the lattice spacing and the $ \b $ quark bare mass.  

To compare our result with the PDG value of $ \overline{m}_b(\overline{m}_b) $ at the scale 
$ \mu = \overline{m}_b $, we solve the equation $ \overline{m}_b = m_b Z_m(\mu=\overline{m}_b) $ 
and obtain
\bea
\label{eq:mb_MS_mb}
\overline{m}_b(\overline{m}_b) = 4.85 \pm 0.04 \mbox{ GeV},  
\eea
which is higher than the PDG value $ (4.18 \pm 0.03) $~GeV for $N_f=2+1+1$ lattice QCD, 
but is closer to the value in the 1S scheme $ m_b^{\text{1S}} = 4.65(3) $~GeV \cite{Tanabashi:2018oca}.
   
Next we turn to the charm quark mass.
Using (\ref{eq:Zs}), the charm quark bare mass $ m_c = 0.200(5) a^{-1} $ is transcribed to 
\BAN
\label{eq:mc_MS_2}
\overline{m}_{c}(2 \mbox{ GeV}) = 1.14 \pm 0.03 \mbox{ GeV},  
\EAN
where the error bar combines (in quadrature) the statistical and the systematic errors 
from the lattice spacing and the charm quark bare mass.
To compare our result with the PDG value of $ \overline{m}_c(\overline{m}_c) $,  
we solve $ \overline{m}_c = m_c Z_m(\mu=\overline{m}_c) $ and obtain
\bea
\label{eq:mc_MS_mc}
\overline{m}_c(\overline{m}_c) = 1.21 \pm 0.03 \mbox{ GeV},  
\eea
which is slightly smaller than the PDG value $ (1.280 \pm 0.025) $~GeV for $N_f=2+1+1$ lattice QCD 
\cite{Tanabashi:2018oca}.

Finally we turn to the strange quark mass.
Using (\ref{eq:Zs}), the strange quark bare mass $ m_s = 0.0150(2) a^{-1} $ is transcribed to 
\bea
\label{eq:ms_MS_2}
\overline{m}_{s}(2 \mbox{ GeV}) = 88.7 \pm 1.3 \mbox{ MeV},  
\eea
where the error bar combines (in quadrature) the statistical and the systematic ones 
from the lattice spacing and the $ \s $ quark bare mass.
Our result of the strange quark mass (\ref{eq:ms_MS_2}) 
is slightly smaller than the PDG value $ (92.9 \pm 0.7) $~MeV 
for $N_f = 2+1+1 $ lattice QCD \cite{Tanabashi:2018oca}.

\section{Concluding remarks}

This study demonstrates that the Dirac $\b$ quark can be simulated dynamically
in lattice QCD, together with the $(\c, \s, \d, \u)$ quarks.  
Even with unphysically heavy $\u$ and $\d $ quarks in the sea,
the low-lying mass spectra of mesons with valence quark contents 
$ \bbar\b $, $ \bbar\c $, $\bbar\s $, and $\cbar\c $ 
are in good agreement with the experimental values. 
Also, we have several predictions which have not been observed in high energy experiments,  
i.e., predicting the mass and the $ J^P $ 
of four $B_c$ meson states (see Table \ref{tab:bbar_c}), 
the $J^P $ of $ B_{sJ}^*(5850) $ to be $ 0^+$,   
and the mass and the $ J^P $ of the pseudovector $B_s$ meson state (see Table \ref{tab:bbar_s}). 
Moreover, we have determined the masses of $(\b, \c, \s) $ quarks, as given in   
(\ref{eq:mb_MS_mb}), (\ref{eq:mc_MS_mc}), and (\ref{eq:ms_MS_2}) respectively. 

These results imply that it is feasible to simulate lattice QCD 
with physical $ (\u,\d,\s,\c,\b) $ 
domain-wall quarks on a large ($ \sim 200^4$) lattice,  
with the Exaflops supercomputers which will be available $\sim 2022$. 
Then physical observables with any ($\u, \d, \s, \c, \b$) quark contents can be computed 
from the first principles of QCD. This will provide a viable way to systematically reduce the  
uncertainties in the theoretical predictions of the Standard Model (SM), which are largely 
stemming from the sector of the strong interaction\footnote{    
The $\t$ quark can be neglected in the strong interaction 
since it is very short-lived and it decays to $W$-boson and $\b/\s/\d$ quarks 
before it can interact with other quarks through the gluons.}. 
This is crucial for unveiling any new physics beyond the standard model (SM),  
by identifying any discrepancies between the high energy experimental results 
and the theoretical values derived from the first principles of the SM with
all quarks (heavy and light) as Dirac fermions, without using nonrelativistic approximation 
or heavy quark effective field theory for $ \b $ and $ \c $ quarks.

\section*{Acknowledgements}

The author is grateful to Academia Sinica Grid Computing Center (ASGC)
and National Center for High Performance Computing (NCHC) for the computer time and facilities.
This work is supported by the Ministry of Science and Technology
(Grant Nos.~108-2112-M-003-005, and 107-2119-M-003-008).

%\begin{appendices}

\eject

\appendix

\section{$C(t)$ and the effective mass of $\bbar \Gamma \b$}
\label{bbar_b}

\begin{figure}[H]
  \centering
  \begin{tabular}{@{}c@{}c@{}}
  \includegraphics[width=7cm,clip=true]{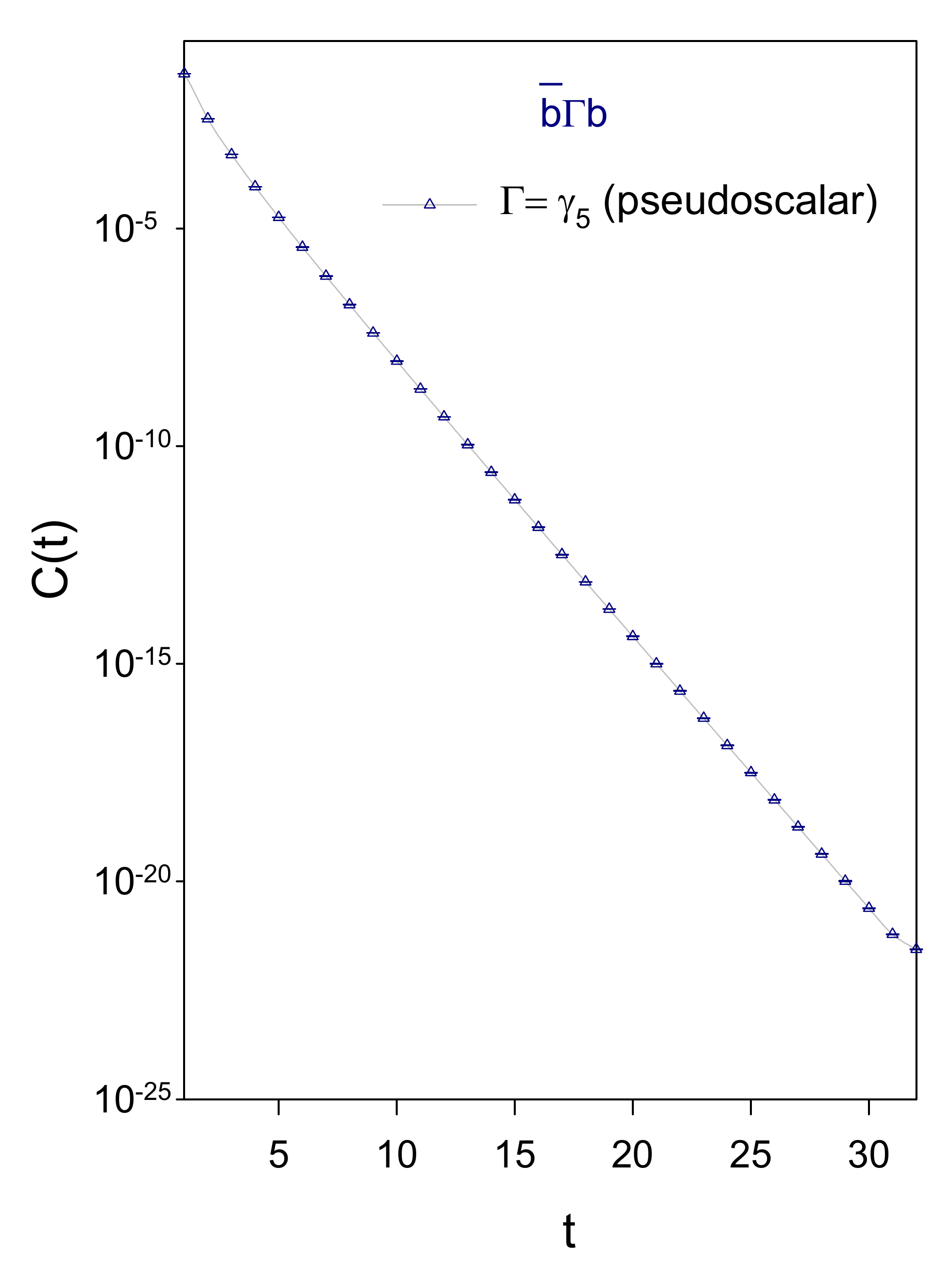}
&
  \includegraphics[width=7cm,clip=true]{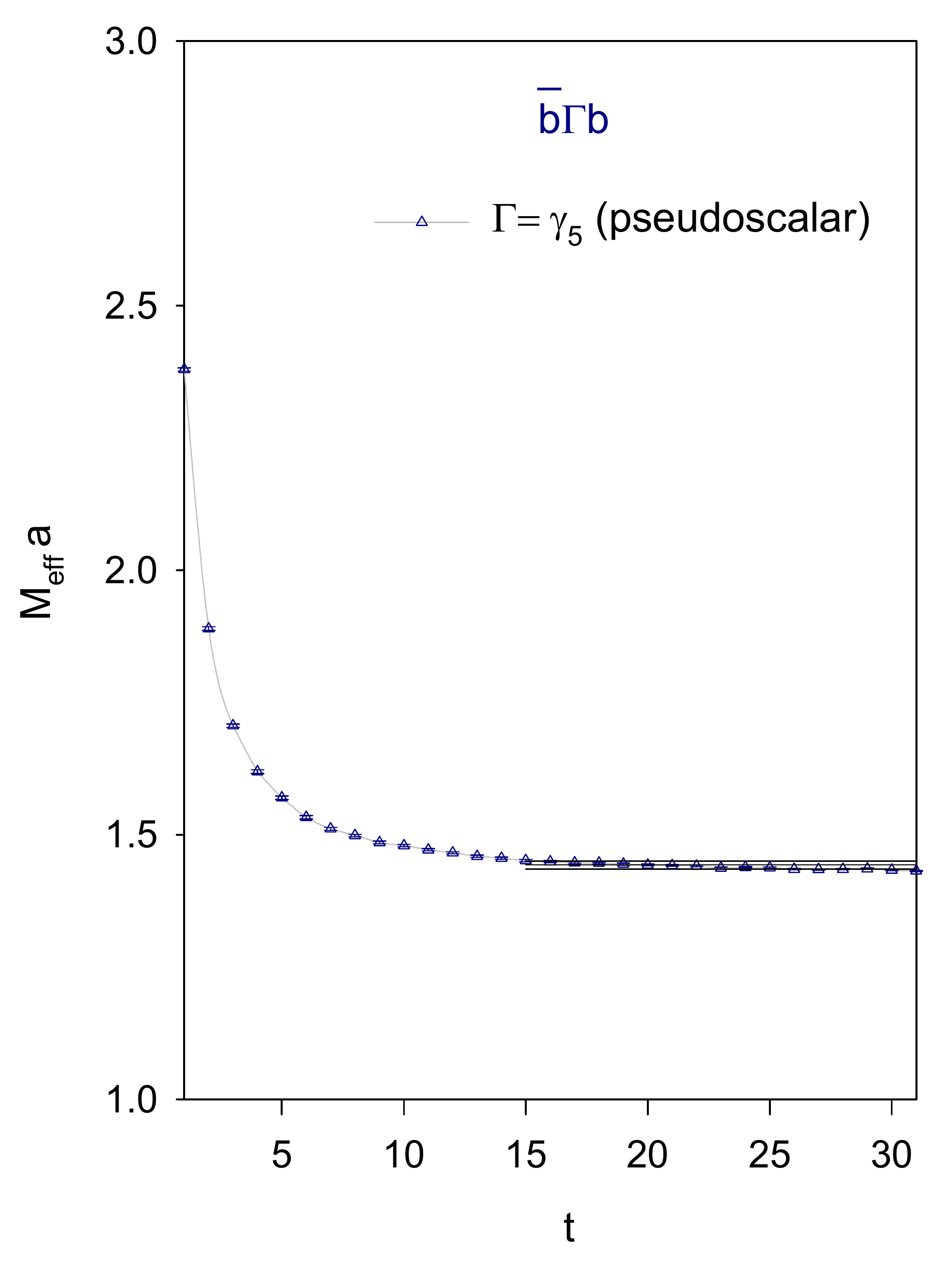}
%  \\ (a) & (b)
  \end{tabular}
  \caption{
    The time-correlation function and   
    the effective mass of the meson interpolator $\bbar \gamma_5 \b $.
  }
  \label{fig:Ct_meff_G55_bb}
\end{figure}

\begin{figure}[H]
  \centering
  \begin{tabular}{@{}c@{}c@{}}
  \includegraphics[width=7cm,clip=true]{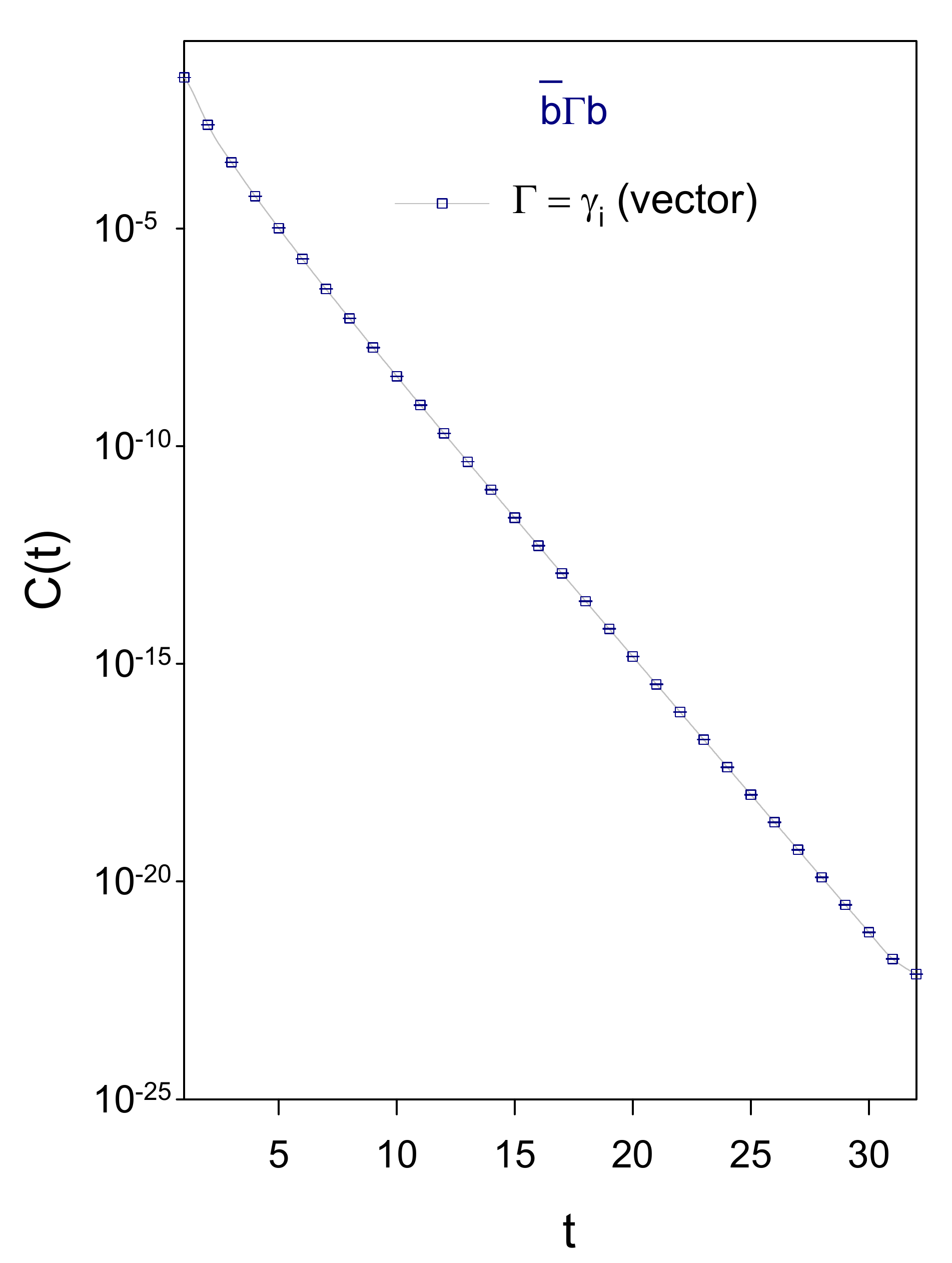}
&
  \includegraphics[width=7cm,clip=true]{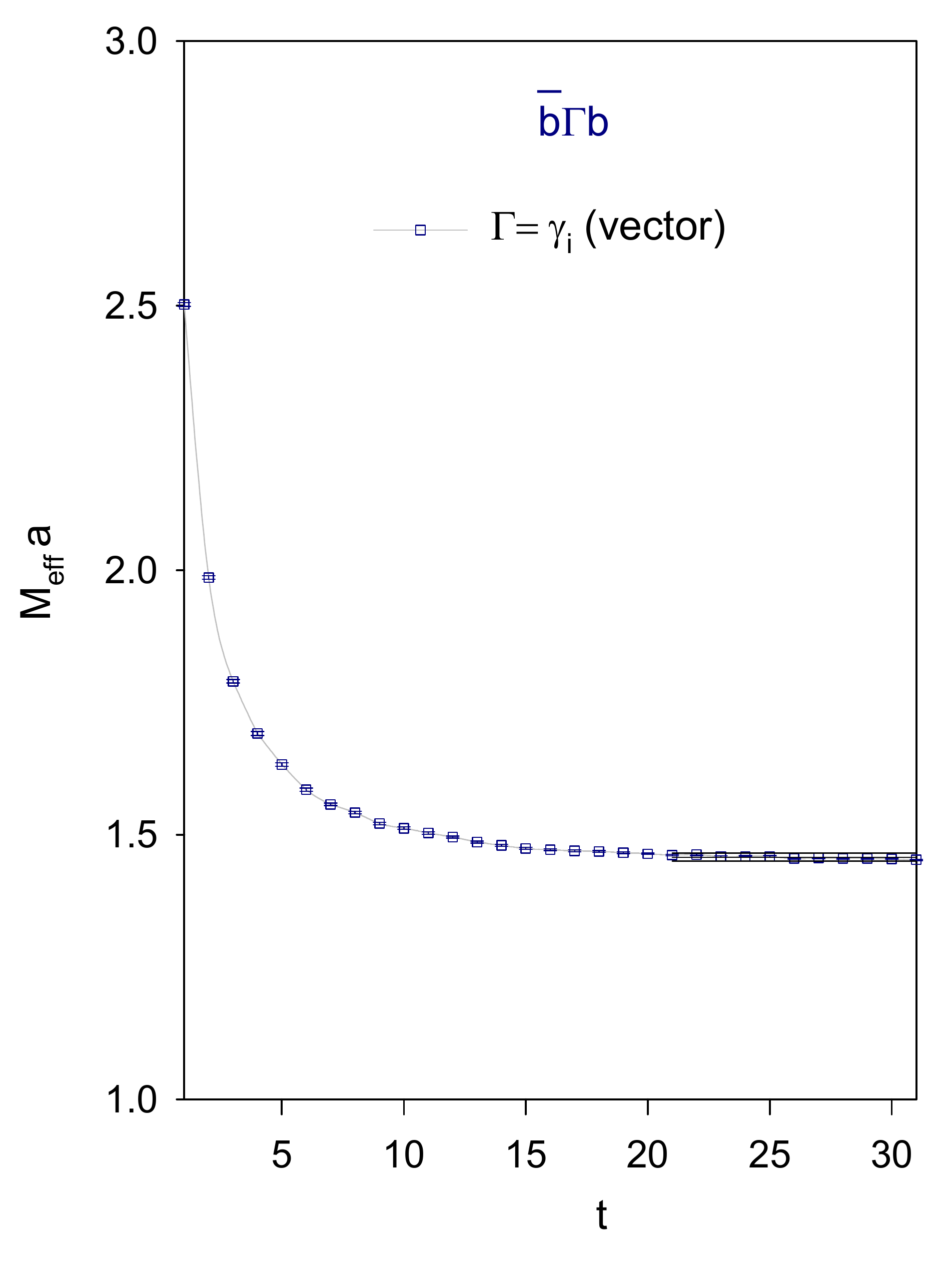}
%  \\ (a) & (b)
  \end{tabular}
  \caption{
    The time-correlation function and   
    the effective mass of the meson interpolator $\bbar \gamma_i \b $.
  }
  \label{fig:Ct_meff_Gv_bb}
\end{figure}

\begin{figure}[H]
  \centering
  \begin{tabular}{@{}c@{}c@{}}
  \includegraphics[width=7cm,clip=true]{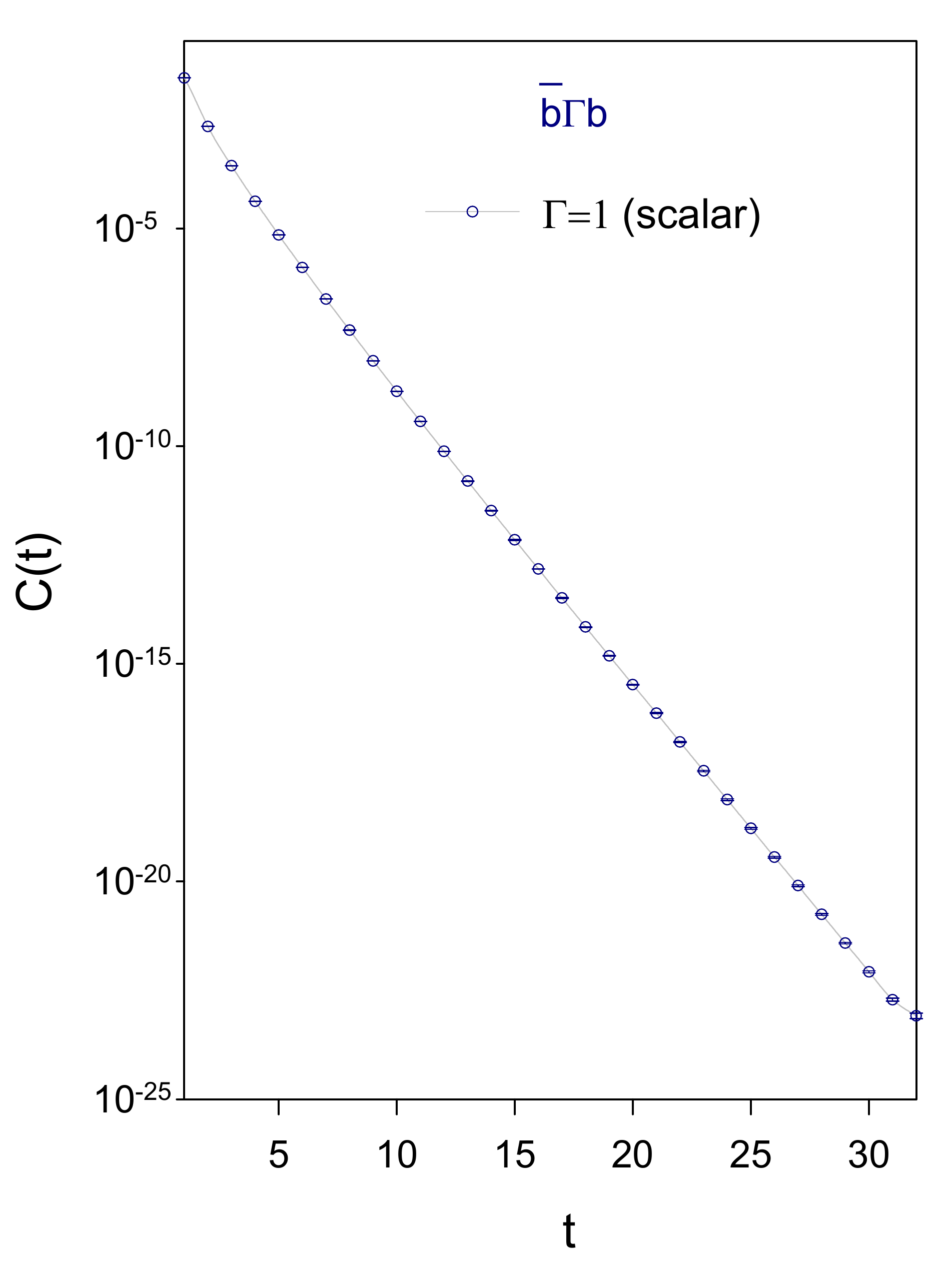}
&
  \includegraphics[width=7cm,clip=true]{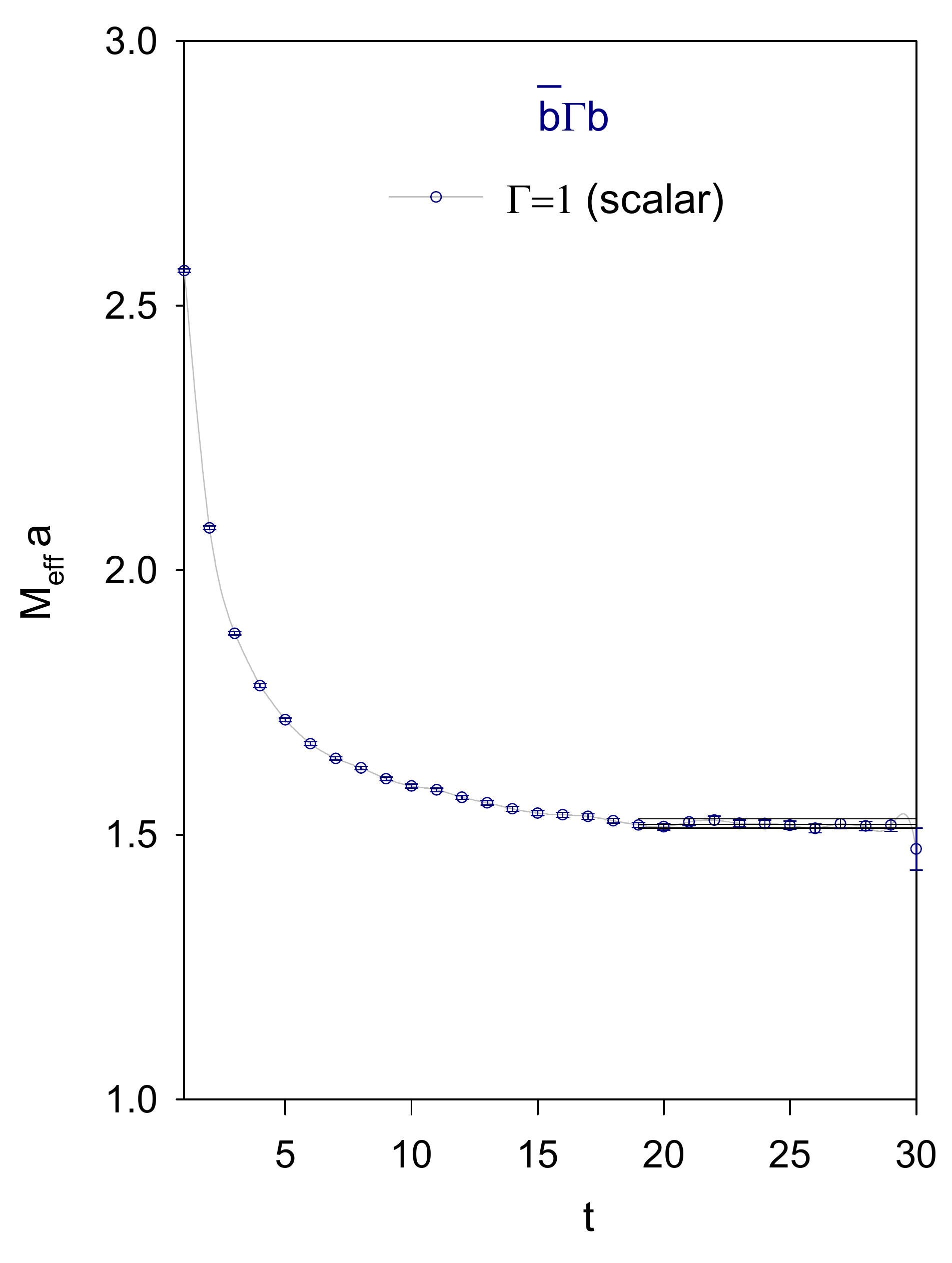}
%  \\ (a) & (b)
  \end{tabular}
  \caption{
    The time-correlation function and   
    the effective mass of the meson interpolator $\bbar \b $.
  }
  \label{fig:Ct_meff_G11_bb}
\end{figure}

\begin{figure}[H]
  \centering
  \begin{tabular}{@{}c@{}c@{}}
  \includegraphics[width=7cm,clip=true]{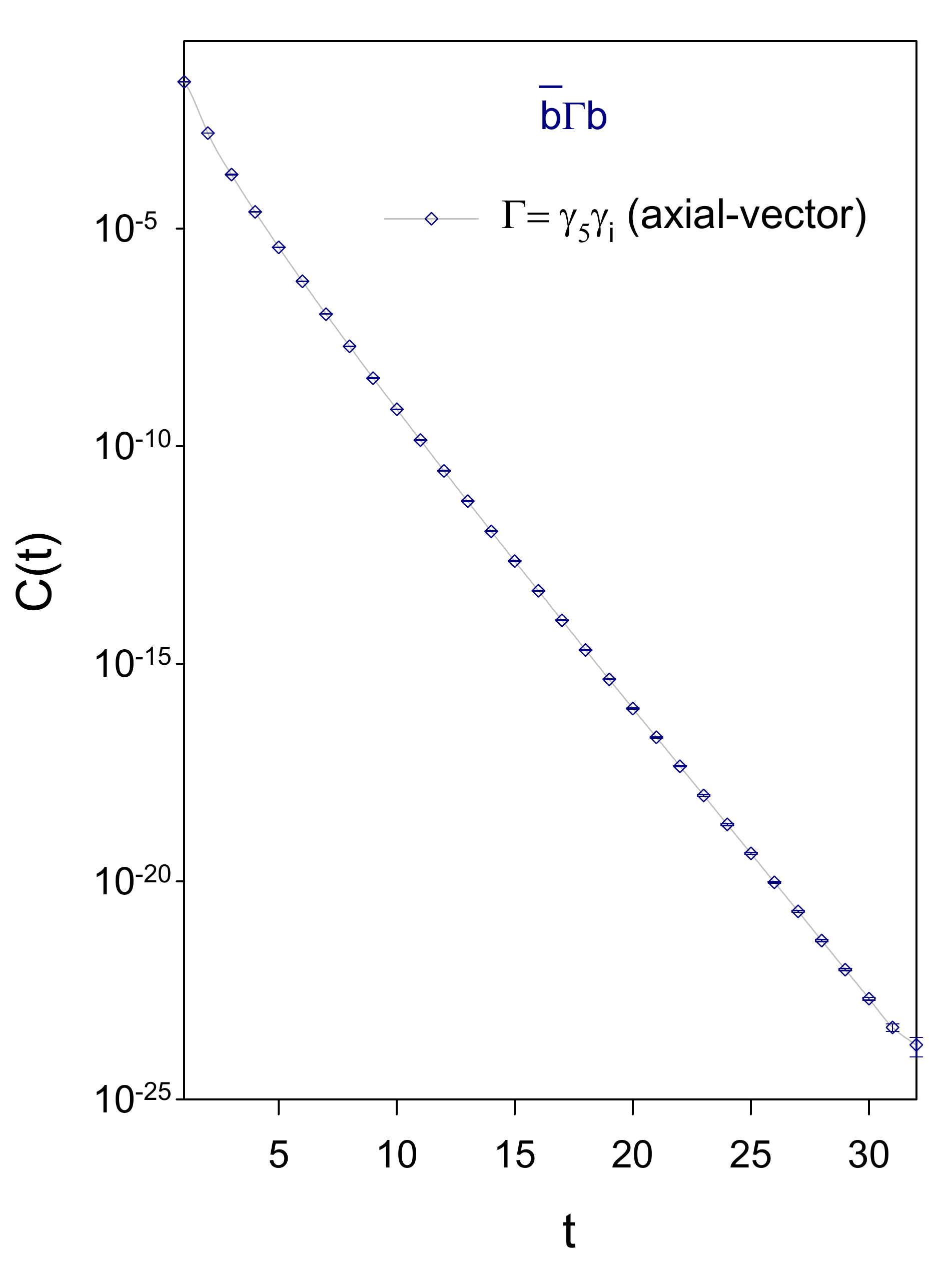}
&
  \includegraphics[width=7cm,clip=true]{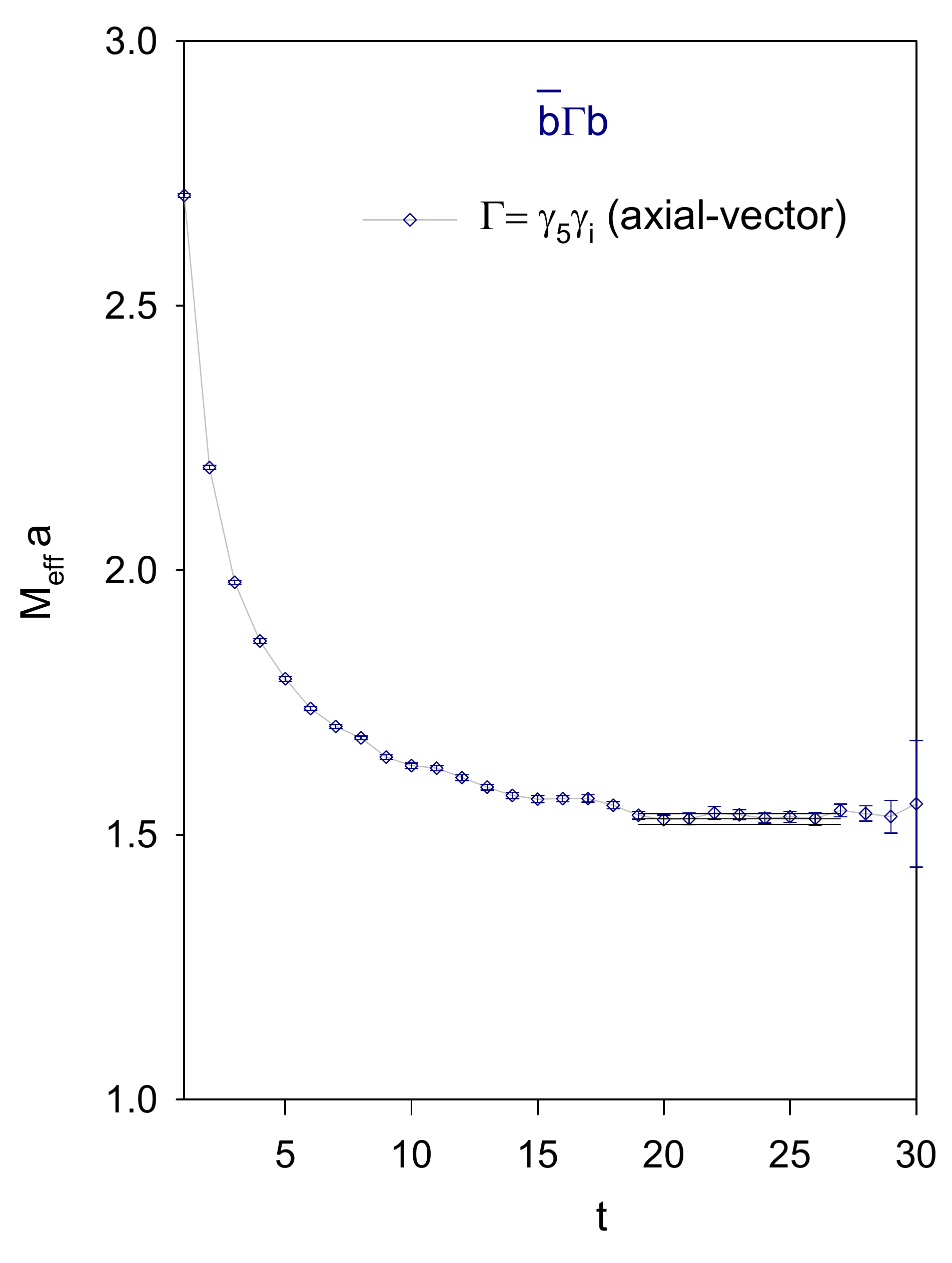}
%  \\ (a) & (b)
  \end{tabular}
  \caption{
    The time-correlation function and   
    the effective mass of the meson interpolator $\bbar \gamma_5 \gamma_i \b $.
  }
  \label{fig:Ct_meff_Ga_bb}
\end{figure}

\eject

\begin{figure}[H]
  \centering
  \begin{tabular}{@{}c@{}c@{}}
  \includegraphics[width=7cm,clip=true]{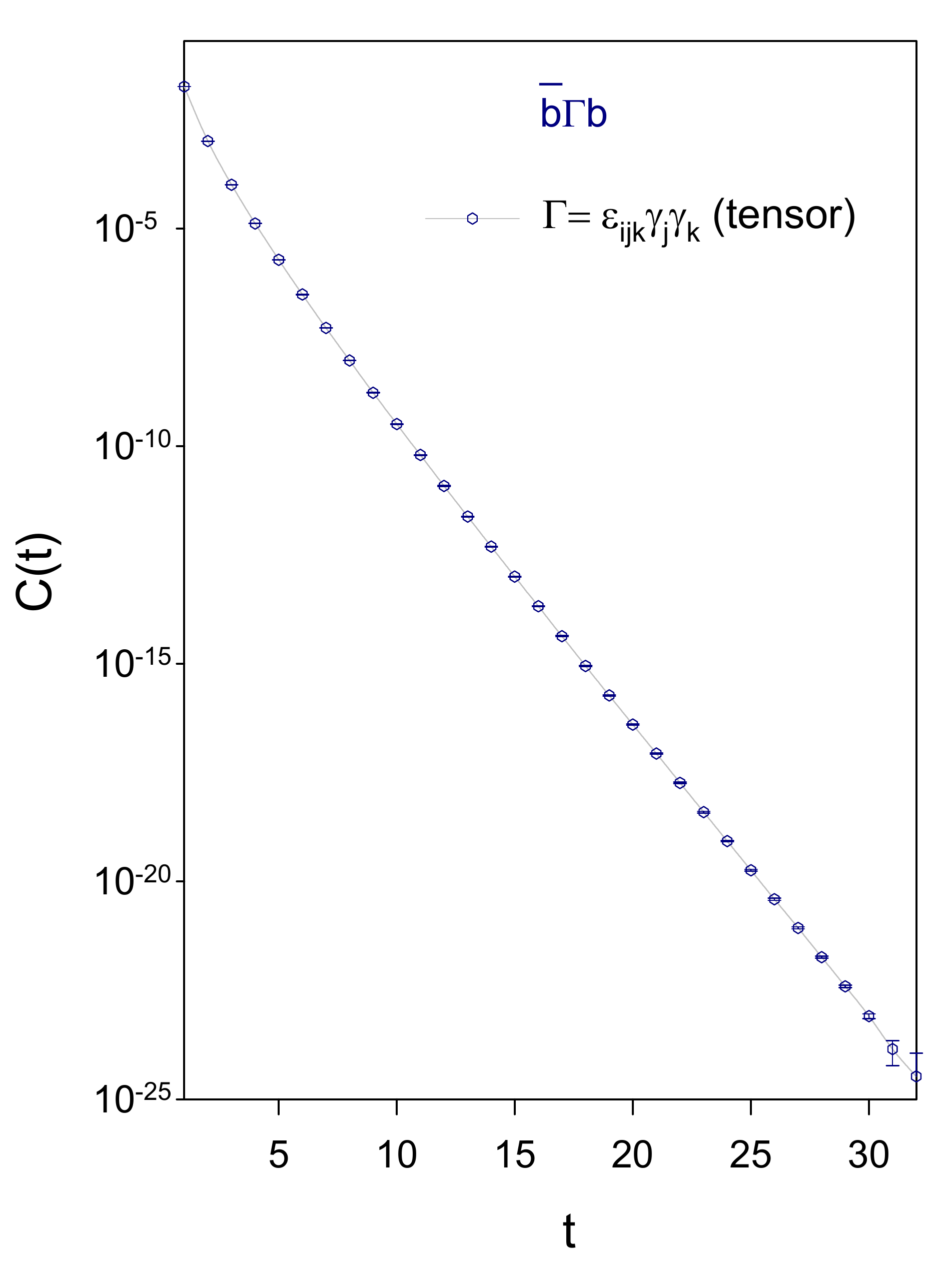}
&
  \includegraphics[width=7cm,clip=true]{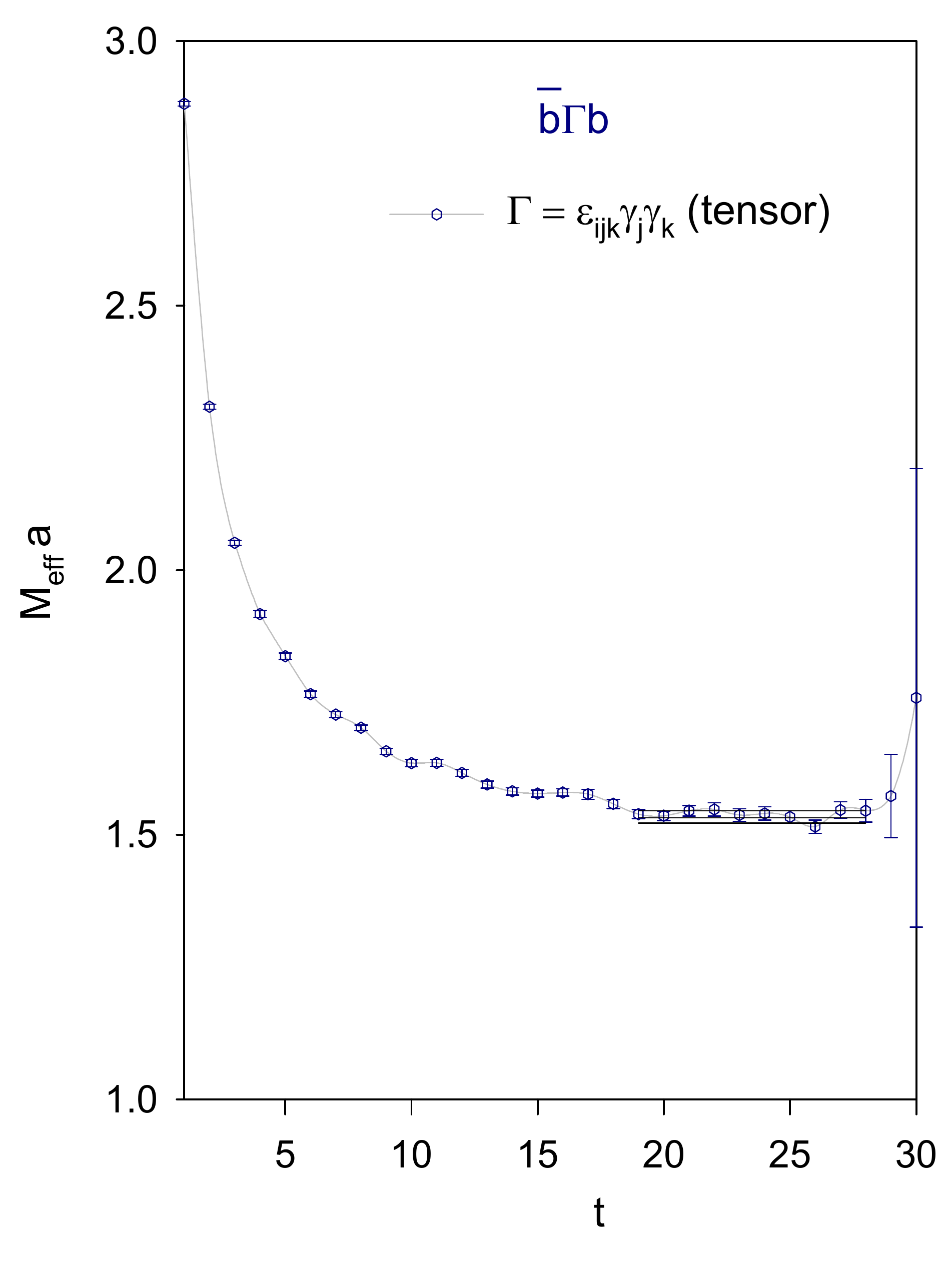}
%  \\ (a) & (b)
  \end{tabular}
  \caption{
    The time-correlation function and   
    the effective mass of the meson interpolator $\bbar \epsilon_{ijk} \gamma_j \gamma_k \b $.
  }
  \label{fig:Ct_meff_Gt_bb}
\end{figure}

\eject

\section{$C(t)$ and the effective mass of $\cbar \Gamma \c$}
\label{cbar_c}

\begin{figure}[H]
  \centering
  \begin{tabular}{@{}c@{}c@{}}
  \includegraphics[width=7cm,clip=true]{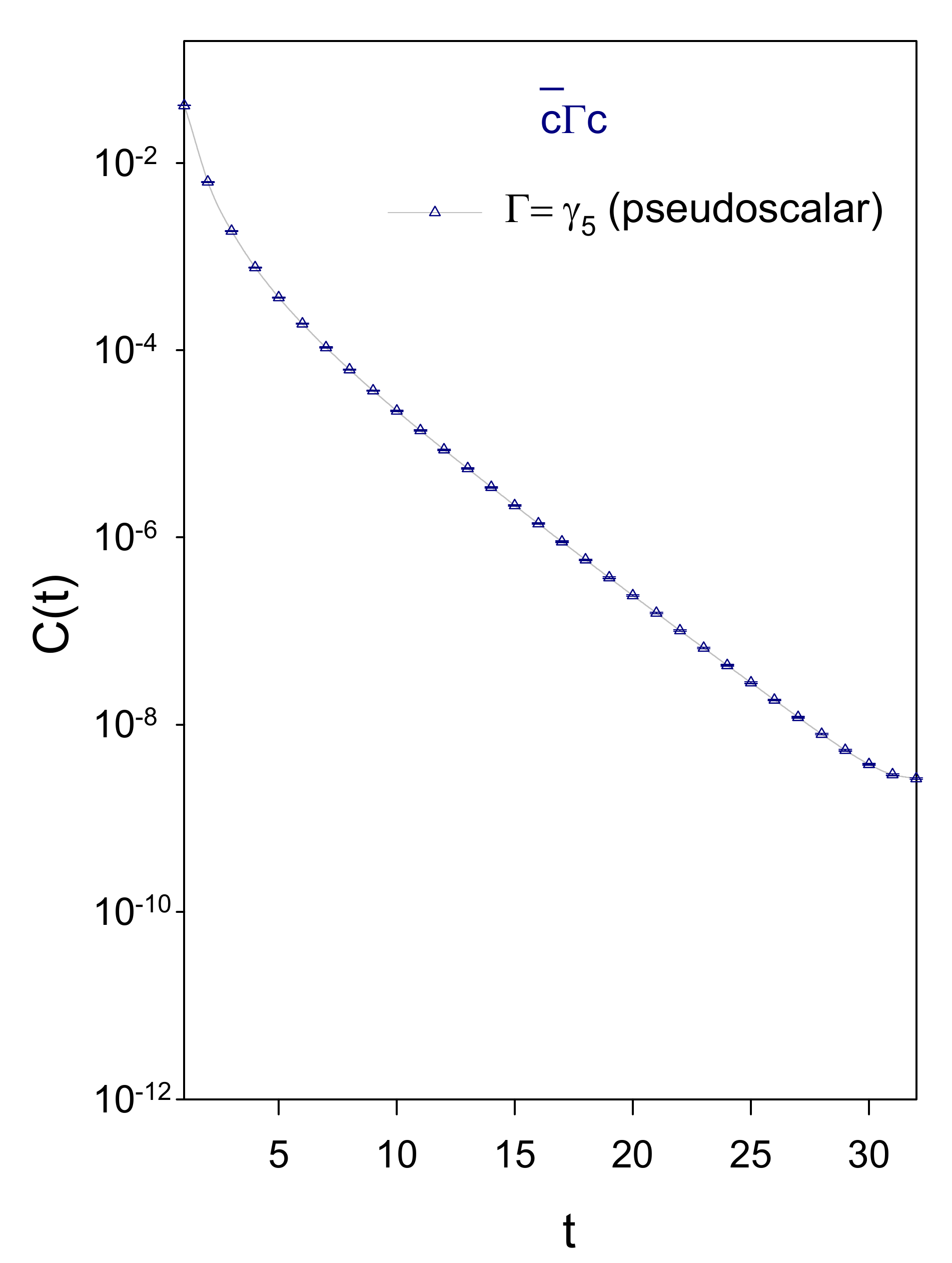}
&
  \includegraphics[width=7cm,clip=true]{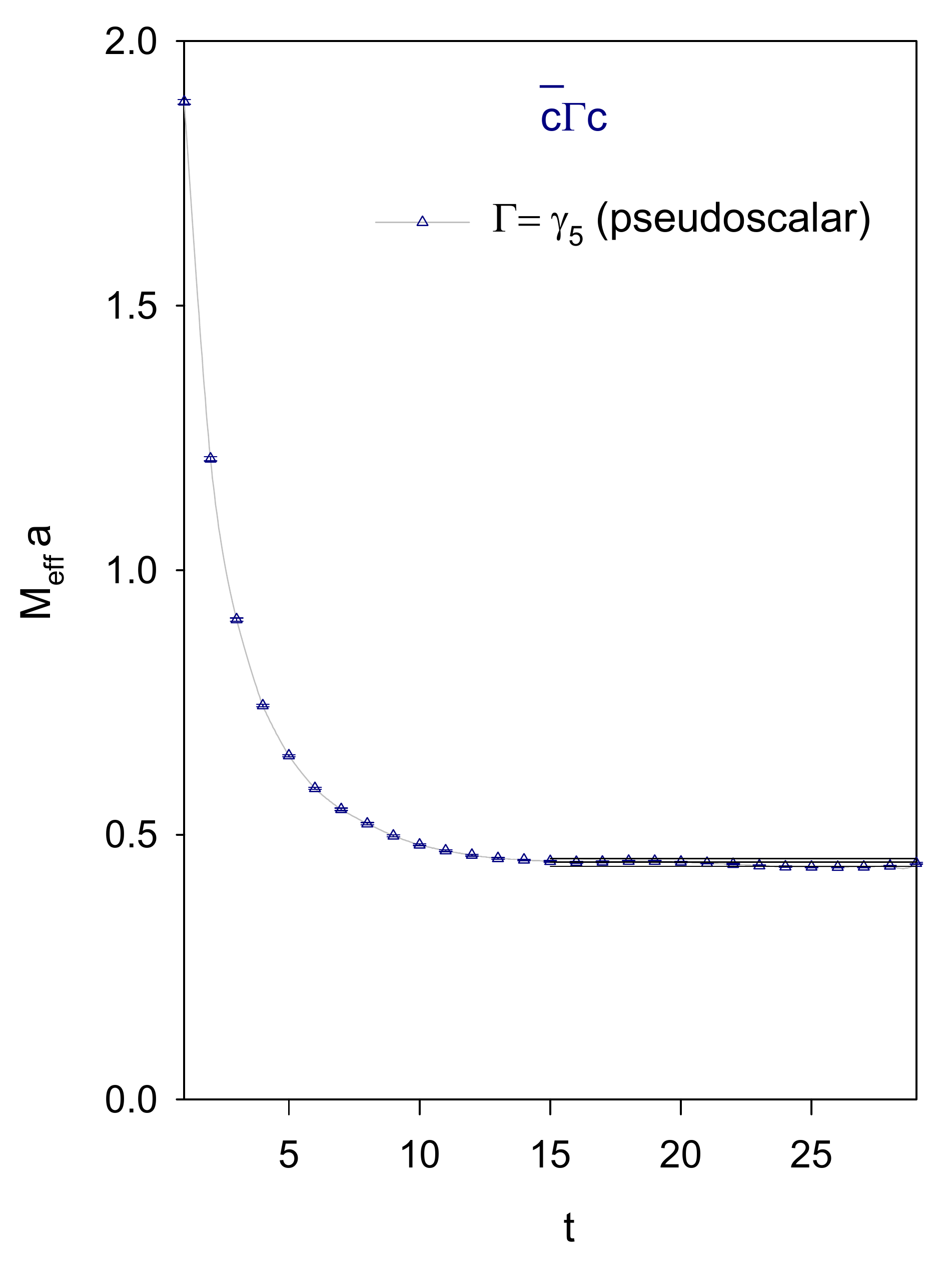}
%  \\ (a) & (b)
  \end{tabular}
  \caption{
    The time-correlation function and   
    the effective mass of the meson interpolator $\cbar \gamma_5 \c $.
  }
  \label{fig:Ct_meff_G55_cc}
\end{figure}

\begin{figure}[H]
  \centering
  \begin{tabular}{@{}c@{}c@{}}
  \includegraphics[width=7cm,clip=true]{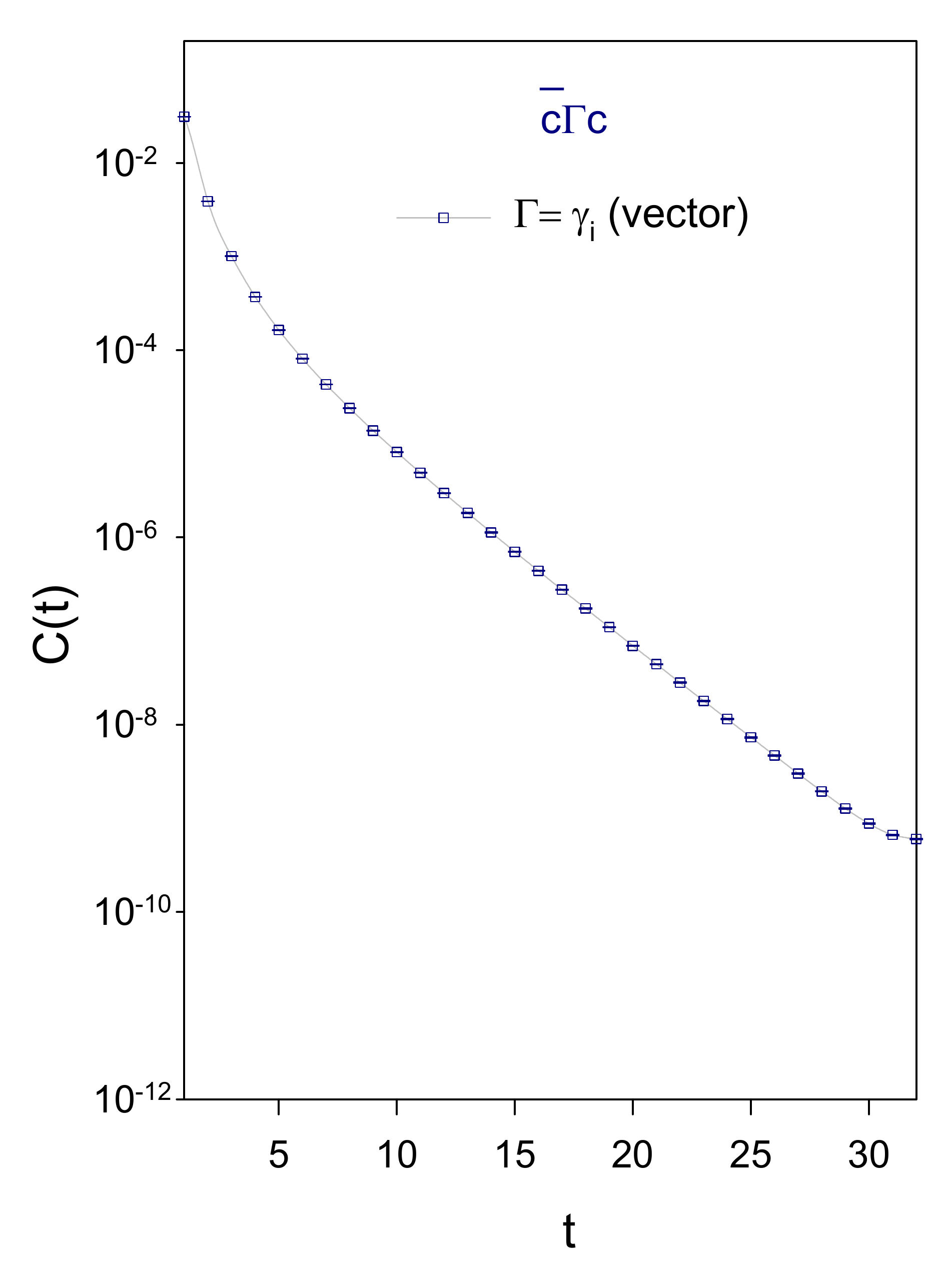}
&
  \includegraphics[width=7cm,clip=true]{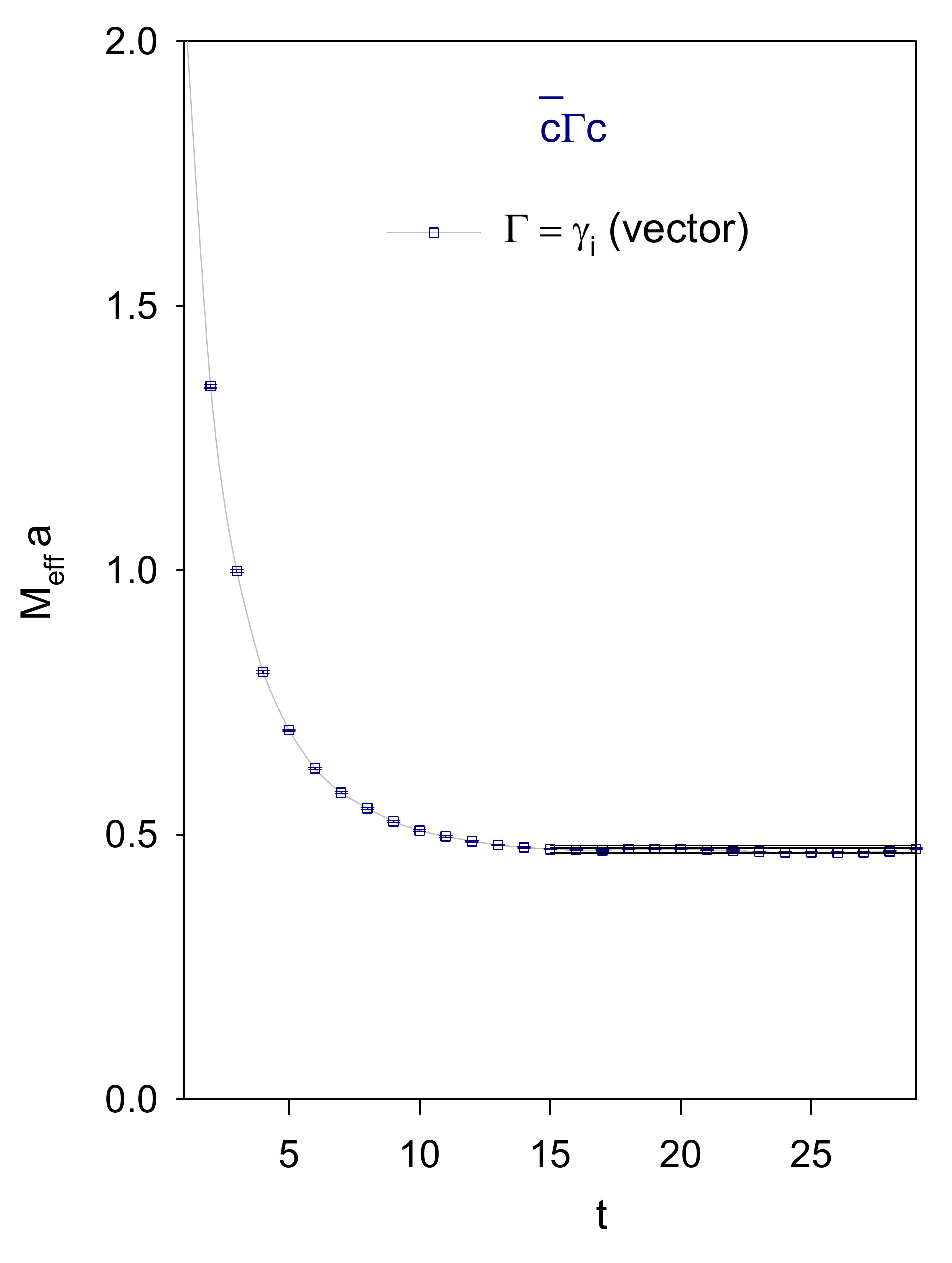}
%  \\ (a) & (b)
  \end{tabular}
  \caption{
    The time-correlation function and   
    the effective mass of the meson interpolator $\cbar \gamma_i \c $.
  }
  \label{fig:Ct_meff_Gv_cc}
\end{figure}

\begin{figure}[H]
  \centering
  \begin{tabular}{@{}c@{}c@{}}
  \includegraphics[width=7cm,clip=true]{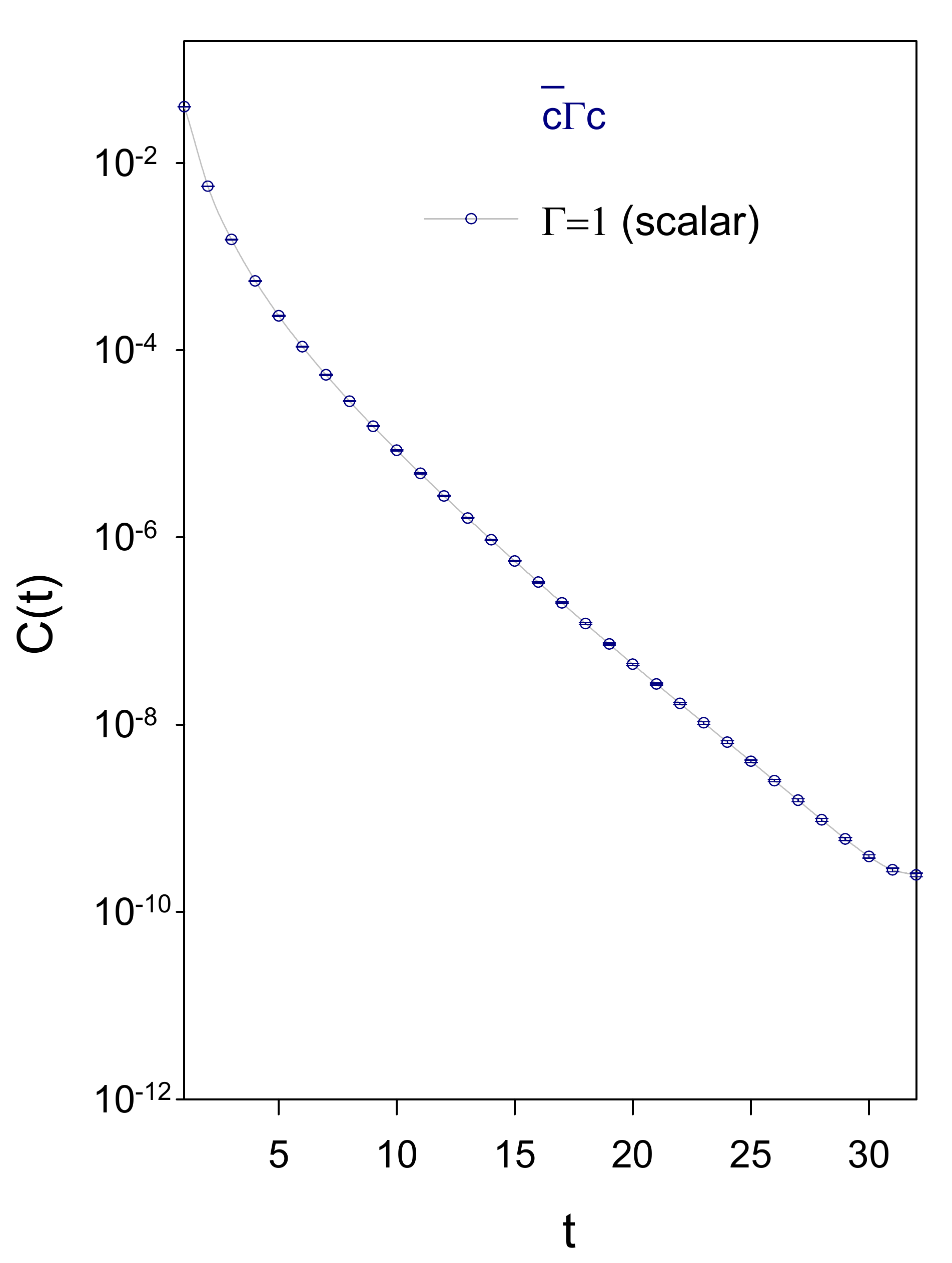}
&
  \includegraphics[width=7cm,clip=true]{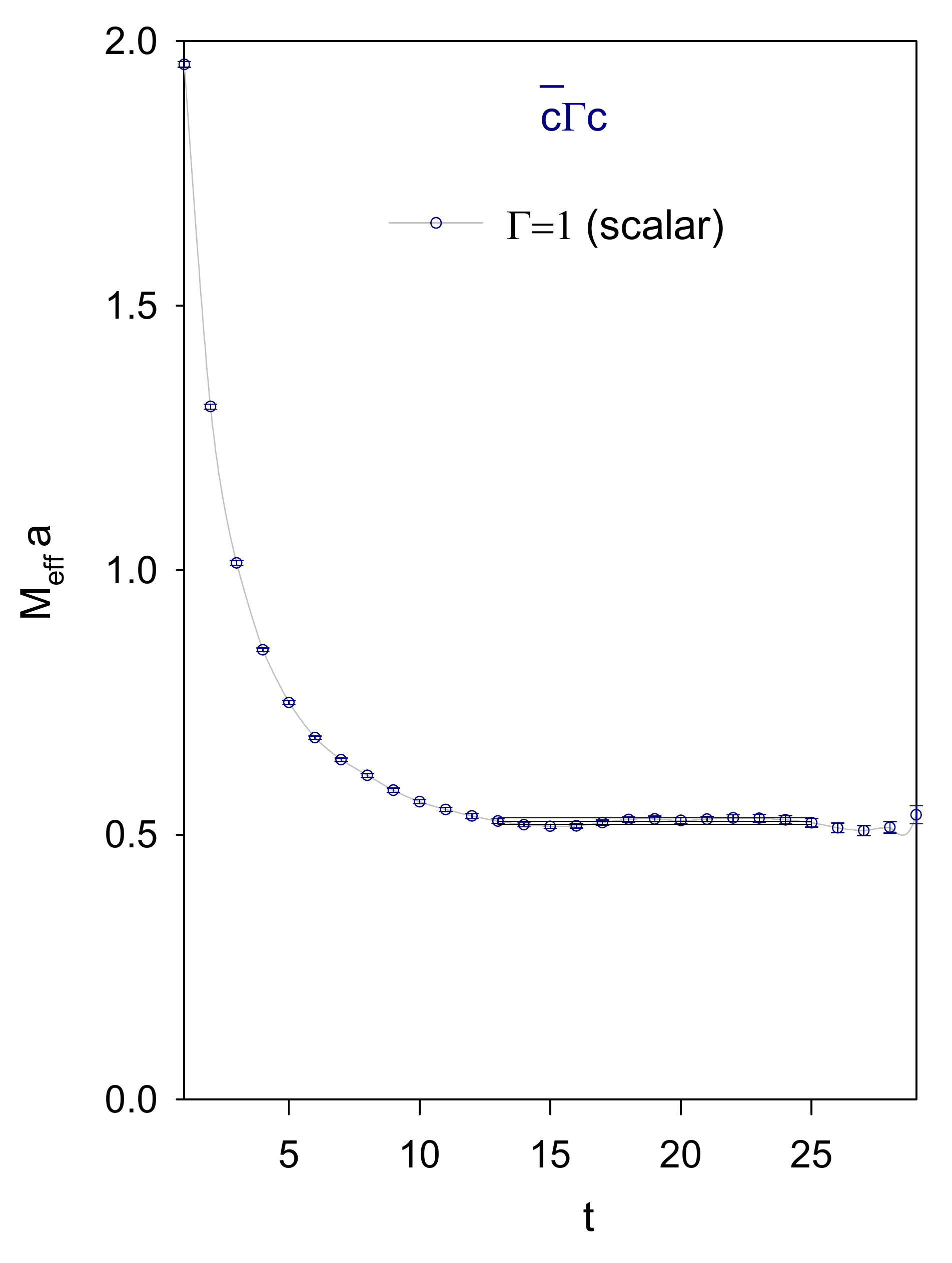}
%  \\ (a) & (b)
  \end{tabular}
  \caption{
    The time-correlation function and   
    the effective mass of the meson interpolator $\cbar \c $.
  }
  \label{fig:Ct_meff_G11_cc}
\end{figure}

\begin{figure}[H]
  \centering
  \begin{tabular}{@{}c@{}c@{}}
  \includegraphics[width=7cm,clip=true]{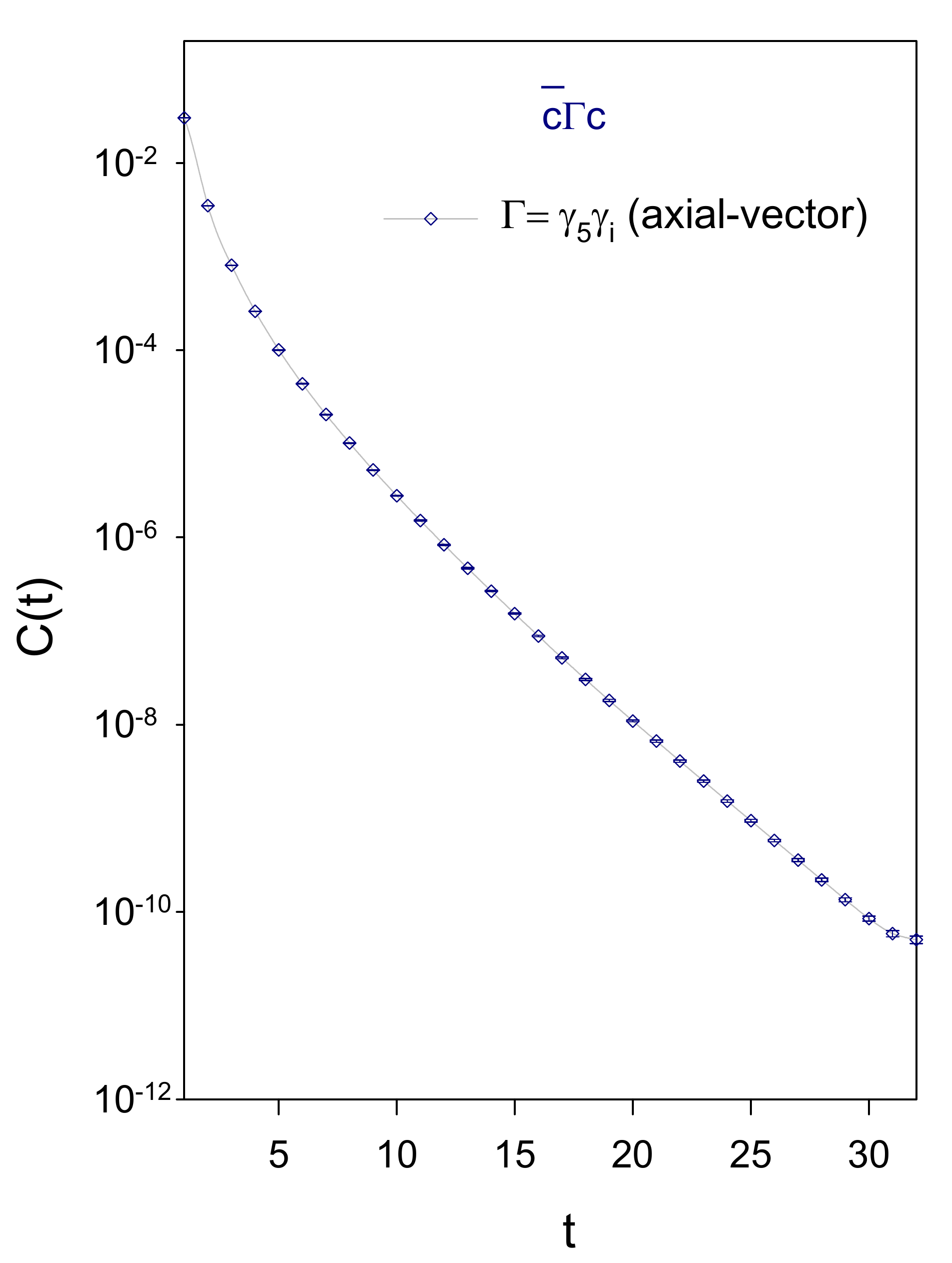}
&
  \includegraphics[width=7cm,clip=true]{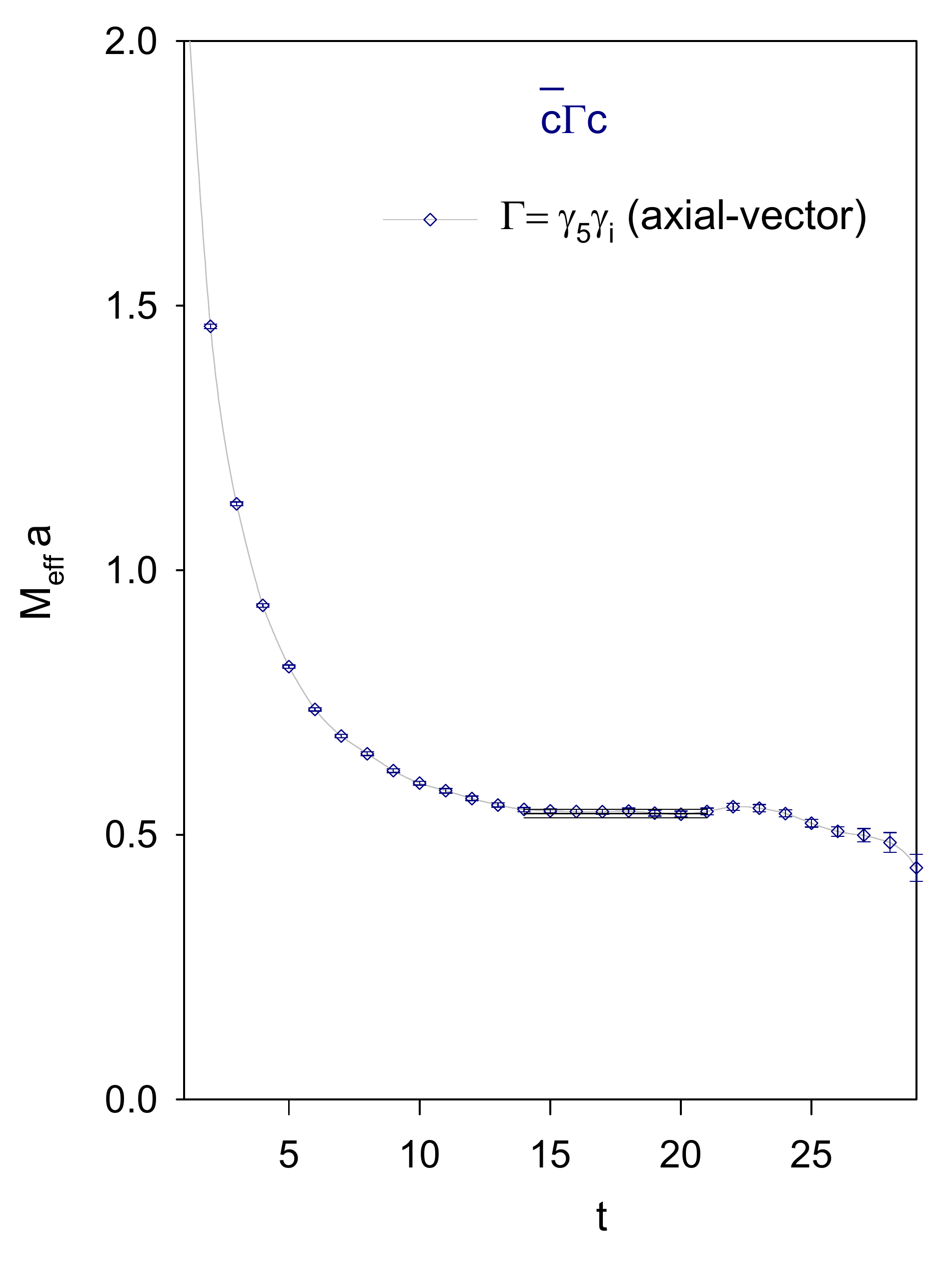}
%  \\ (a) & (b)
  \end{tabular}
  \caption{
    The time-correlation function and   
    the effective mass of the meson interpolator $\cbar \gamma_5 \gamma_i \c $.
  }
  \label{fig:Ct_meff_Ga_cc}
\end{figure}

\eject

\begin{figure}[H]
  \centering
  \begin{tabular}{@{}c@{}c@{}}
  \includegraphics[width=7cm,clip=true]{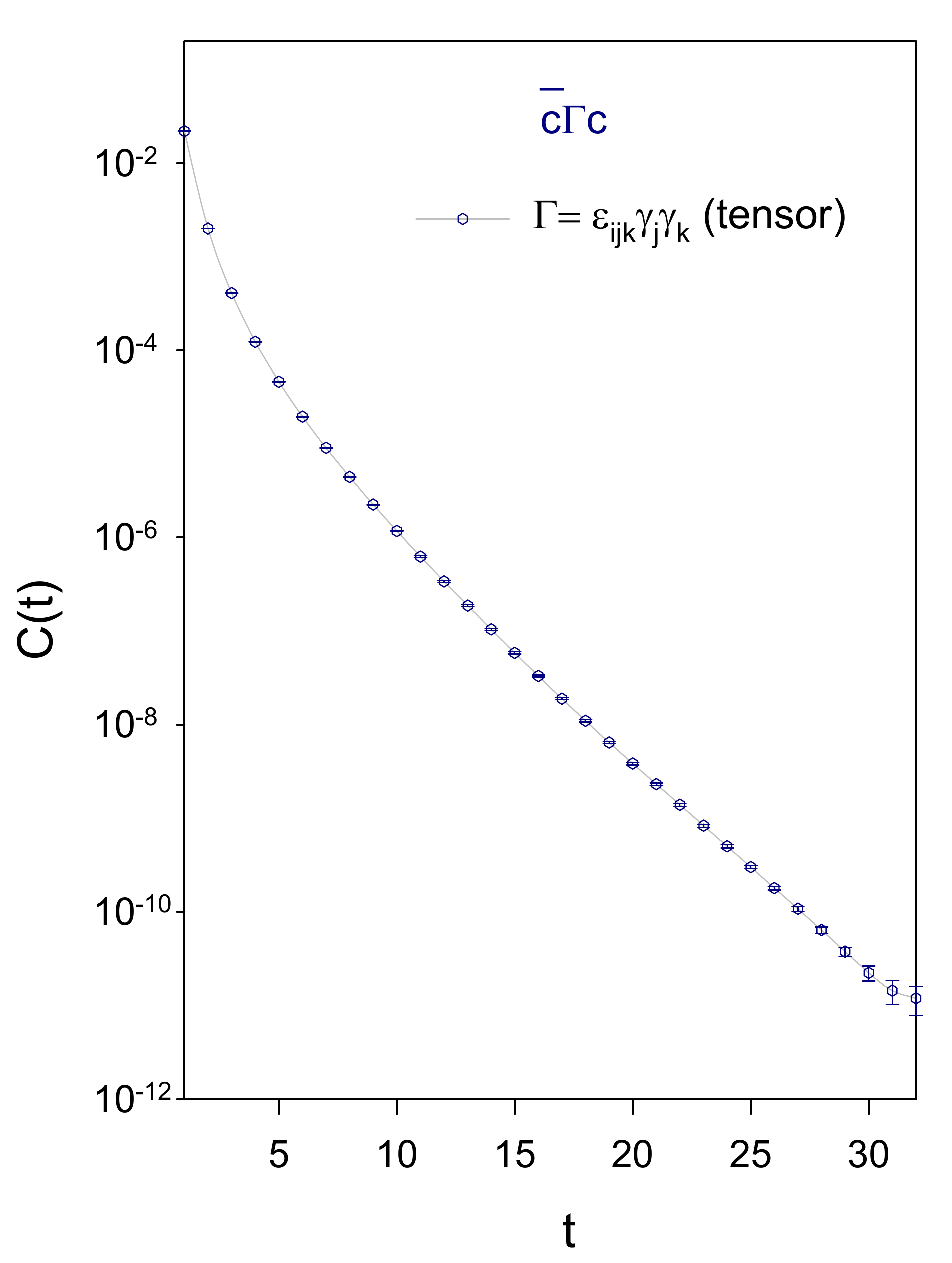}
&
  \includegraphics[width=7cm,clip=true]{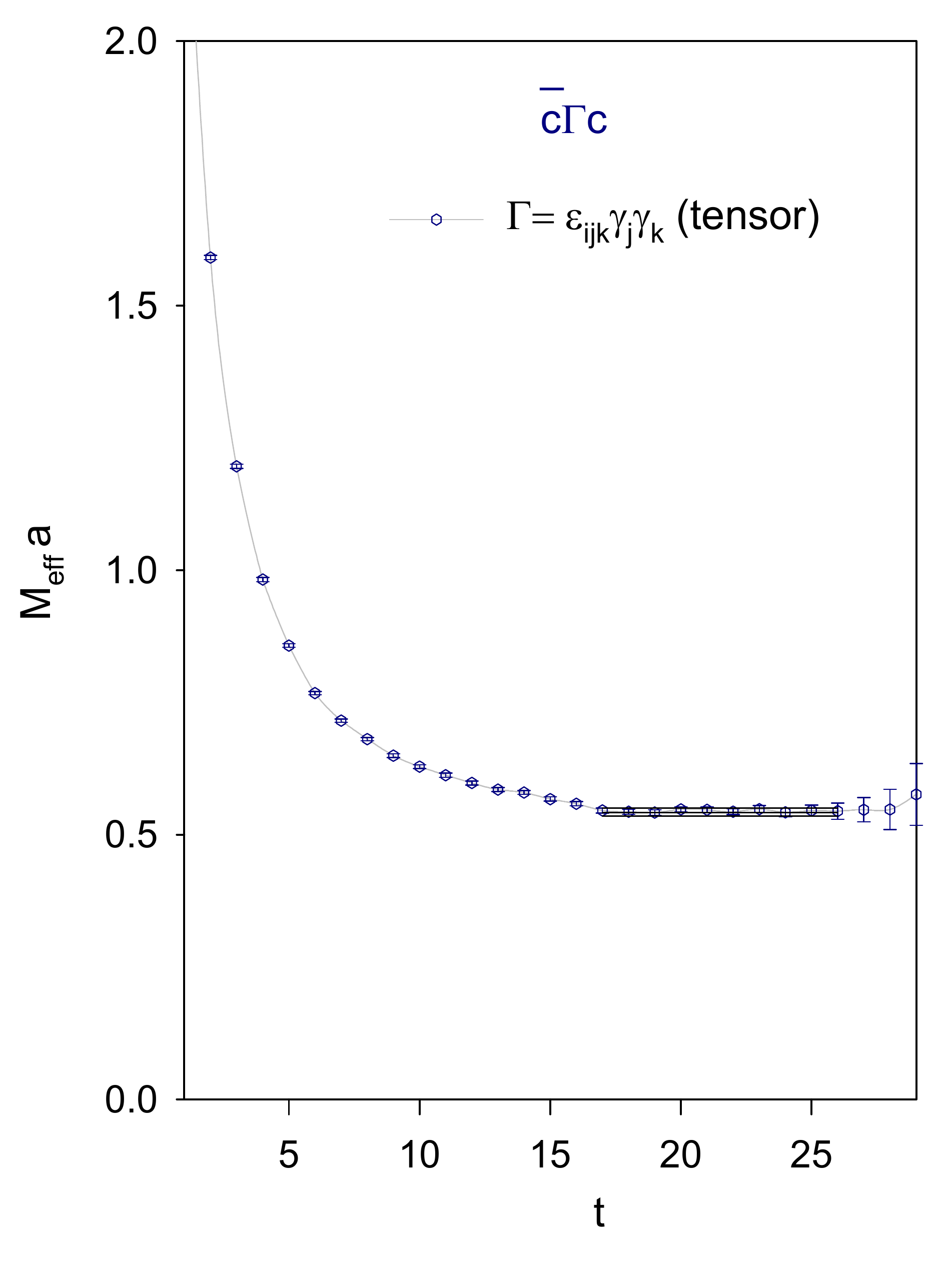}
%  \\ (a) & (b)
  \end{tabular}
  \caption{
    The time-correlation function and   
    the effective mass of the meson interpolator $\cbar \epsilon_{ijk} \gamma_j \gamma_k \c $.
  }
  \label{fig:Ct_meff_Gt_cc}
\end{figure}

\eject

\section{$C(t)$ and the effective mass of $\bbar \Gamma \c$}
\label{bbar_c}

\begin{figure}[H]
  \centering
  \begin{tabular}{@{}c@{}c@{}}
  \includegraphics[width=7cm,clip=true]{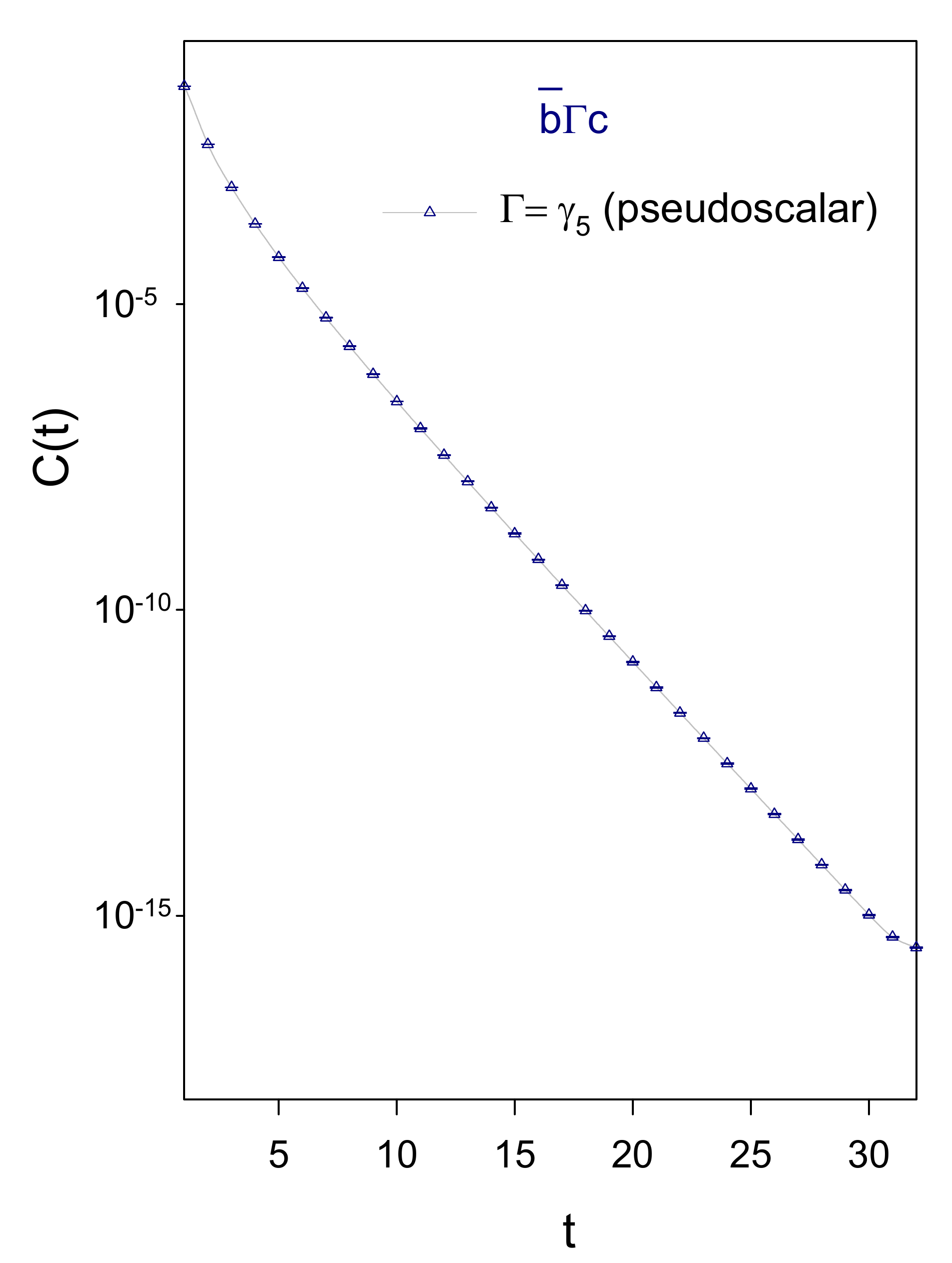}
&
  \includegraphics[width=7cm,clip=true]{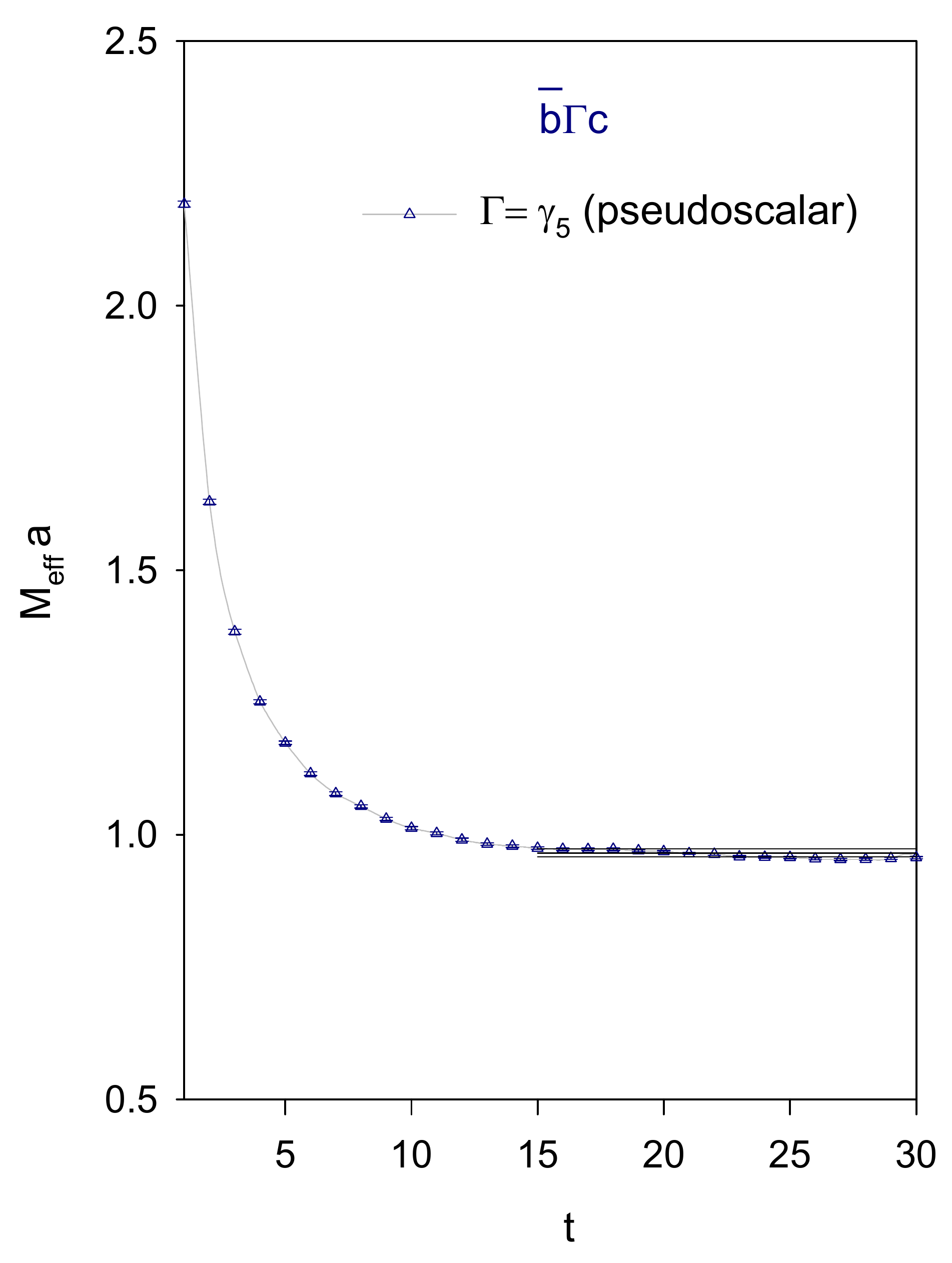}
%  \\ (a) & (b)
  \end{tabular}
  \caption{
    The time-correlation function and
    the effective mass of the meson interpolator $\bbar \gamma_5 \c $.
  }
  \label{fig:Ct_meff_G55_bc}
\end{figure}

\begin{figure}[H]
  \centering
  \begin{tabular}{@{}c@{}c@{}}
  \includegraphics[width=7cm,clip=true]{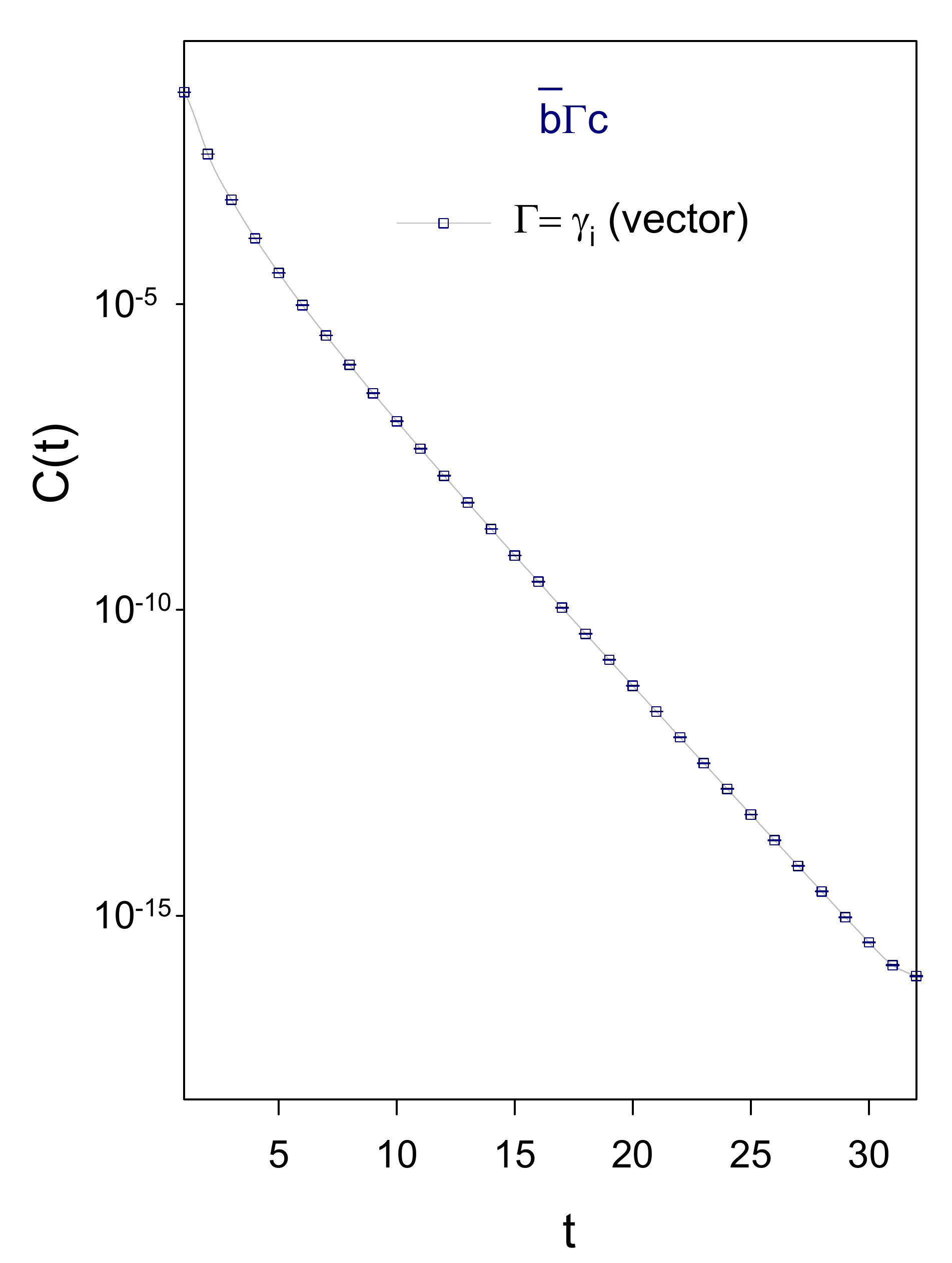}
&
  \includegraphics[width=7cm,clip=true]{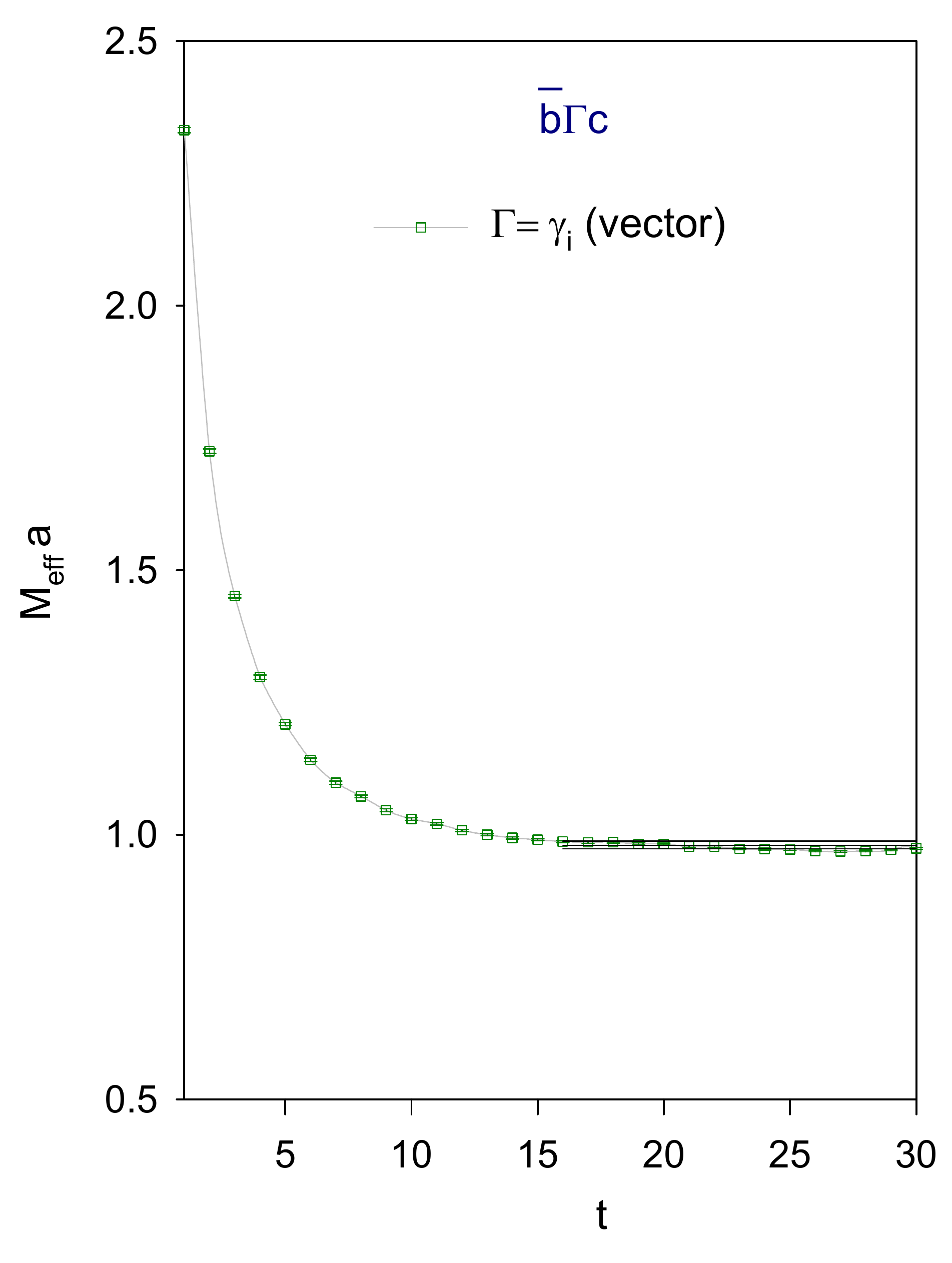}
%  \\ (a) & (b)
  \end{tabular}
  \caption{
    The time-correlation function and
    the effective mass of the meson interpolator $\bbar \gamma_i \c $.
  }
  \label{fig:Ct_meff_Gv_bc}
\end{figure}

\begin{figure}[H]
  \centering
  \begin{tabular}{@{}c@{}c@{}}
  \includegraphics[width=7cm,clip=true]{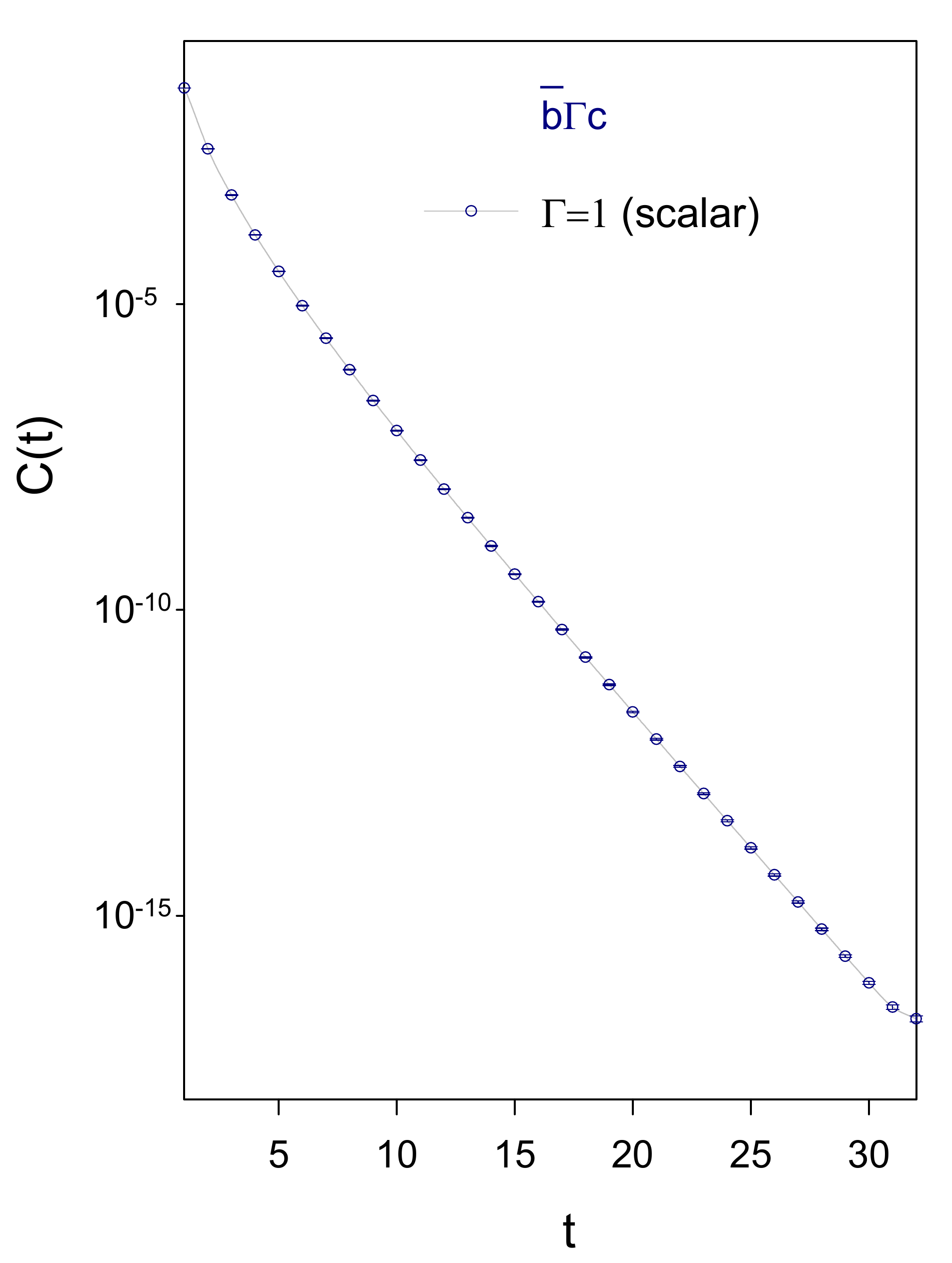}
&
  \includegraphics[width=7cm,clip=true]{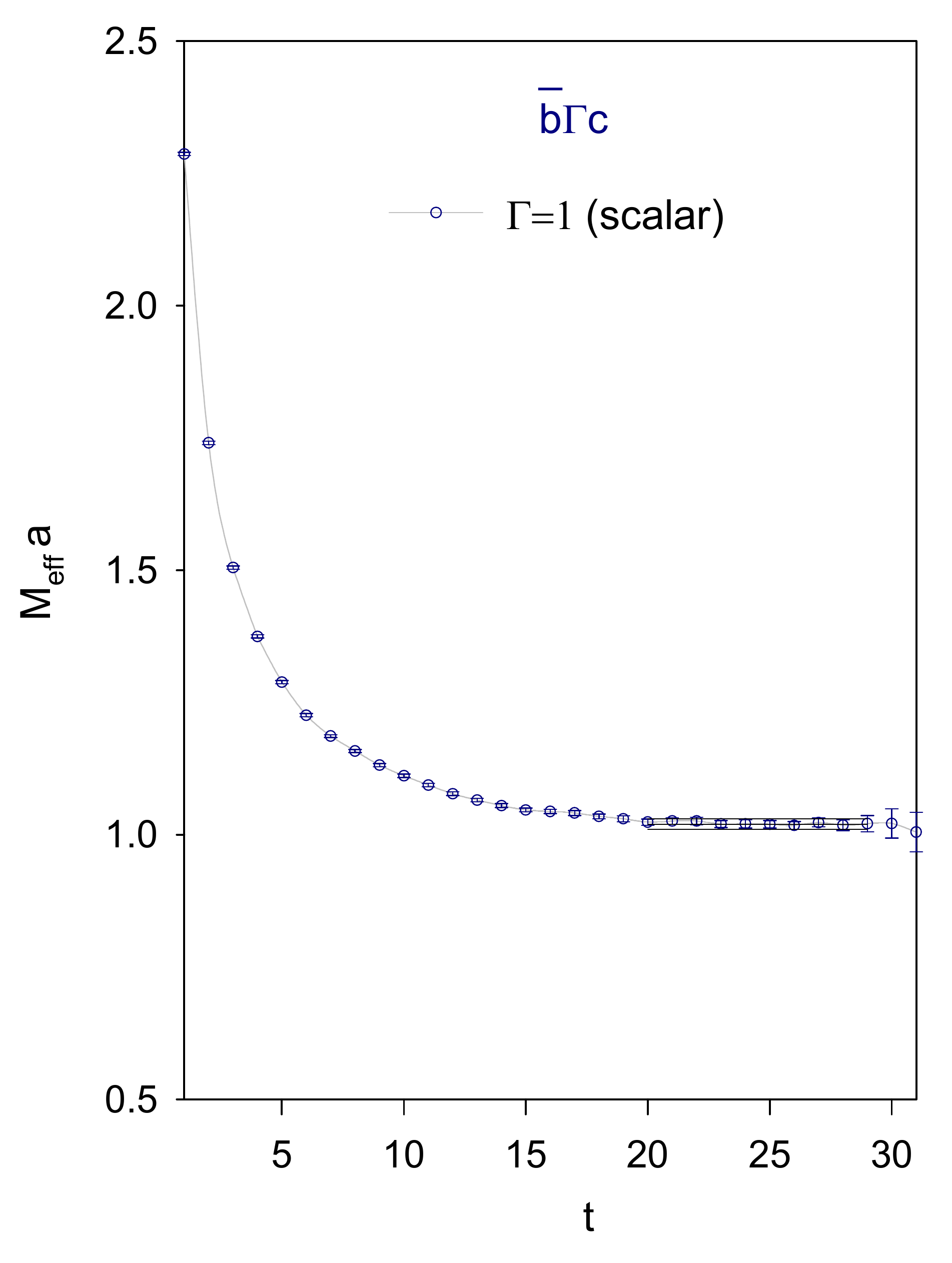}
%  \\ (a) & (b)
  \end{tabular}
  \caption{
    The time-correlation function and   
    the effective mass of the meson interpolator $\bbar \c $.
  }
  \label{fig:Ct_meff_G11_bc}
\end{figure}

\begin{figure}[H]
  \centering
  \begin{tabular}{@{}c@{}c@{}}
  \includegraphics[width=7cm,clip=true]{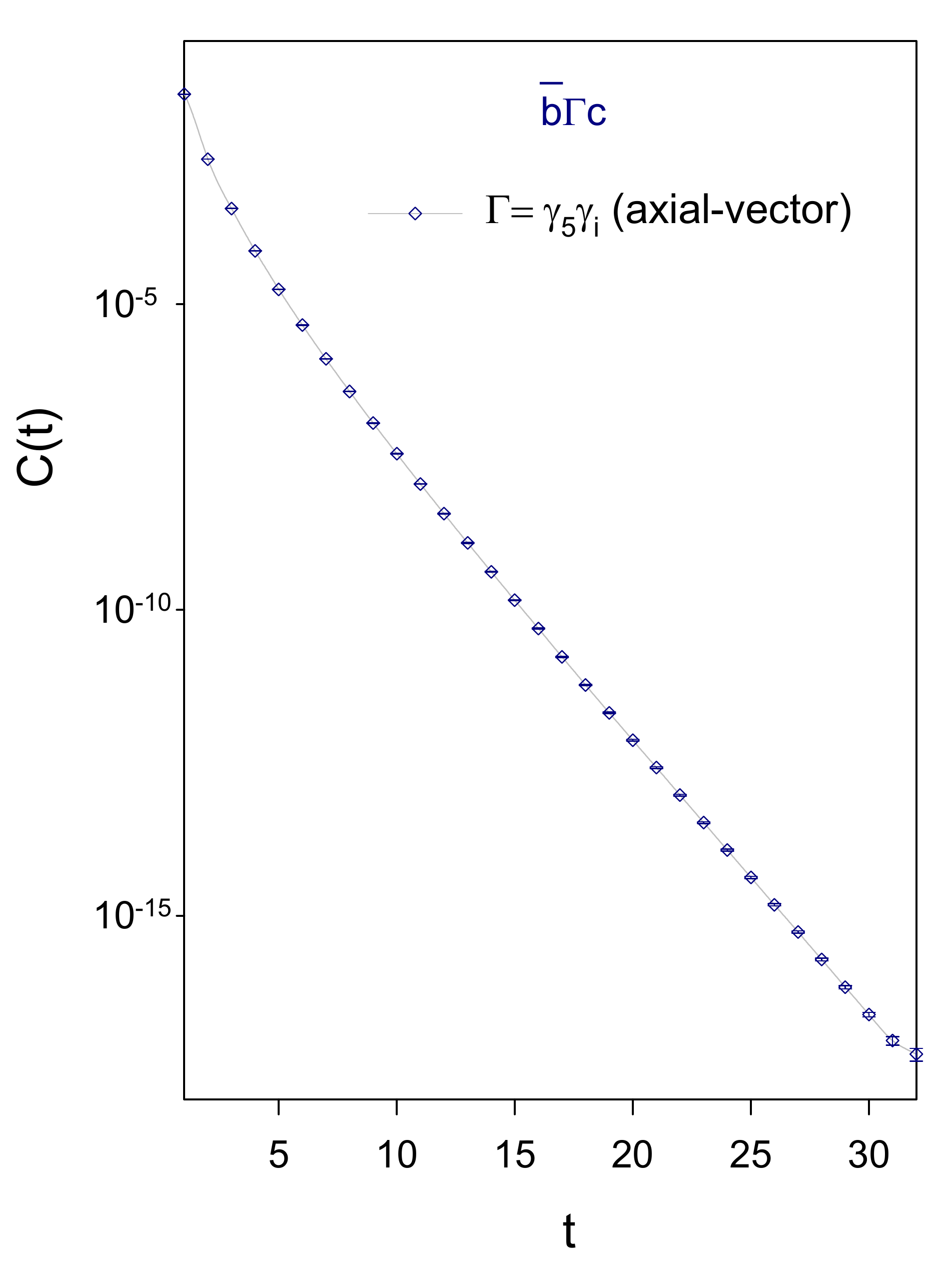}
&
  \includegraphics[width=7cm,clip=true]{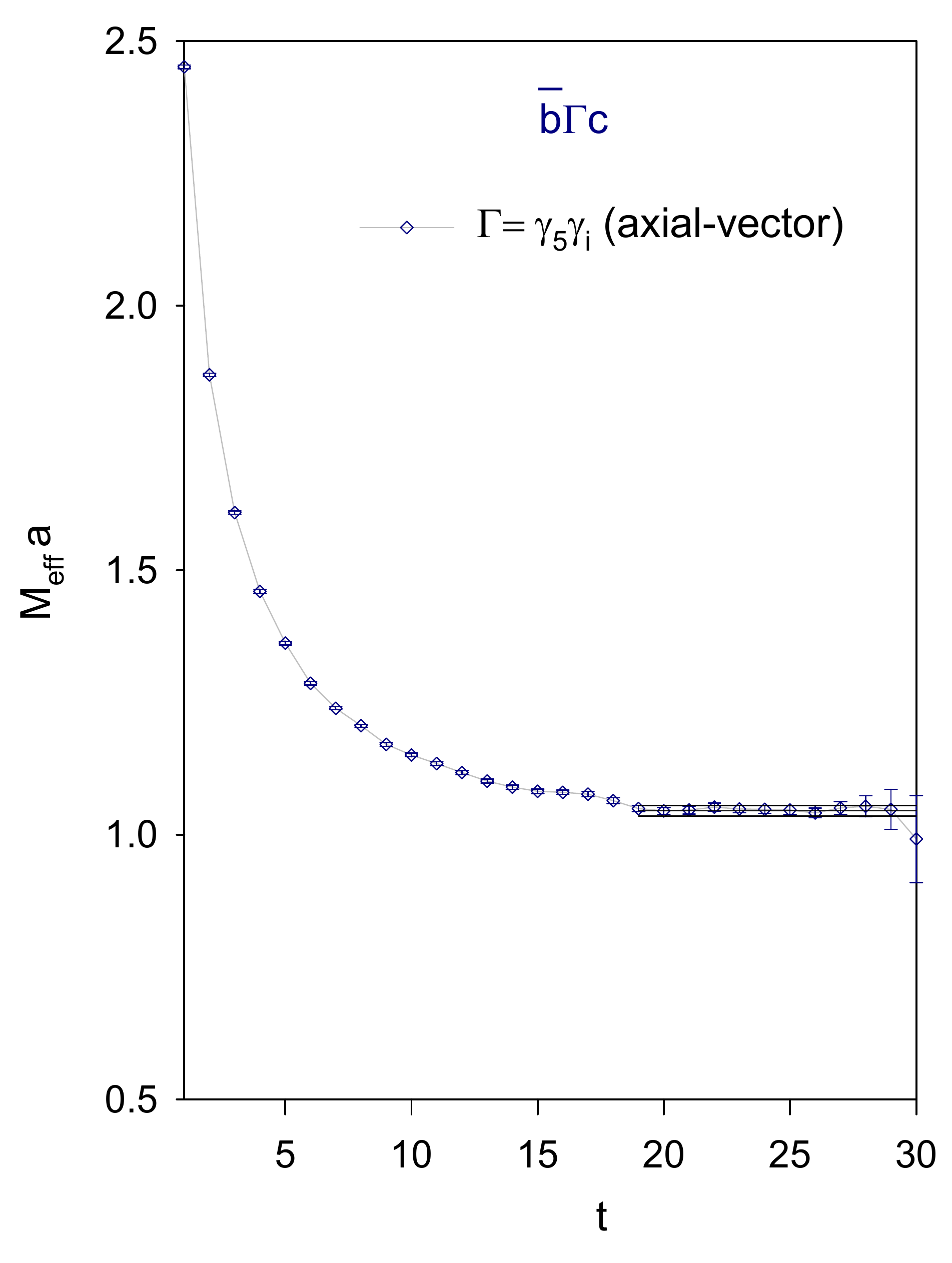}
%  \\ (a) & (b)
  \end{tabular}
  \caption{
    The time-correlation function and   
    the effective mass of the meson interpolator $\bbar \gamma_5 \gamma_i \c $.
  }
  \label{fig:Ct_meff_Ga_bc}
\end{figure}

\begin{figure}[H]
  \centering
  \begin{tabular}{@{}c@{}c@{}}
  \includegraphics[width=7cm,clip=true]{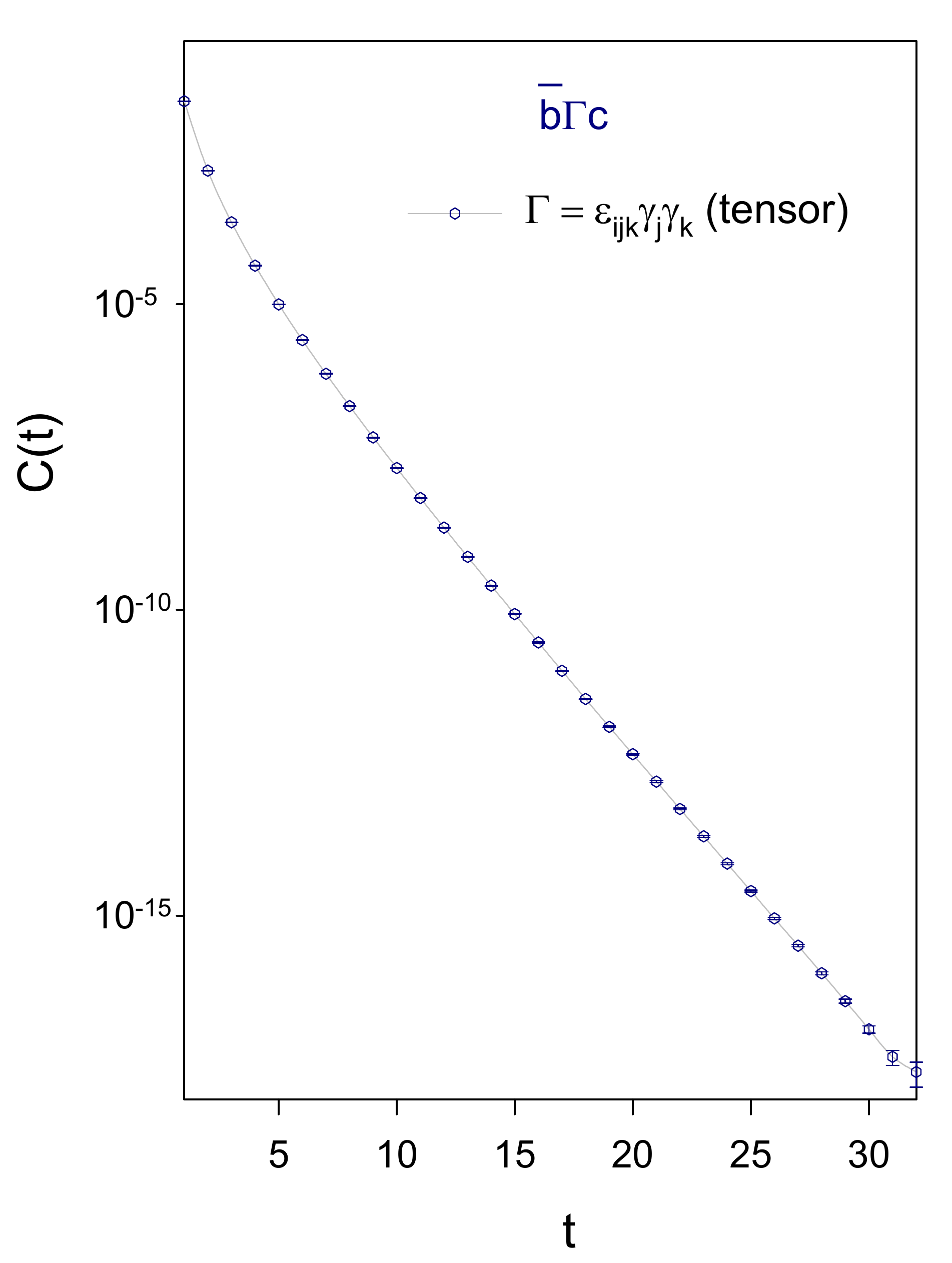}
&
  \includegraphics[width=7cm,clip=true]{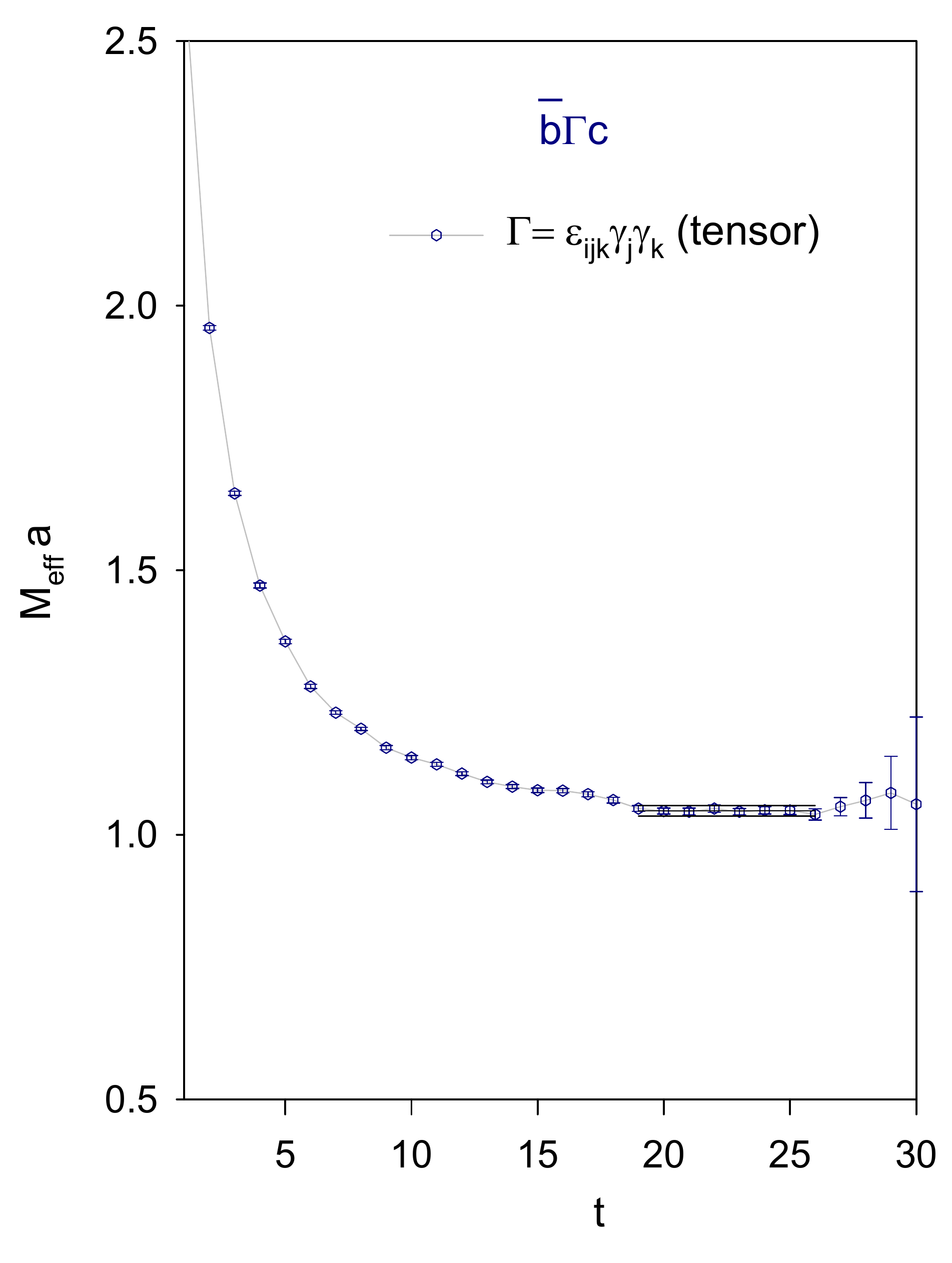}
%  \\ (a) & (b)
  \end{tabular}
  \caption{
    The time-correlation function and   
    the effective mass of the meson interpolator $\bbar \epsilon_{ijk} \gamma_j \gamma_k \c $.
  }
  \label{fig:Ct_meff_Gt_bc}
\end{figure}

\eject

\section{$C(t)$ and the effective mass of $\bbar \Gamma \s$}
\label{bbar_s}

\begin{figure}[H]
  \centering
  \begin{tabular}{@{}c@{}c@{}}
  \includegraphics[width=7cm,clip=true]{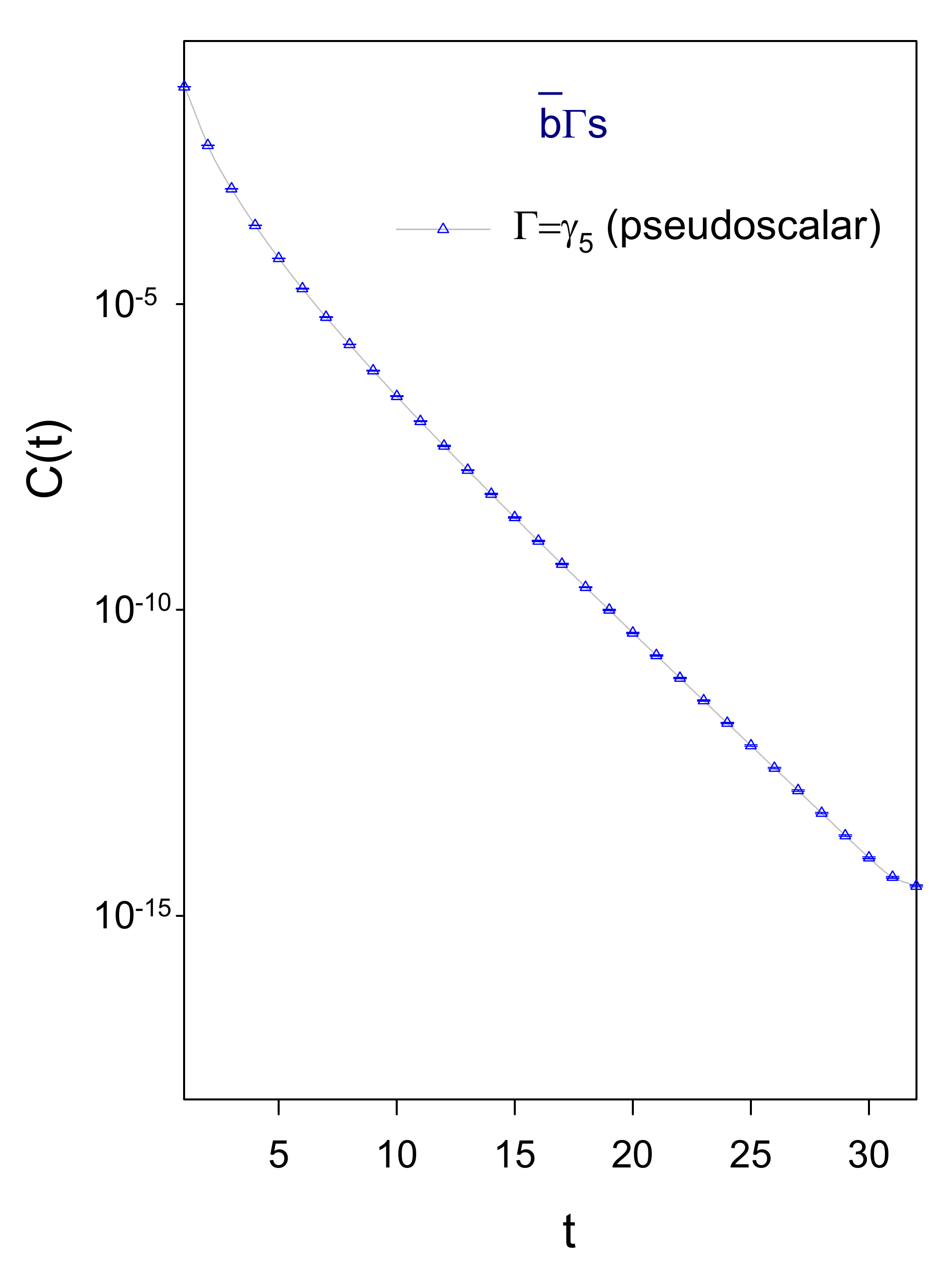}
&
  \includegraphics[width=7cm,clip=true]{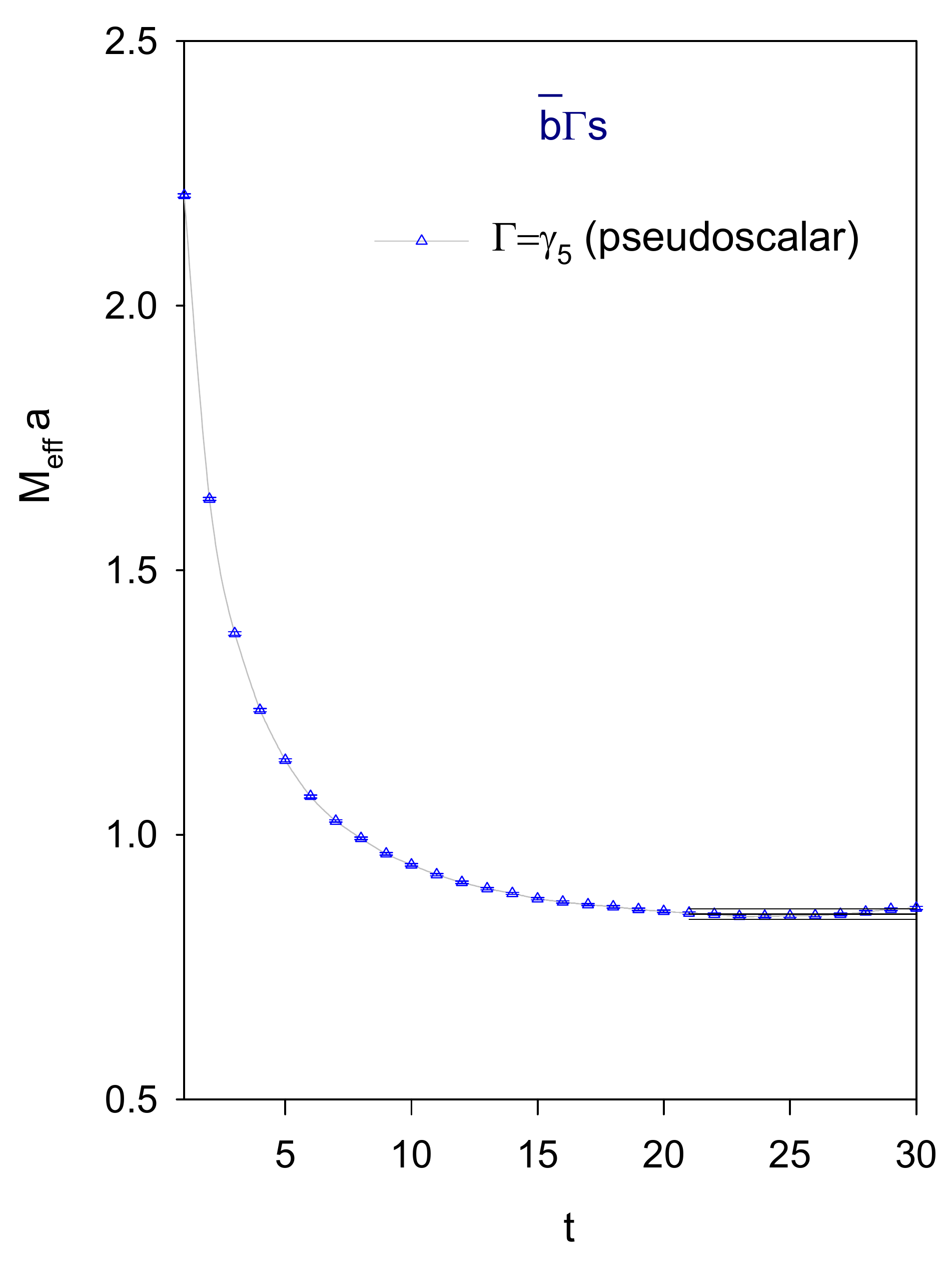}
%  \\ (a) & (b)
  \end{tabular}
  \caption{
    The time-correlation function and
    the effective mass of the meson interpolator $\bbar \gamma_5 \s $.
  }
  \label{fig:Ct_meff_G55_bs}
\end{figure}

\begin{figure}[H]
  \centering
  \begin{tabular}{@{}c@{}c@{}}
  \includegraphics[width=7cm,clip=true]{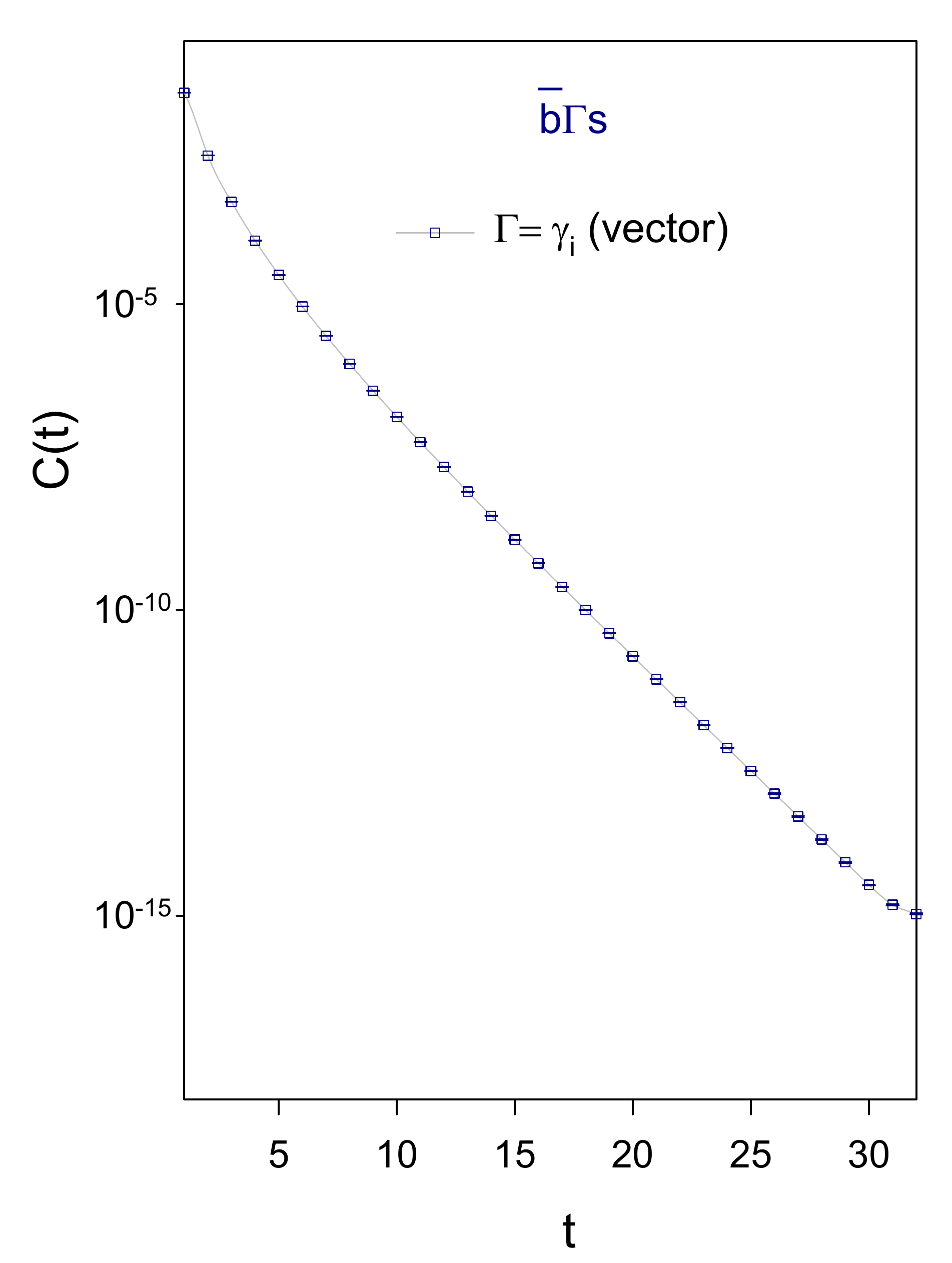}
&
  \includegraphics[width=7cm,clip=true]{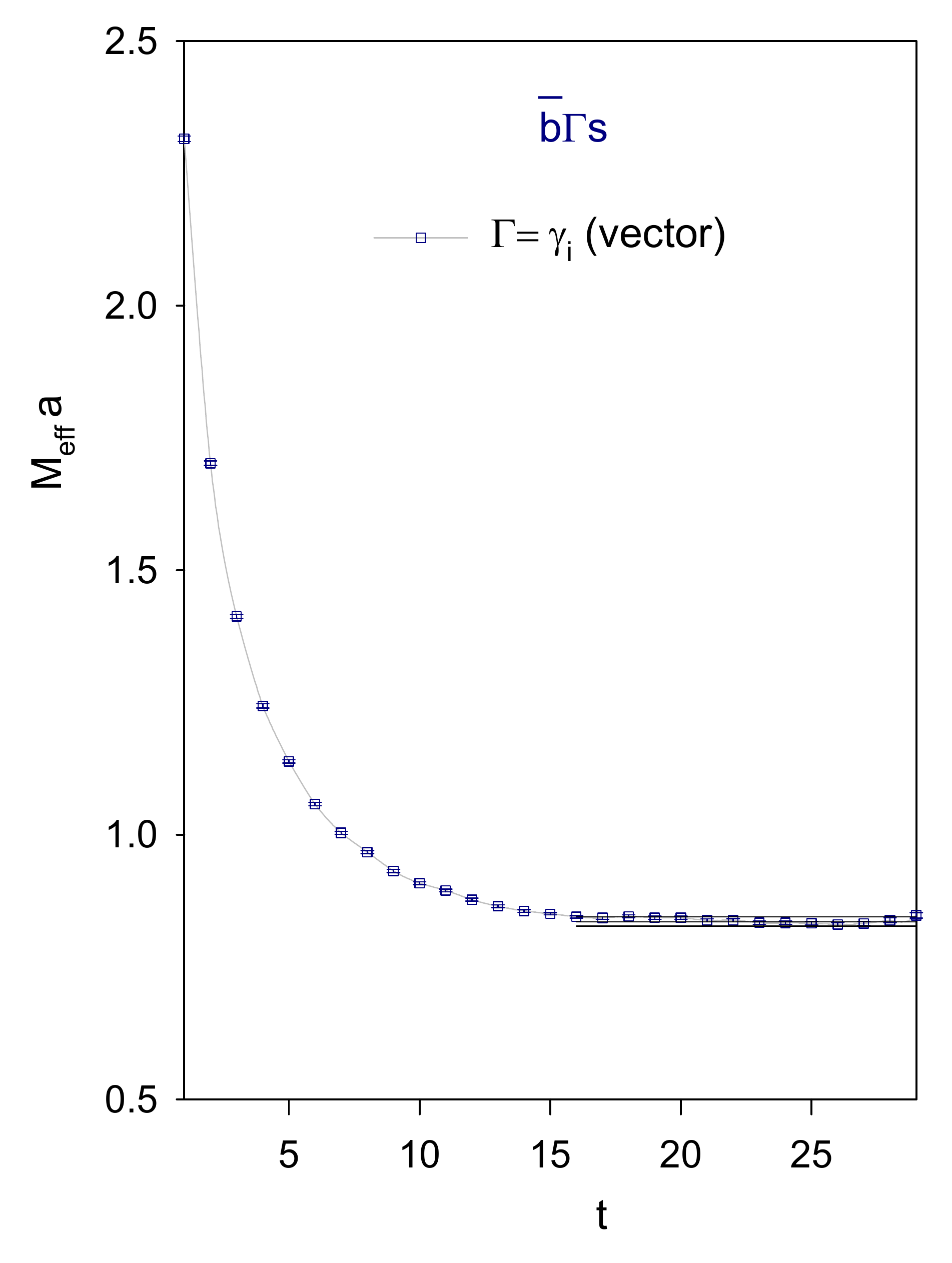}
%  \\ (a) & (b)
  \end{tabular}
  \caption{
    The time-correlation function and   
    the effective mass of the meson interpolator $\bbar \gamma_i \s $.
  }
  \label{fig:Ct_meff_Gv_bs}
\end{figure}

\begin{figure}[H]
  \centering
  \begin{tabular}{@{}c@{}c@{}}
  \includegraphics[width=7cm,clip=true]{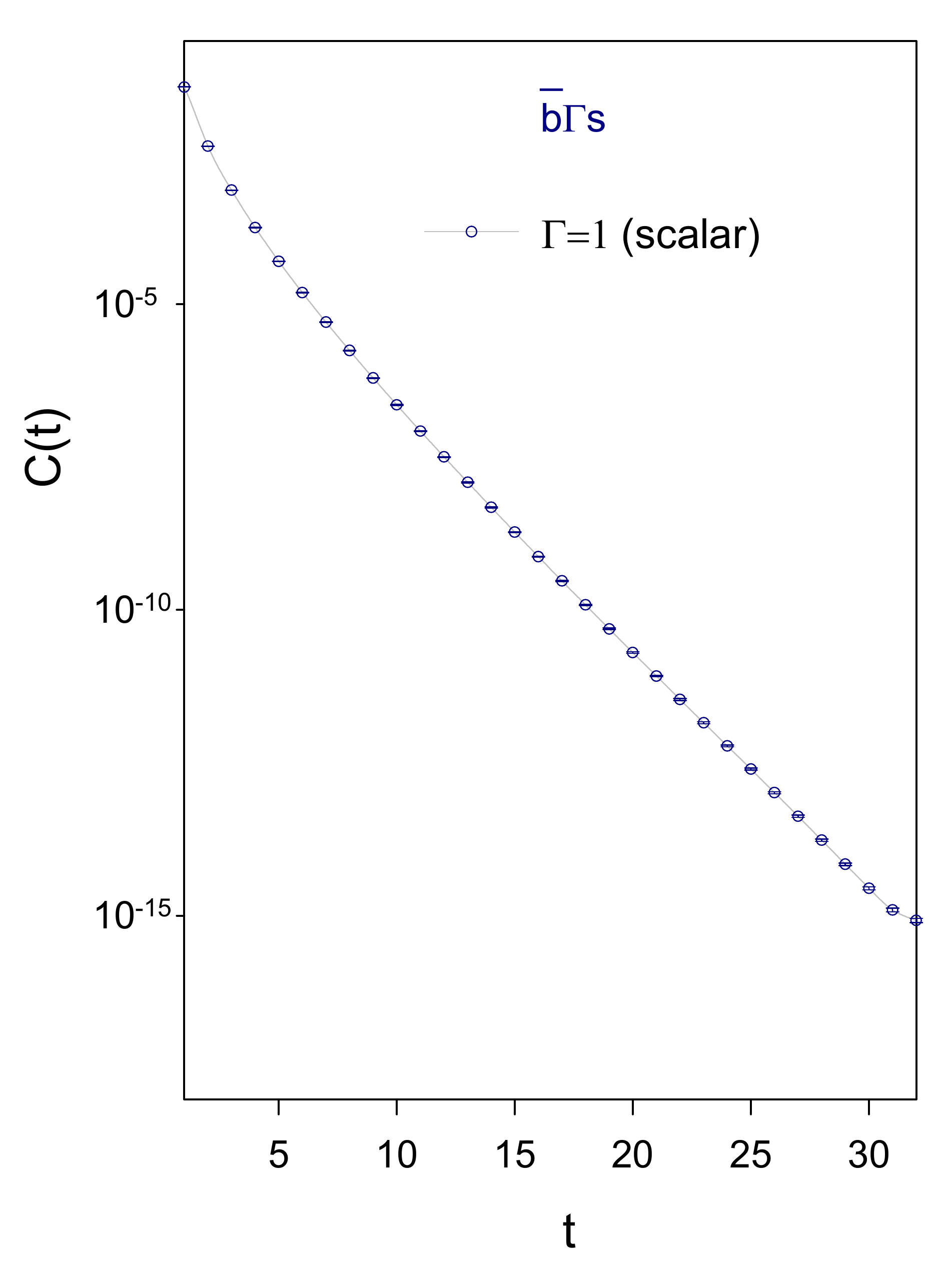}
&
  \includegraphics[width=7cm,clip=true]{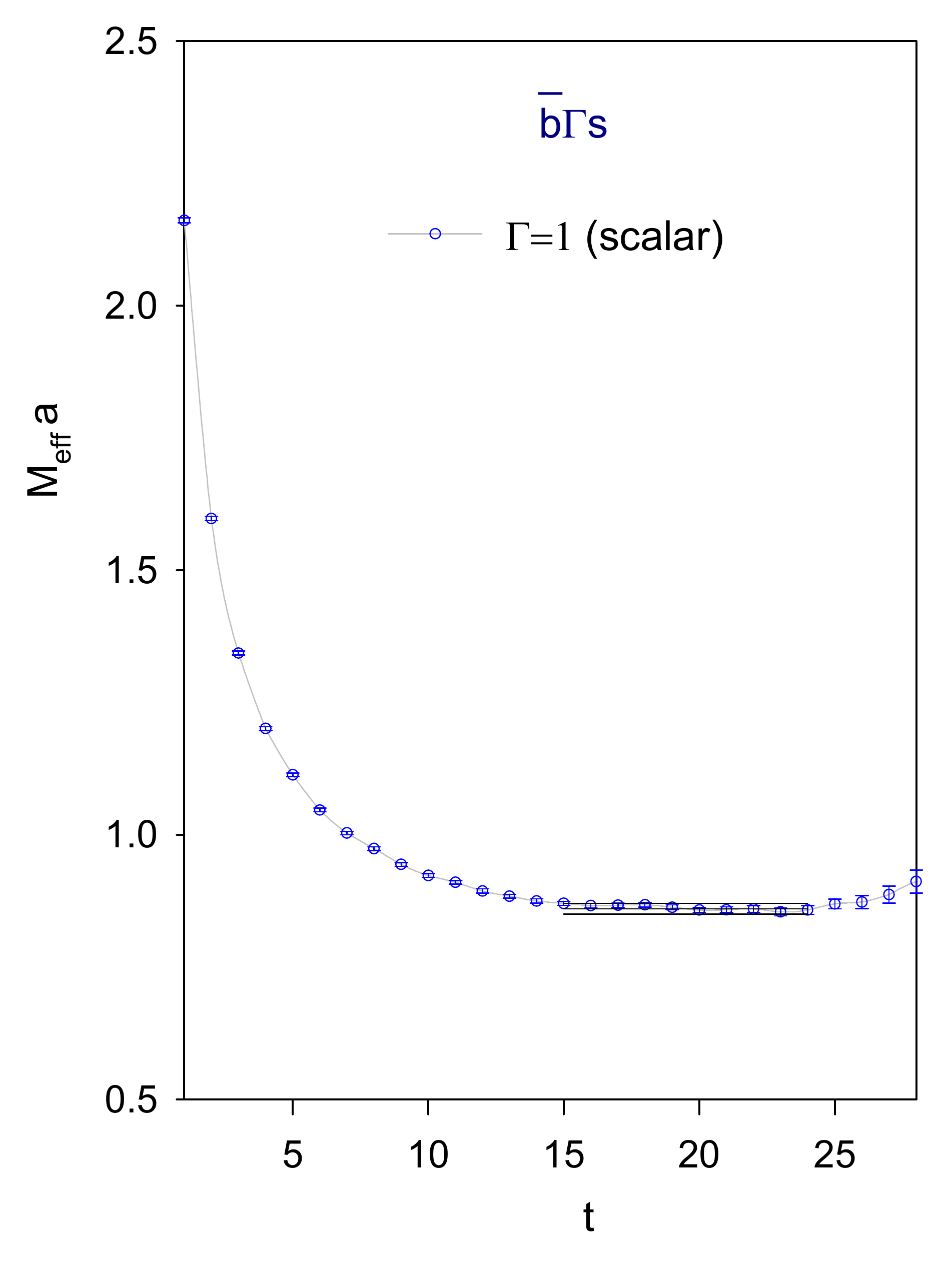}
%  \\ (a) & (b)
  \end{tabular}
  \caption{
    The time-correlation function and   
    the effective mass of the meson interpolator $\bbar \s $.
  }
  \label{fig:Ct_meff_G11_bs}
\end{figure}

\begin{figure}[H]
  \centering
  \begin{tabular}{@{}c@{}c@{}}
  \includegraphics[width=7cm,clip=true]{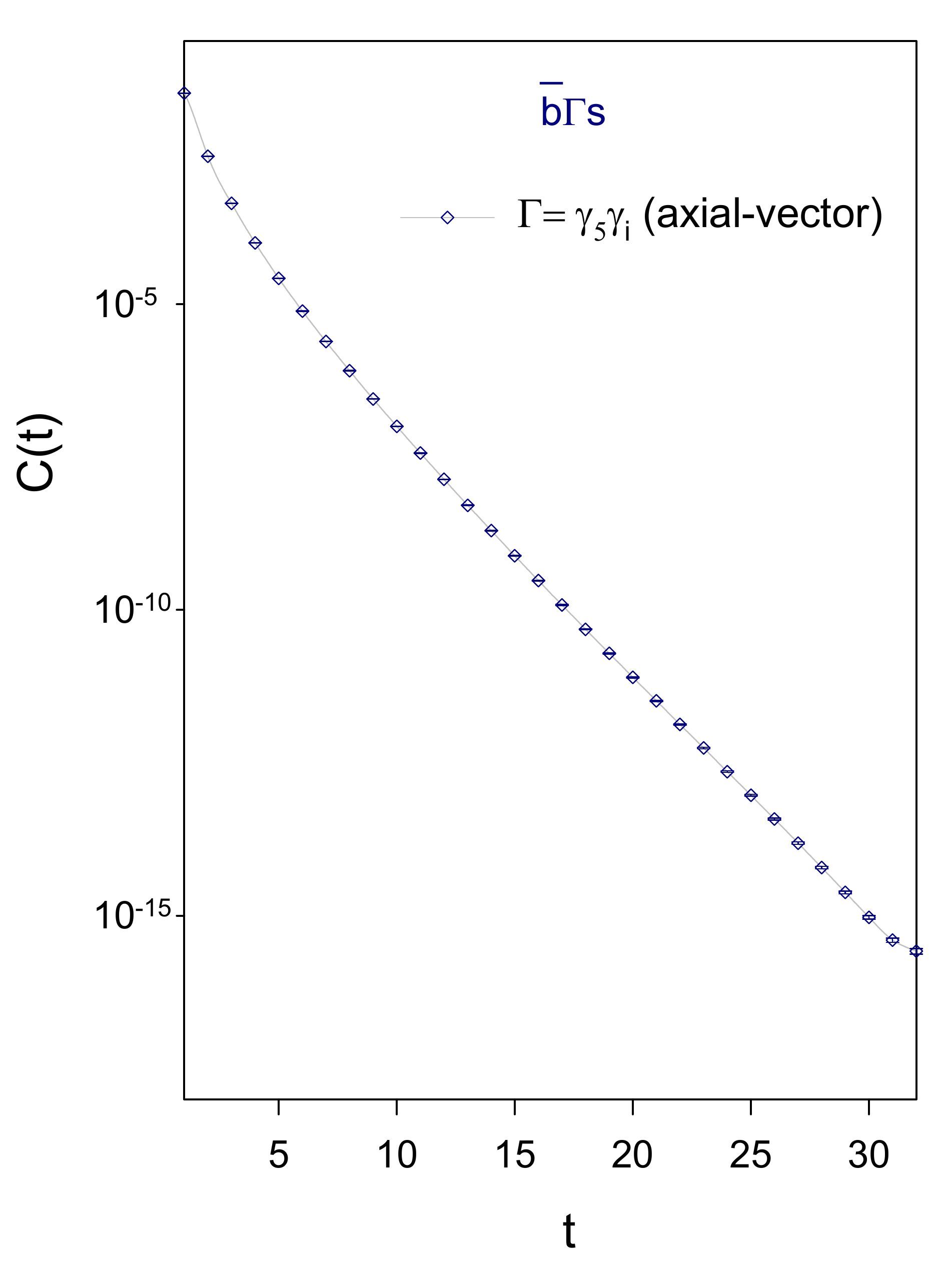}
&
  \includegraphics[width=7cm,clip=true]{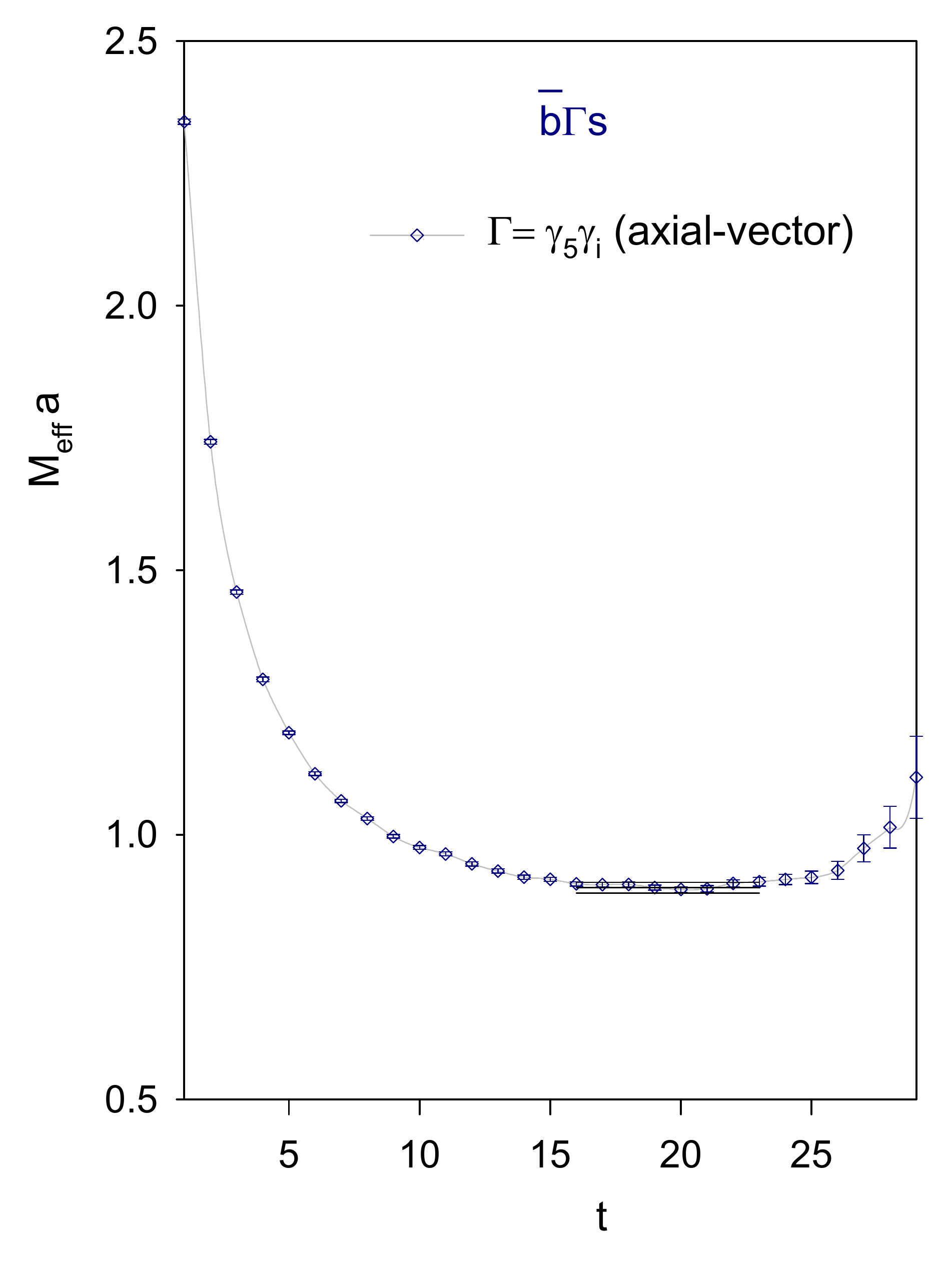}
%  \\ (a) & (b)
  \end{tabular}
  \caption{
    The time-correlation function and   
    the effective mass of the meson interpolator $\bbar \gamma_5 \gamma_i \s $.
  }
  \label{fig:Ct_meff_Ga_bs}
\end{figure}

\begin{figure}[H]
  \centering
  \begin{tabular}{@{}c@{}c@{}}
  \includegraphics[width=7cm,clip=true]{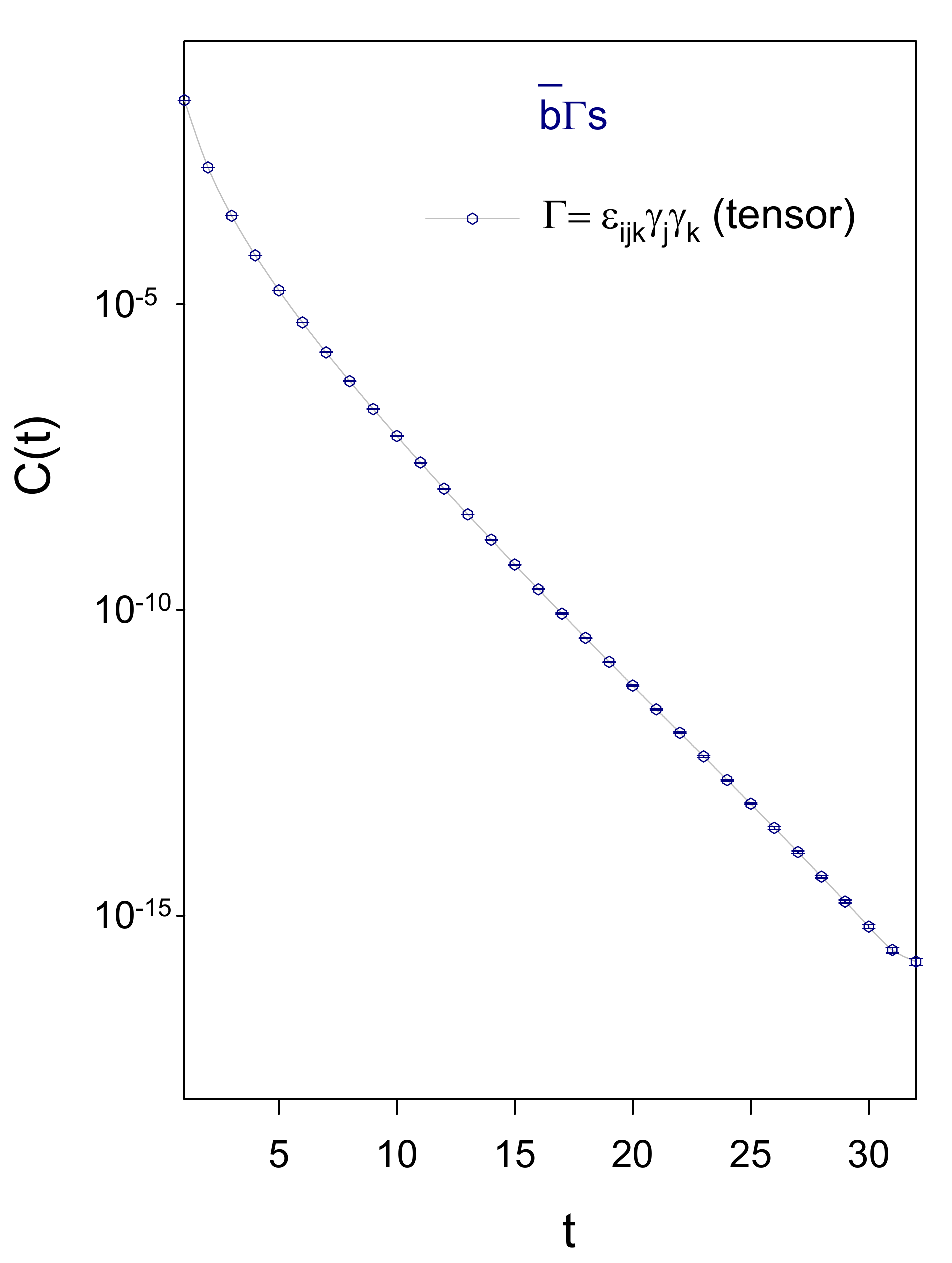}
&
  \includegraphics[width=7cm,clip=true]{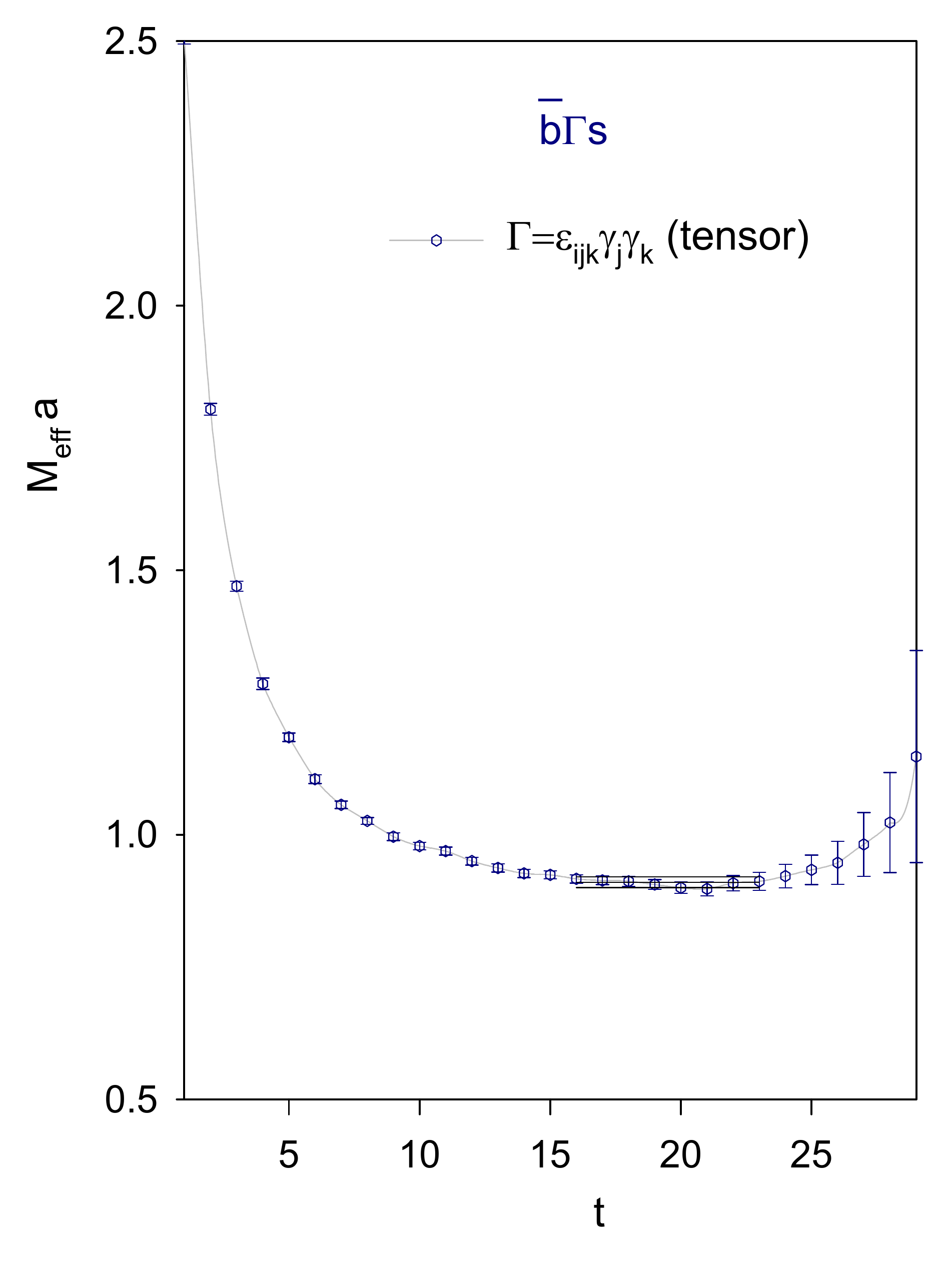}
%  \\ (a) & (b)
  \end{tabular}
  \caption{
    The time-correlation function and   
    the effective mass of the meson interpolator $\bbar \epsilon_{ijk} \gamma_j \gamma_k \s $.
  }
  \label{fig:Ct_meff_Gt_bs}
\end{figure}

%\end{appendices}

\end{document}